\documentclass[10pt,english]{article}
\usepackage[T1]{fontenc}
\usepackage[latin9]{inputenc}
\usepackage{geometry}
\usepackage{color}
\usepackage{float}
\usepackage{amsmath}
\usepackage{amssymb}
\usepackage{graphicx}
\usepackage{esint}

\makeatletter

\providecommand{\tabularnewline}{\\}
\floatstyle{ruled}
\newfloat{algorithm}{tbp}{loa}
\providecommand{\algorithmname}{Algorithm}
\floatname{algorithm}{\protect\algorithmname}

\usepackage{amsmath}
\usepackage{amssymb}

\usepackage{cite}

\usepackage{hyperref}
\usepackage[T1]{fontenc}
\usepackage{lmodern}

\usepackage{lineno}

\usepackage{microtype}
\DisableLigatures[f]{encoding = *, family = * }

\usepackage[labelfont=bf,labelsep=period,justification=raggedright]{caption}

\makeatletter
\renewcommand{\@biblabel}[1]{\quad#1.}
\makeatother

\date{}

\makeatother

\usepackage{babel}
\begin{document}
\global\long\def\E{\mathbb{E}}

\begin{flushleft} 
{\Large \textbf{A shotgun sampling solution for the common input problem in neural connectivity inference}} 
\\ Daniel Soudry*$^{1}$, Suraj Keshri$^{2}$, Patrick Stinson$^{1}$, Min-hwan Oh$^{2}$, Garud Iyengar$^{2}$, Liam Paninski$^{1}$
\\ \bf{1} Department of Statistics, Department of Neuroscience, the Center for Theoretical Neuroscience, the Grossman Center for the Statistics of Mind, the Kavli Institute for Brain Science, and the NeuroTechnology Center, Columbia University, New York, NY, USA
\\ \bf{2} Department of Industrial Engineering and Operations Research, Columbia University, New York, NY, USA
\\ $\ast$ E-mail: daniel.soudry@gmail.com
\end{flushleft}

\section*{Abstract}

Inferring connectivity in neuronal networks remains a key challenge
in statistical neuroscience. The ``common input'' problem presents
the major roadblock: it is difficult to reliably distinguish causal
connections between pairs of observed neurons from correlations induced
by common input from unobserved neurons. Since available recording
techniques allow us to sample from only a small fraction of large
networks simultaneously with sufficient temporal resolution, naive
connectivity estimators that neglect these common input effects are
highly biased. 

This work proposes a ``shotgun'' experimental design, in which we
observe multiple sub-networks briefly, in a serial manner. Thus, while
the full network cannot be observed simultaneously at any given time,
we may be able to observe most of it during the entire experiment.
Using a generalized linear model for a spiking recurrent neural network,
we develop scalable approximate Bayesian methods to perform network
inference given this type of data, in which only a small fraction
of the network is observed in each time bin. 

We demonstrate in simulation that, using this method: (1) The shotgun
experimental design can eliminate the biases induced by common input
effects. (2) Networks with thousands of neurons, in which only a small
fraction of the neurons is observed in each time bin, could be quickly
and accurately estimated. (3) Performance can be improved if we exploit
prior information about the probability of having a connection between
two neurons, its dependence on neuronal cell types (e.g., Dale's law),
or its dependence on the distance between neurons.

\section*{Author Summary}

Optically imaging the activity in a neuronal circuit is limited by
the scanning speed of the imaging device. Therefore, commonly, only
a small fixed part of the circuit is observed during the entire experiment.
However, in such an experiment it can be hard to infer from the observed
activity patterns whether (1) a neuron A directly affects neuron B,
or (2) another, unobserved neuron C affects both A and B. To deal
with this issue, we propose a \textquotedblleft shotgun\textquotedblright{}
observation scheme, in which, at each time point, we randomly observe
a small percentage of the neurons from the circuit. And so, no neuron
remains completely hidden during the entire experiment and we can
eventually distinguish between cases (1) and (2). Previous inference
algorithms cannot handle efficiently so many missing observations.
We therefore develop a scalable algorithm, for data acquired using
shotgun observation scheme - in which only a small fraction of the
neurons are observed in each time bin. Using this kind of simulated
data, we show the algorithm is able to quickly infer connectivity
in spiking recurrent networks with thousands of neurons.

\section{Introduction}

The simultaneous activity of hundreds - and even thousands - of neurons
is now being routinely recorded in a wide range of preparations. The
number of recorded neurons is expected to grow exponentially over
the years \cite{Stevenson2011}. This, in principle, provides the
opportunity to infer the ``functional connectivity'' of neuronal
networks, \emph{i.e. }a statistical estimate of how neurons are affected
by each other, and by stimulus. The ability to accurately estimate
large, possibly time-varying, neural connectivity diagrams would open
up an exciting new range of fundamental research questions in systems
and computational neuroscience \cite{Alivisatos2012}. Therefore,
the task of estimating connectivity from neural activity can be considered
one of the central problems in statistical neuroscience - attracting
much attention in recent years (e.g., see \cite{NYK03b,Paninski2004,PILL05a,Rigat06,Nykamp2007,Pillow2007,Pillow2008,Mishchenko2011c,Stetter2012,Paninski2012,Pakman2012,Gerhard2013,Lutcke2013,Memmesheimer2014,Fletcher2014}
and references therein).

Perhaps the biggest challenge for inferring neural connectivity from
activity --- and in more general network analysis --- is the presence
of hidden nodes which are not observed directly \cite{Pillow2007,Nykamp2008,Vidne08,Mishchenko2011b,Turaga2013a,Romano2014,Tyrcha2014}.
Despite swift progress in simultaneously recording activity in massive
populations of neurons, it is still beyond the reach of current technology
to simultaneously monitor a complete network of spiking neurons at
high temporal resolution (though see \cite{Ahrens2013} for some impressive
recent progress in that direction). Since estimation of functional
connectivity relies on the analysis of the inputs to target neurons
in relation to their observed spiking activity, the inability to monitor
all inputs can result in persistent errors in the connectivity estimation
due to model miss-specification. More specifically, ``common input\textquotedbl{}
errors --- in which correlations due to shared inputs from unobserved
neurons are mistaken for direct, causal connections --- plague most
naive approaches to connectivity estimation. Developing a robust approach
for incorporating the latent effects of such unobserved neurons remains
an area of active research in connectivity analysis \cite{Pillow2007,Nykamp2008,Vidne08,Mishchenko2011b,Turaga2013a,Romano2014,Tyrcha2014}.

In this paper we propose an experimental design which greatly ameliorates
these common-input problems. The idea is simple: if we cannot observe
all neurons in a network simultaneously, maybe we can instead observe
many overlapping sub-networks in a serial manner over the course of
a long experiment. Then we use statistical techniques to patch the
full estimated network back together, analogously to ``shotgun\textquotedbl{}
genetic sequencing \cite{Venter1998}. Obviously, it is not feasible
to purposefully sample from many distinct sub-networks at many different
overlapping locations using multi-electrode recording arrays, since
multiple re-insertions of the array would lead to tissue damage. However,
fluorescence-based imaging of neuronal calcium \cite{RKFS08,Grewe2010}
or voltage \cite{Hochbaum2014} dynamics make this approach experimentally
feasible. In the ideal experiment, we would target our observations
so they fully cover a neuronal circuit together with all its inputs
(Figure \ref{fig:observation schemes}). However, in each time step,
we would only observe a random fraction of all targeted neurons.

In this shotgun approach only a small fraction of the network is observed
at any single time. However, connectivity estimation with missing
observations is particularly challenging since exact Bayesian inference
with unobserved spikes is generally intractable. Therefore, Markov
Chain Monte-Carlo (MCMC) methods have been previously used to infer
the unobserved activity (spikes) by sampling \cite{Pillow2007,Mishchenko2011c,Mishchenko2011a}.
However, such methods typically do not scale well for large networks.
MCMC methods are computationally expensive (though, sometimes highly
parallelizable), and usually take a long time to converge. Variational
approaches \cite{Fletcher2014}, may speed up inference, but have
not been shown to be robust to missing observations.

Fortunately, as we show here, given a shotgun sampling scheme, it
is not actually required to infer the unobserved spikes. We considerably
simplify the loglikelihood using the expected loglikelihood approximation\cite{Park2011,Sadeghi12,Ramirez2013},
and a generalized Central Limit Theorem (CLT) \cite{Diaconis1984}
argument to approximate neuronal input as a Gaussian variable when
the size of the network is large. The simplified loglikelihood depends
only on the empiric second order statistics of the spiking process.
Importantly, these sufficient statistics can be calculated even with
partial observations, by simply ``ignoring'' any unobserved activity. 

In order to obtain an accurate estimation of the connectivity, this
simplified loglikelihood can be very efficiently maximized --- together
with various types of prior information. An abundance of such prior
information is available on both connection probabilities and synaptic
weight distributions as a function of cell location and identity \cite{Song2005}.
In addition, cutting edge labeling and tissue preparation methods
such as Brainbow \cite{Lichtman2008} and CLARITY \cite{Chung2013}
are beginning to provide rich anatomical data about ``potential connectivity\textquotedbl{}
(e.g., the degree of coarse spatial overlap between a given set of
dendrites and axons) that can be incorporated into these priors. Exploiting
this prior information can greatly improve inference quality, as was
already demonstrated in previous network inference papers \cite{Pillow2007,Mishchenko2011c,Jonas2014}.

We present numerical results which demonstrate the effectiveness of
our approach on a on a simulated recurrent network of spiking neurons.
First, we demonstrate that the shotgun experimental design can largely
eliminate the biases induced by common input effects, as originally
intended (Figure \ref{fig: Common Input Problem}). Specifically,
we show that we can quickly infer the network connectivity for large
networks, with a low fraction of neurons observed in each time bin.
For example, our algorithm can be used to infer the connectivity of
a sparse network with $O\left(10^{3}\right)$ neurons and $O\left(10^{5}\right)$
connections, given $O\left(10^{5}\right)$ timebins of spike data
in which only $10\%-20\%$ of the neurons are observed in each time
bin, after running less than a minute on a standard laptop (Figure
\ref{fig:Sparsity-1}). This is achieved by assuming only a sparsity
inducing prior on the weights. Our parameter scans suggest that our
method could be used for arbitrarily low observation ratios (Figure
\ref{fig:Parameter-scans P_obs T}) and arbitrarily high number of
neurons (Figure \ref{fig:Parameter-scans N-T}), given long enough
experiments. Then, we demonstrate (Figure \ref{fig:Prior Results})
the usefulness of the following additional pieces of prior information:
(1) Dale's Law - all outgoing synaptic connections from the same neuron
have the same sign. (2) Neurons have several types - and connection
strength between two types is affected by the identity of these types.
(3) The probability of connection between neurons is distance dependent.
Performance can also be improved by using external stimuli (Figure
\ref{fig:The-effect-of-stimulus}), similar to the stimulus that can
be induced by persistent light sensitive ion channels \cite{Berndt2009}.

\section{Results \label{sec:-Results}}

Suppose we wish to perform an experiment in order to measure the functional
connectivity in a neuronal circuit by simply observing its activity.
In this experiment, we optically capture the neuronal spikes, visible
through the use of genetically encoded calcium \cite{Wallace2008}
or voltage \cite{Hochbaum2014} indicators. Current imaging methods
(\emph{e.g.}, two-photon or light sheet microscopy) record this neuronal
activity by scanning through the neuron tissue. The scanning protocol,
and consequently, the identity of the observed neurons, have various
constraints. An important constraint is the maximal scanning speed
of the recording device. Since the scanning speed is limited, we cannot
observe all the neurons in the circuit all the time with sufficient
spatio-temporal resolution. We must decide where to focus our observations. 

Commonly, observations are focused on a fixed subset of the neurons
in the circuit. However, as we will show here, this procedure can
generate persistent errors due to unobserved common inputs, that generate
spurious correlations in the network activity. To prevent this, we
propose a shotgun observation approach, in which we scan the network
at a random order (Figure \ref{fig:observation schemes}, \emph{top}).
Thus, at each time bin of experimental data, only a random fraction
of the neurons in the network are observed. Intuitively, the main
goal of such an approach is to evenly distribute our observations
on all pairs of neurons, so that all the relevant spike correlations
could be eventually estimated. Note that simple serial scanning of
the circuit in contiguous blocks does not accomplish this purpose
(Figure \ref{fig:observation schemes}, \emph{bottom}). However, several
other scanning schemes do (random or deterministic) - including the
random shotgun scheme, on which we focus here.

In this section we test the shotgun scheme on simulated $T$-long
spike data from a recurrent spiking neural network with $N$ neurons.
Specifically, we use a Generalized Linear network Model (GLM \cite{BRIL88,Rigat06,Pillow2007},
see Eqs. \ref{eq: logistic}-\ref{eq: U}) with synaptic connectivity
matrix $\mathbf{W}$. Typically, $\mathbf{W}$ is sparse, so $p_{\mathrm{conn}}$,
the average probability that two neurons are directly connected to
each other, is low. To infer $\mathbf{W}$ we use the inference method
described in section \ref{sec:Bayesian Inference}. This approximate
Bayesian method can exploit various priors (section \ref{sub:Priors}),
such as the sparsity of $\mathbf{W}$, to improve estimation quality.
We define $p_{\mathrm{obs}}$ as the fraction of neurons observed
at each timebin, \emph{i.e.}, the observation probability in the shotgun
scheme. For a general list of basic notation, see Table \ref{tab:Basic-notation}.

We start (section \ref{sub:Dealing-with-common}) with a qualitative
demonstration that the shotgun approach can be used without the usual
persistent bias resulting from common inputs. Then, in section \ref{sub:Connectivity-estimation-with},
we perform quantitative tests to show that our estimation method is
effective for various network sizes, observation probabilities, and
stimulus input amplitudes. This is done using only a sparsity prior.
Lastly, in section \ref{sub:Additional-priors}, we show how more
advanced priors can improve estimation quality.

We conclude (section \ref{sec:Discussion}) that the limited scanning
speed of the any imaging device recording neuronal activity is not
a fundamental barrier which prevents consistent estimation of functional
connectivity.

\subsection{Dealing with common input\label{sub:Dealing-with-common}}

In this section we use a toy network with $N=50$ neurons to visualize
the common input problem, and its suggested solution - the ``shotgun''
approach. 

\begin{figure}[H]
\begin{centering}
\begin{center}
\includegraphics[width=0.8\columnwidth]{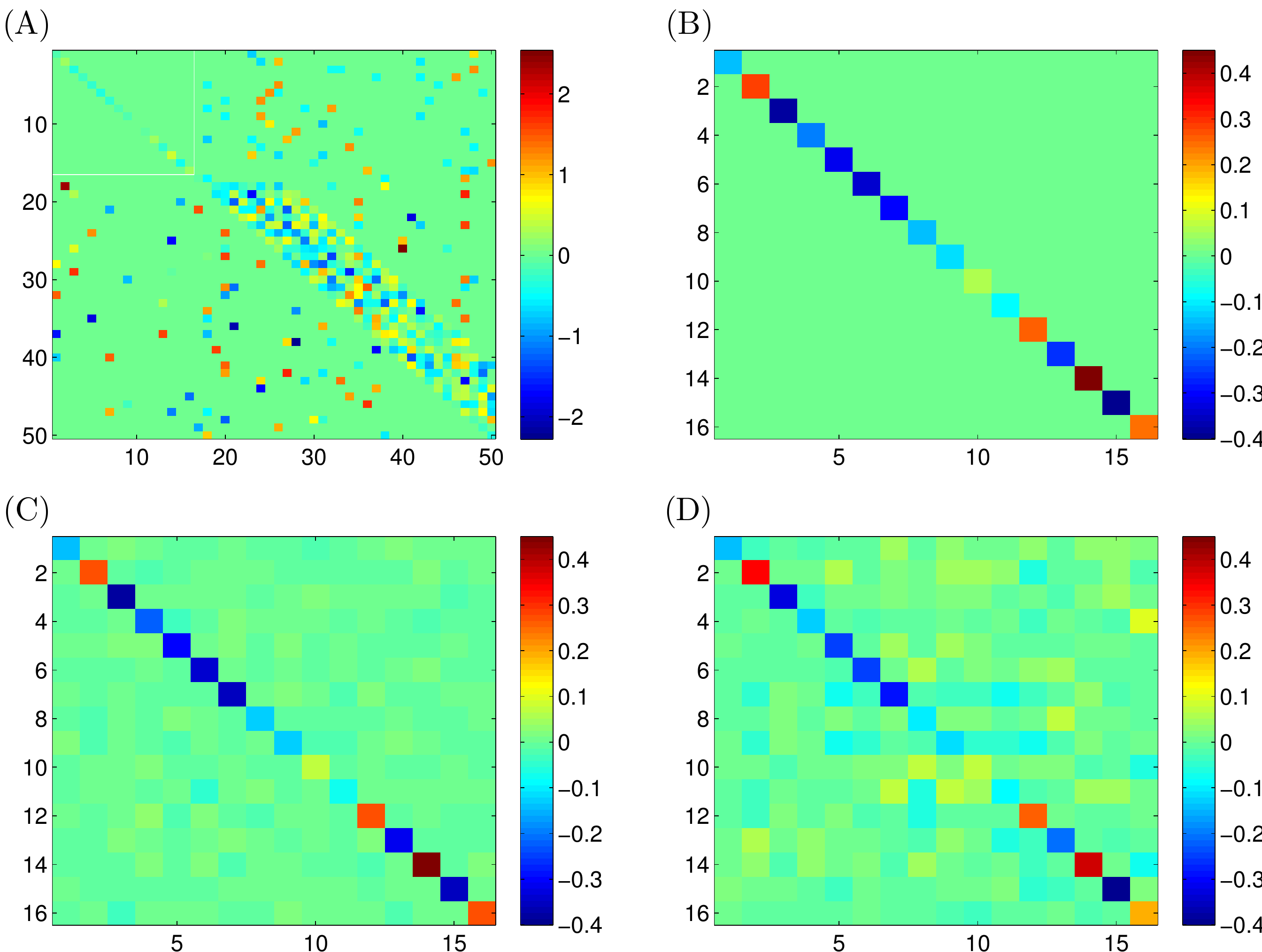}
\par\end{center}
\par\end{centering}

\protect\caption{Visualization of the persistence of the common input problem, despite
a large amount of spiking data ($T=5\cdot10^{6}$), and its suggested
solution - the shotgun approach. \textbf{(A)} The true connectivity
- the weight matrix $\mathbf{W}$ of a network with $N=50$ neurons.\textbf{
(B)} A zoomed-in view of the top 16 neurons in A (white rectangle
in A). \textbf{(C)} The same zoomed-in view of the top 16 neurons
in the ML estimate of the weight matrix $\mathbf{W}$ (Eq. \ref{eq: W_MLE}),
where we used shotgun observation scheme on the whole network, with
a random observation probability of $p_{\mathrm{obs}}=16/50$. \textbf{(D)}
The ML estimator of the weight matrix $\mathbf{W}$ of the top 16
neurons if we observe only these neurons. Note the unobserved neurons
cause false positives in connectivity estimation. These ``spurious
connections'' do not vanish even though we have a large amount of
spike data. In contrast, the shotgun approach (C), does not have these
persistent errors, since it spreads the same number of observations
evenly over the network. $b_{i}\sim\mathcal{N}\left(-0.5,0.1\right)$.\label{fig: Common Input Problem}}
\end{figure}

Errors caused by common inputs are particularly troublesome for connectivity
estimation, since they can persist even when $T\rightarrow\infty$.
Therefore, for simplicity, in this case we work in a regime where
the experiment is long and data is abundant ($T=5\cdot10^{6}$). In
this regime, any prior information we have on the connectivity becomes
unimportant so we simply use the Maximum Likelihood (ML) estimator
(Eq. \ref{eq: W_MLE}). We chose the weight matrix $\mathbf{W}$ to
illustrate a ``worst-case\textquotedbl{} common input condition (Figure
\ref{fig: Common Input Problem}A). Note that the upper-left third
of $\mathbf{W}$ is diagonal (Figure \ref{fig: Common Input Problem}B):
\emph{i.e.}, neurons $i=1,\ldots,16$ share no connections to each
other, other than the self-connection terms $W_{i,i}$. However, we
have seeded this $\mathbf{W}$ with many common-input motifs, in which
neurons $i$ and $j$ (with $i,j\leq16$) both receive common input
from neurons $k$ with $k\geq17$. 

If we use a ``shotgun'' approach and observe the whole network with
$p_{\mathrm{obs}}=16/50$, we obtain a good ML estimate of the network
connectivity, including the $16\times16$ upper-left submatrix (Figure
\ref{fig: Common Input Problem}C). Now suppose instead we concentrate
all our observations on these $16$ neurons, so $p_{\mathrm{obs}}=1$
within that sub-network, but all the other neurons are unobserved.
If common input was not a problem then our estimation quality should
improve on that submatrix (since we have more measurements per neuron).
However, if common noise is problematic then we will ``hallucinate\textquotedbl{}
many nonexistent connections (i.e., off-diagonal terms) in this submatrix.
Figure \ref{fig: Common Input Problem}D illustrates exactly this
phenomenon. In contrast to the shotgun case, the resulting estimates
are significantly corrupted by the common input effects.

\subsection{Connectivity estimation with a sparsity inducing prior\label{sub:Connectivity-estimation-with}}

Next, we quantitatively test the performance of Maximum A Posteriori
(MAP) estimate of the network connectivity matrix $\mathbf{W}$, using
a sparsity prior (section \ref{sub:Sparse-prior}) on a simulated
neural network model.

\subsubsection{Network model \label{sub:Network-model}}

The neurons are randomly placed on a 1D lattice ring in locations
$x\in\left[0,1\right)$. To determine the connection probability (the
probability that $W_{i,j}$ is not zero) we used $f\left(d\right)=\exp\left(-ad\right)$,
where $d$ is the distance between two neurons, and $a$ was chosen
so that the network sparsity is $p_{\mathrm{conn}}=0.25$. For self
connectivity, we simply used $W_{i,i}=-1$ to account for the refractory
period. 

All the non-zero off-diagonal weights $W_{i,j}$ are sampled uniformly
from the range $\left[0,1\right]$. Also, all outgoing weights from
a neuron have the same sign, following Dale's law, where $50\%$ of
the neurons are inhibitory (the rest excitatory). These parameters
were chosen to obtain a balanced spontaneous network activity (\emph{i.e.},
without `epileptic' activity where all the neurons are extremely synchronous).
Initially, we do not have any external inputs.

The neurons are firing spontaneously, with no external input. We sampled
each neuronal bias from $b_{i}\sim\mathcal{N}\left(-1.2,0.1\right)$
to obtain a mean firing probability of $0.2-0.25$ in each time bin.
The rest of the parameters are individually set for each figure -
$N$, the number of neurons, $T$, the number of timebins and $p_{\mathrm{obs}}$,
the observation probability.

\subsubsection{Quality Measures}

We use four measures to assess the quality of the estimated matrix
$\hat{\mathbf{W}}$ in comparison with the ``ground truth'' matrix
$\mathbf{W}$. First, we define 
\[
\left\langle \left\langle W\right\rangle \right\rangle \triangleq\frac{1}{N^{2}}\sum_{i,j}W_{i,j}\,.
\]
The measures we use are:
\begin{itemize}
\item The square root of the coefficient of determination ($R^{2}$):
\begin{equation}
R\triangleq\sqrt{1-\frac{\sum_{i,j}\left(W_{i,j}-\hat{W}\right)^{2}}{\sum_{i,j}\left(W_{i,j}-\left\langle \left\langle W\right\rangle \right\rangle \right)^{2}}}\label{eq:R}
\end{equation}

\item Correlation:
\begin{equation}
C\triangleq\frac{\sum_{i,j}\left(W_{i,j}-\left\langle \left\langle W\right\rangle \right\rangle \right)\left(\hat{W}_{i,j}-\left\langle \left\langle \hat{W}\right\rangle \right\rangle \right)}{\sqrt{\sum_{i,j}\left(W_{i,j}-\left\langle \left\langle W\right\rangle \right\rangle \right)^{2}\sum_{i,j}\left(\hat{W}_{i,j}-\left\langle \left\langle \hat{W}\right\rangle \right\rangle \right)^{2}}}\label{eq:C}
\end{equation}

\item Zero matching: 
\begin{equation}
Z\triangleq1-\frac{1}{2}\frac{\sum_{i,j}\left(\mathcal{I}\left(W_{i,j}=0\right)\mathcal{I}\left(\hat{W}_{i,j}\neq0\right)+\mathcal{I}\left(W_{i,j}\neq0\right)\mathcal{I}\left(\hat{W}_{i,j}=0\right)\right)}{\sum_{i,j}\mathcal{I}\left(W_{i,j}=0\right)}\label{eq:Z}
\end{equation}

\item Sign matching: 
\begin{equation}
S\triangleq1-\frac{1}{2}\frac{\sum_{i,j}\left|\mathrm{sign}\left(W_{i,j}\right)-\mathrm{sign}\left(\hat{W}_{i,j}\right)\right|\mathcal{I}\left(W_{i,j}\neq0\right)\mathcal{I}\left(\hat{W}_{i,j}\neq0\right)}{\sum_{i,j}\mathcal{I}\left(W_{i,j}\neq0\right)\mathcal{I}\left(\hat{W}_{i,j}\neq0\right)}\label{eq:S}
\end{equation}

\end{itemize}
High values indicate high quality of estimation, with $1$ indicating
zero error. To simplify the presentation scale, if some measure becomes
negative or imaginary, we just set it to zero.

\subsubsection{Visualizing the inferred connectivity}

\begin{figure}[H]
\begin{centering}
\begin{center}
\includegraphics[width=0.8\columnwidth]{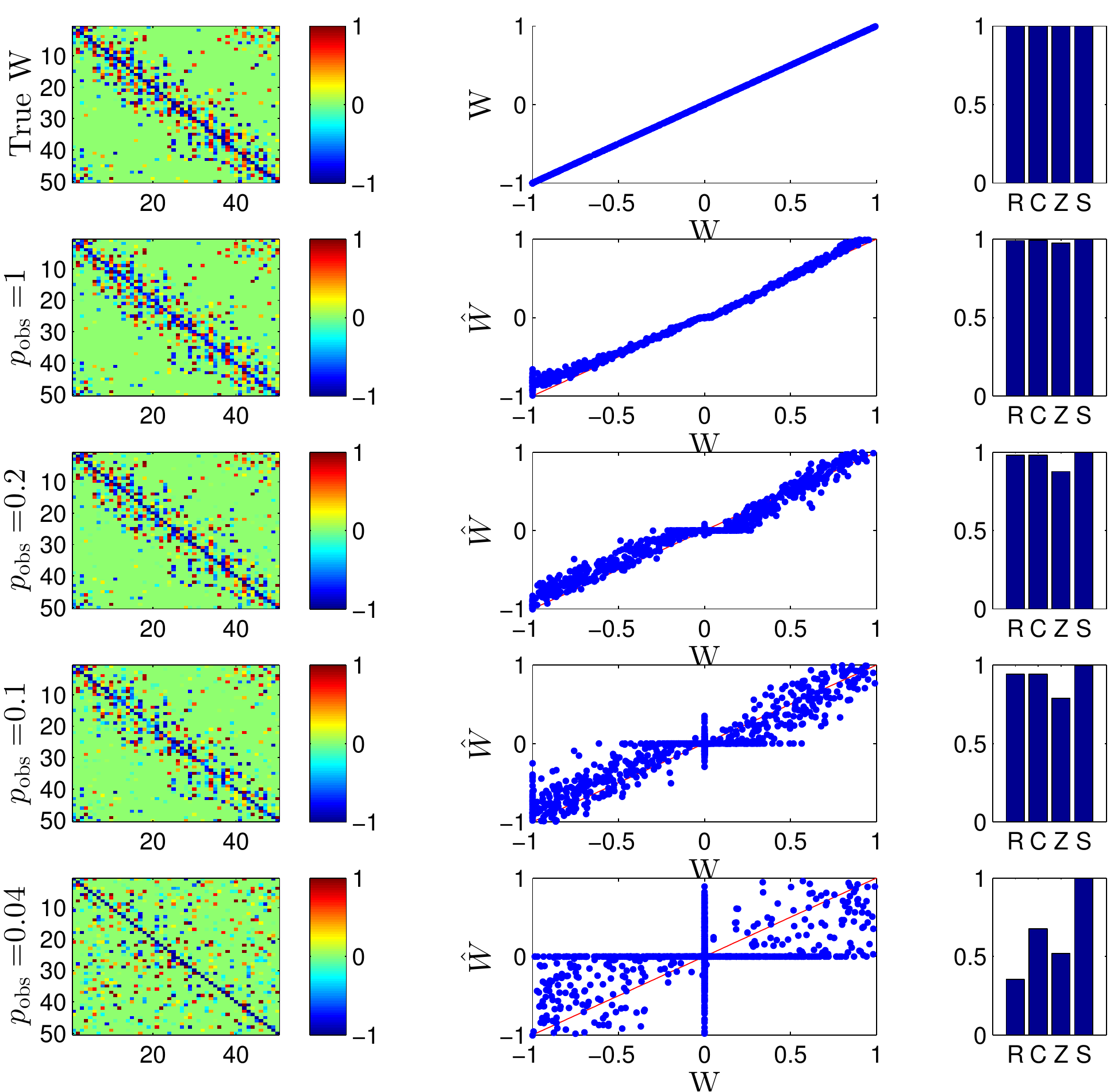}
\par\end{center}
\par\end{centering}

\protect\caption{\textbf{Network connectivity can be well estimated even with low observation
ratios.} The network has $N=50$ neurons, the experiment length is
$T=5\cdot10^{5}$, and we examine various observation probabilities:
$p_{\mathrm{obs}}=1,0.2,0.1,0.04$. Left column - weight matrix, middle
column - inferred weight vs. true weight, last column - quality of
estimation (based on the measures in Eqs. \ref{eq:R}-\ref{eq:S}).
In the first row we have the true weight matrix $\mathbf{W}$. In
the other rows we have $\hat{\mathbf{W}}$ - the MAP estimate of the
weight matrix with L1 prior (section \ref{sub:Sparse-prior}), with
$\lambda$ chosen so that the sparsity of $\hat{\mathbf{W}}$ matches
that of $\mathbf{W}$. Estimation is possible even with very low observation
ratios. \label{fig:Sparsity}}
\end{figure}

\begin{figure}[H]
\begin{centering}
\begin{center}
\includegraphics[width=0.8\columnwidth]{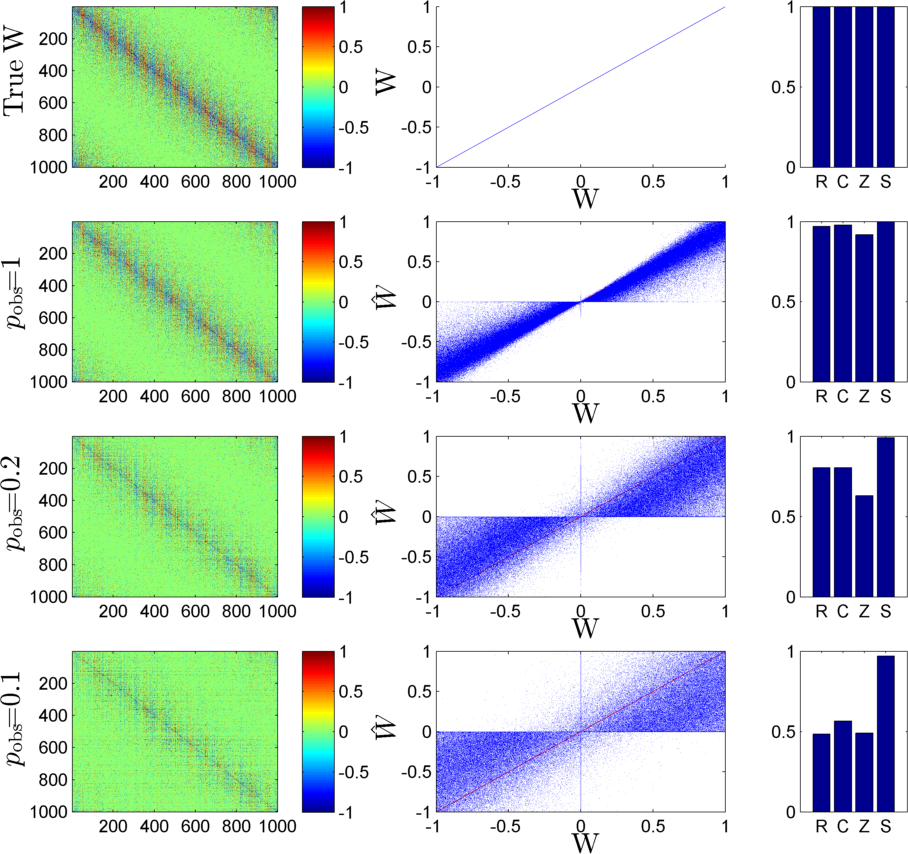}
\par\end{center}
\par\end{centering}

\protect\caption{\textbf{Network connectivity can be well estimated even with large
networks.} This is the same figure as Figure \ref{fig:Sparsity},
except now $N=1000$, $T=5\cdot10^{5}$, and $p_{\mathrm{obs}}=1,0.2,0.1$.
Since this estimate can be produced in less then a minute on a standard
laptop, this demonstrates that our algorithm is scalable to networks
with thousands of neurons. \label{fig:Sparsity-1}}
\end{figure}

In Figure \ref{fig:Sparsity}, for visualization purposes, we examine
a toy network with $N=50$ neurons. In Figure \ref{fig:Sparsity-1},
we examine another network with $N=1000$ neurons, which is closer
to the scale of the number of recorded neurons in current calcium
imaging experiments. 

\begin{figure}[H]
\begin{centering}
\begin{center}
\includegraphics[width=0.8\columnwidth]{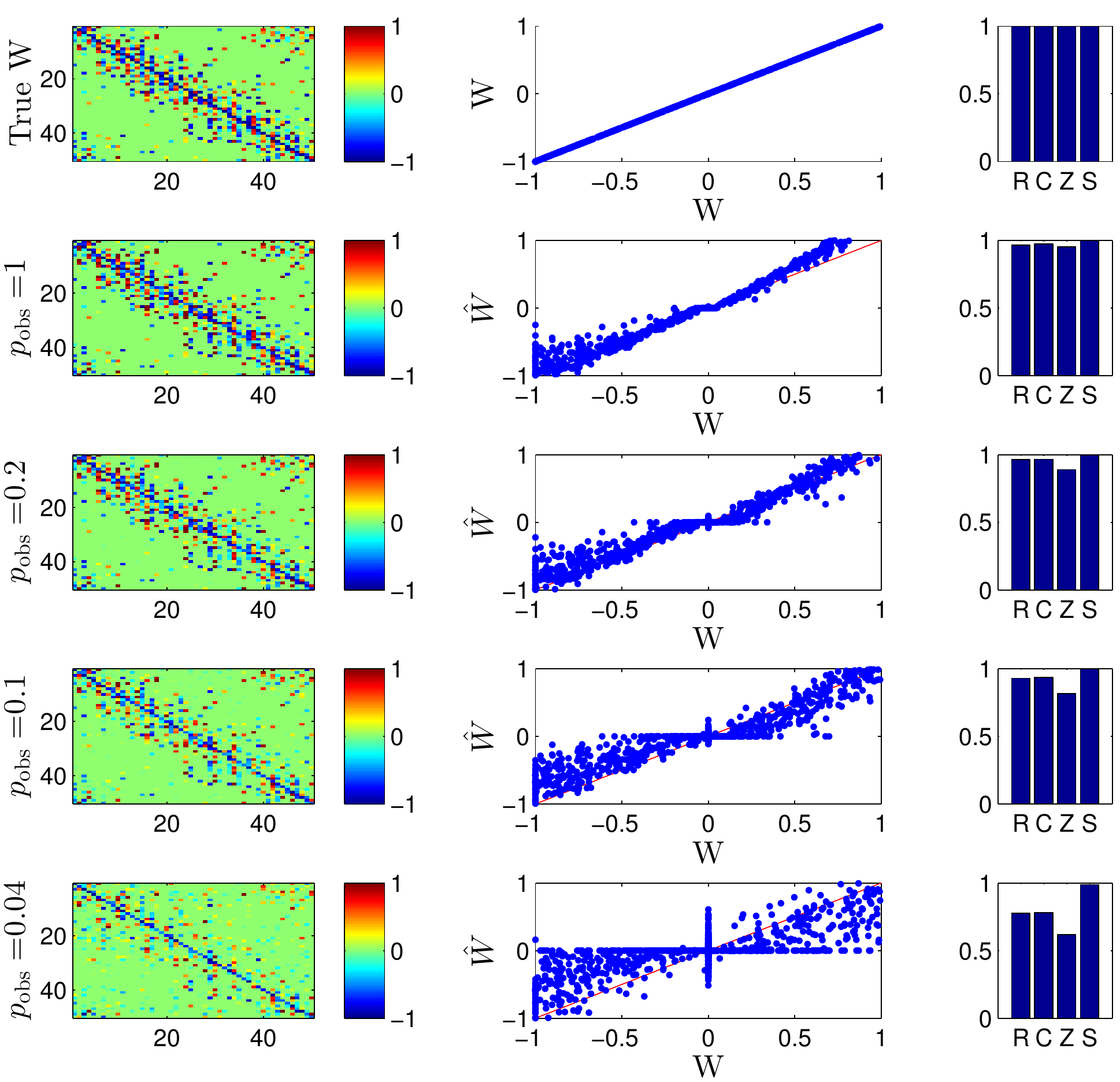}
\par\end{center}
\par\end{centering}

\protect\caption{\textbf{Network connectivity can be reasonably estimated, even with
model mismatch. }The same as Figure \ref{fig:Sparsity}, where instead
of a logistic GLM (Eq. \ref{eq: logistic}), we used a stochastic,
discrete time, leaky integrate and fire neuron model. In this model,
$V_{i,t}=\left(\gamma V_{i,t-1}+U_{i,t}+\epsilon_{i,t}\right)\mathcal{I}\left[S_{i,t-1}=0\right]$
($\mathbf{U}$ defined in Eq. \ref{eq: U}), $S_{i,t+1}=\mathcal{I}\left[V_{i,t}>0\right]$.
We used $\epsilon_{i,t}\sim\mathcal{N}\left(0,1\right)$ as a white
noise source. Also, we set $\gamma=0.5$ so that the neuronal timescale
would be restricted to a few time bins, as we assumed for simplicity
in the current GLM model. We conclude that our estimation method is
robust to modeling errors. \label{fig: LIF}}
\end{figure}

As can be seen in both figures, the weight matrix can be estimated
accurately for high values of observation probability $p_{\mathrm{obs}}$,
and reasonably well even for low value of $p_{\mathrm{obs}}$. For
example, in the network with $N=50$ neurons (Figure \ref{fig:Sparsity})
we see a quite reasonable performance when $p_{\mathrm{obs}}=0.1$,
\emph{i.e.}, only five neurons are observed in each timestep. Even
if $p_{\mathrm{obs}}=0.04$, i.e., only two neurons observed in each
timestep, we can still infer quite well the sign of inferred connections.
Note that our method works well even if the neuron model is not a
GLM, as we assumed. We demonstrate this using a leaky integrate and
fire neuron model (Figure \ref{fig: LIF}).

\subsubsection{Parameter scans}

\begin{figure}[H]
\begin{centering}
\begin{center}
\includegraphics[width=0.8\columnwidth]{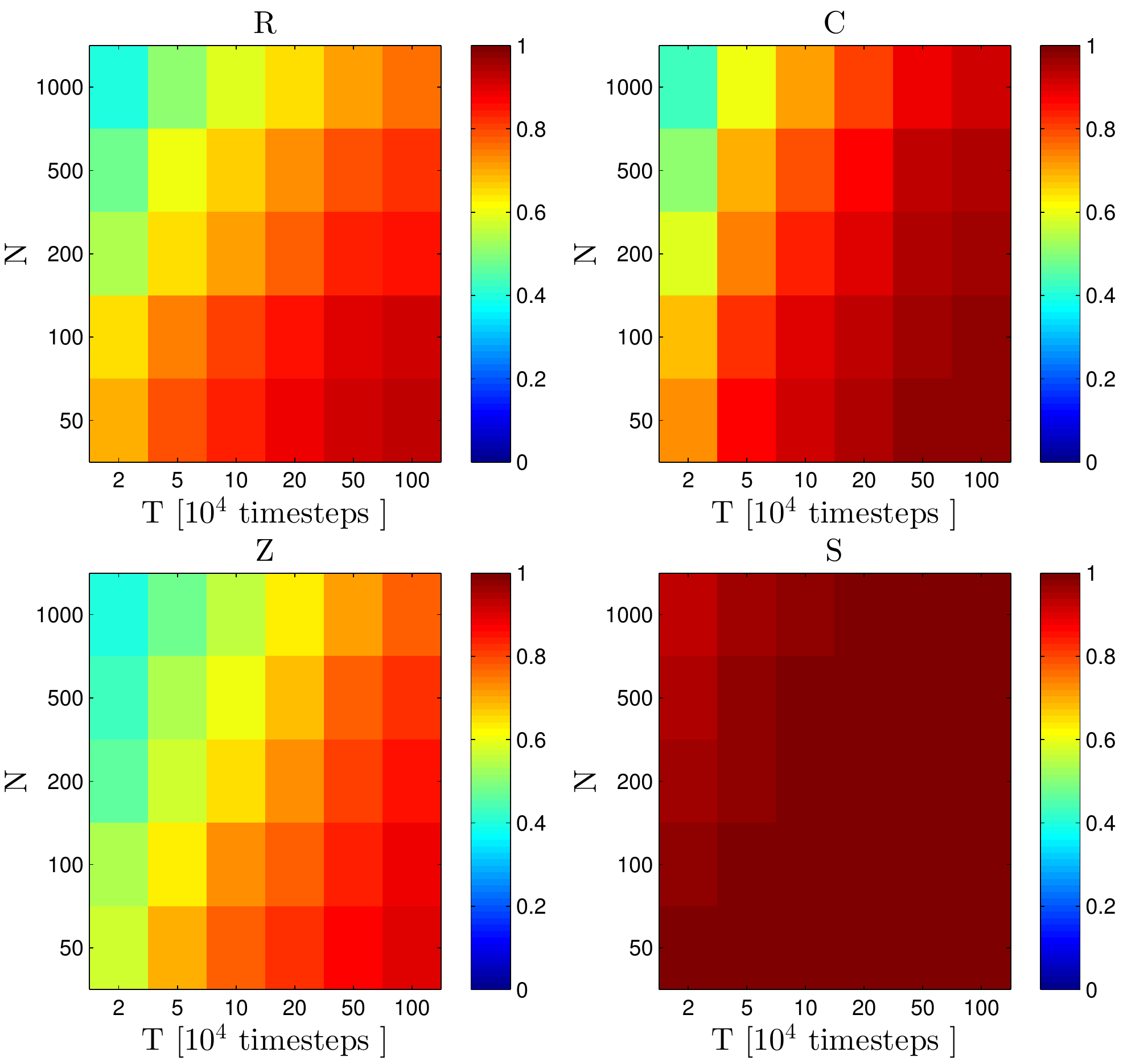}
\par\end{center}
\par\end{centering}

\protect\caption{\textbf{Parameter scans for network size $N$ and observation time
$T$. }We examine the quality (based on the measures in Eqs. \ref{eq:R}-\ref{eq:S})
of the MAP estimate of the weight matrix with L1 prior (section \ref{sub:Sparse-prior}).
Closely inspecting these figures, we find that performance is approximately
maintained constant when $N/T$ is constant (but less so when $T=2$).
We used an observation probability of $p_{\mathrm{obs}}=0.2$, and
different values of\textbf{ $N$} and $T$. Performance was averaged
over three repetitions. \label{fig:Parameter-scans N-T}}
\end{figure}

Next, we aimed to quantify how inference performance changes with
the number of neurons, $N$ (Figure \ref{fig:Parameter-scans N-T})
and observation probability $p_{\mathrm{obs}}$ (Figure \ref{fig:Parameter-scans P_obs T}).
On all measures, the performance changed qualitatively the same (therefore,
in subsequent figures, we will focus the correlation measure). For
any given fixed parameter ($N$ or $p_{\mathrm{obs}}$), performance
smoothly improves when $T$ increases. These scans suggest we can
maintain a good quality of connectivity estimation for arbitrarily
values of $N$ or $p_{\mathrm{obs}}$ - as long as we sufficiently
increase $T$. Specifically, if we closely examine these figures,
we find that, approximately, $T$ should be scaled as 
\begin{equation}
T\propto\frac{N}{p_{\mathrm{obs}}^{2}}\,\label{eq: T to N-P scaling}
\end{equation}
in order to maintain good estimation quality. This scaling can be
explained intuitively. Suppose first that $p_{\mathrm{obs}}=1$. Then
the total number of spike measurements is $NT$, and the number of
non-zero network parameters (synaptic weights) is $N^{2}p_{\mathrm{conn}}$
- where we recall that $p_{\mathrm{conn}}$ is the connection probability
between two neurons. Therefore, in order to maintain a fixed ratio
of measurements per parameter we must scale $T\propto Np_{\mathrm{conn}}$
(to first order in $N$). 

Now suppose $p_{\mathrm{obs}}<1$. Then the total number of spike
measurements is $NTp_{\mathrm{obs}}$, and the number of non-zero
parameters is still $N^{2}p_{\mathrm{conn}}$. Therefore, it might
seem that in order to maintain a fixed ratio of measurements per parameter
we must scale $T\propto Np_{\mathrm{conn}}/p_{\mathrm{obs}}$. However,
this ignores the adverse affect of the unobserved common input. Let's
examine the neuronal input (Eq. \ref{eq: U}) 
\begin{equation}
U_{i,t}=U_{i,t}^{\mathrm{obs}}+U_{i,t}^{\mathrm{unobs}}+b_{i}\,,
\end{equation}
where $U_{i,t}^{\mathrm{obs}}\triangleq\sum_{k:O_{k,t}=1}W_{i,k}S_{k,t-1}$
is the input current arriving from the observed neurons and $U_{i,t}^{\mathrm{unobs}}\triangleq\sum_{k:O_{k,t}=0}W_{i,k}S_{k,t-1}$
is the input current arriving from the unobserved neurons. The latter,
$U_{i,t}^{\mathrm{unobs}}$, can be effectively considered as ``noise'',
which makes it harder to infer the weights in $U_{i,t}^{\mathrm{obs}}$.
Since the number of terms in $U_{i,t}^{\mathrm{unobs}}$ is $1/p_{\mathrm{obs}}$
times more than the number in $U_{i,t}^{\mathrm{obs}}$, we have,
approximately $\mathrm{Var}\left(U_{i,t}^{\mathrm{obs}}\right)/\mathrm{Var}\left(U_{i,t}^{\mathrm{unobs}}\right)\propto p_{\mathrm{obs}}$,
if the neurons are not highly correlated and $p_{\mathrm{obs}}$ is
small. Therefore, the signal to noise ratio for estimating the weights
should also scale approximately as $p_{\mathrm{obs}}$. In order to
compensate for this, we should increase $T$ by $1/p_{\mathrm{obs}}$
. Taking this into account gives the quadratic scaling in $1/p_{\mathrm{obs}}$
that we observe (Eq. \ref{eq: T to N-P scaling}). For exact analytical
results on this issue see \cite{Mishchenko2015}.

\subsubsection{The effect of stimulus inputs\label{sub:Stimulus}}

Next, we demonstrate how stimulus inputs can be used to improve the
quality of estimation. Here we use a simple periodic pulse stimulus,
similar to that generated by light activated persistent ion channels
\cite{Berndt2009}. Since only a brief optical stimulus is required
to activate these channels, this minimizes cross-talk between optical
recording (\emph{e.g.}, calcium imaging) and optical stimulation.
Therefore, such type of stimulus is generally beneficial in all-optical
experiments (but may not be necessary, given recent advances \cite{Rickgauer2014}).
Here, we used a single external input $X\left(t\right)$ which similarly
stimulates all the neurons in the circuit (\emph{i.e.}, \textbf{$\mathbf{G}=\boldsymbol{1}^{N\times1}$}
in Eq. \ref{eq: U}). The stimulus $X\left(t\right)$ has a period
of $T_{0}=10^{3}$ time-steps, where the pulse is ``ON'' $50\%$
of the time. We tested various pulse magnitudes, and added white noise
to the pulse with a variance which is $10\%$ of the pulse magnitude.
We used the same simulation parameters as in section \ref{sub:Network-model},
with $N=50$, $T=10^{5}$ and $p_{\mathrm{obs}}=0.2$. The only exception
is that we decreased $\mathbf{b}$ by $4$ to achieve a low firing
rate regime at zero stimulus.

As can be seen in Figure \ref{fig:The-effect-of-stimulus}, this type
of stimulus has a strong effect on reconstruction performance, and
there is an ``optimal magnitude'' for improving reconstruction.
A main reason for this effect is that the stimulus magnitude directly
determines $m\triangleq\frac{1}{N}\sum_{i=1}^{N}m_{i}$, the mean
spike probability (\emph{i.e.}, the firing rate). This directly affects
the number of spike pair combinations observed during the experiment.
Such spike pairs are necessary for a high quality of inference, since
they constitute the sufficient statistics in our posterior (Eq. \ref{eq:P(S|W,B*)}).
Therefore if the stimulus (and therefore, firing rate $m$) is too
low or high, we observe fewer spike pair combinations, and this decreases
inference quality. However, introducing strong variable stimulus can
also reduce inference quality, by masking the effect of weak connections
on spiking. Therefore, the optimal inference quality is achieved here
at for $m\approx0.13$, and not at $m=0.25$ - the mean spike probability
in which the maximal number of spike pair combinations is observed%
\footnote{This happens if the neurons fire with rate $0.5$ when the pulse is
``ON''. Therefore, the average firing rate in this case is $m=0.25$,
since the neurons hardly fire when the pulse is ``OFF'' (which is
$50\%$ of the time)%
}.

\subsection{Additional priors\label{sub:Additional-priors}}

Next, we incorporate into our inference procedure additional prior
information. We show that incorporating stronger prior information
improves inference performance.

\subsubsection{Network model}

The network is the same as in section \ref{sub:Network-model}, except
we now choose the weight magnitude differently, as described in Figure
\ref{sub:Priors}. We use $f\left(d\right)=\left(1+\exp\left(15d-1.3\right)\right)^{-1}$
to select non-zero weights ($f\left(d\right)$ in Figure \ref{fig:Schematic-description}),
and choose connection amplitudes according to $W_{ij}\sim\mathcal{N}(V_{ij},\sigma^{2}),$
where $\sigma=0.07$ and $\mathbf{V}$ is a block matrix (with a $2\times2$
block structure), corresponding to the two types of neurons inhibitory
and excitatory (each covers half the network). We choose $V_{i,j}=1.1$
for excitatory to excitatory connections, $V_{i,j}=0.9$ for excitatory
to inhibitory connections, $V_{i,j}=-1.8$ for inhibitory to excitatory
connections, and $V_{i,j}=-2.2$ for inhibitory to inhibitory connections.
Again, these specific parameters were chosen in order to create a
balanced spontaneous network activity.

\subsubsection{Results}

We tested various combinations of priors:
\begin{enumerate}
\item That most connections are zero (sparsity). We use either $L0$ or
$L1$ penalties (section \ref{sub:Sparse-prior}). \label{enu:spar}
\item That neurons are either excitatory or inhibitory (Dale's law). We
use the method in \cite[section 2.5.2]{Mishchenko2011c}. \label{enu:Dale}
\item That neuronal types affect connection amplitude. We use the stochastic
block model (section \ref{sub:Stochastic-block-model}), in which
we either know the number of these types (\emph{i.e.}, we just know
the rank of $\mathbf{V}$), or we know exactly how the types affect
connection amplitude (\emph{i.e.}, we know $\mathbf{V}$).\label{enu:sbm}
\item That the connection probability is distance dependent. We modify the
sparsity penalty (section \ref{sub:Distance-dependence}) using $f\left(d\right)=1/\left(1+\mathrm{exp}\left(ad+b\right)\right)$,
where we either do not know $a$ and $b$, or do know them. \label{enu:dd}
\end{enumerate}
In Figure \ref{fig:Prior Results}, we show the correlation quality
measure (Eq. \ref{eq:C}) for different prior combinations. As can
be seen, each additional prior incorporated into our connectivity
estimate improves its quality. We find that usually the prior on the
distance dependence of the connectivity has a more significant impact
then the prior on type-dependence. Also, both sparsity-inducing priors,
$L0$ and $L1$, give similar performance. Typically $L0$ is slightly
better when $T$ is low, while $L1$ is slightly better when $T$
is high (not shown). The $L1$ penalty allows somewhat faster optimization
then $L0$. However, the $L0$ penalty allows us to more easily incorporate
neuronal type-related connectivity information. Therefore when combining
prior \ref{enu:spar} with priors \ref{enu:sbm}-\ref{enu:dd}, we
only use $L0$.

\section{Discussion\label{sec:Discussion}}

Current technology limits the number of neurons that can be simultaneously
observed. Therefore, common approaches to infer functional connectivity
of a neural circuit focus all the observations in one experiment on
a small part of the circuit - in which all neurons are fully observed
at all timebins. However, unobserved input into this sub-circuit can
generate significant error in the estimation - and this error does
not vanish with longer experiments. 

To deal with this ``common input'' problem, we propose a ``shotgun''
observation scheme, in which we randomly observe neurons from the
circuit, so only a small percentage of the network is observed at
each time point. Therefore, despite the limited scanning speed of
the imaging device, using this method we can arbitrary expand the
field of view to include an entire circuit, together with any input
to the circuit, so no neuron is completely hidden. However, existing
inference algorithms cannot handle efficiently so many missing observations.

For this purpose, we develop (section \ref{sec:Bayesian Inference})
a new scalable Bayesian inference method. As we demonstrate numerically
(section \ref{sec:-Results}) this method can be used to estimate
the synaptic connectivity of a spiking neural network from spike data
sub-sampled at arbitrarily low observation ratios (e.g., $10\%$).
Previously, the lowest observation ratio demonstrated was $50\%$,
in a two-neuron network \cite{Pillow2007}, or in a simple linear
dynamical model for neuronal activity \cite{Turaga2013a}. Moreover,
our method is very fast computationally, and be can be easily used
on networks with thousands of neurons. Previous works, which used
standard inference methods, in which the unobserved spikes are treated
as latent variables (Section \ref{sec:Other-Inference-Methods}),
are slower by several orders of magnitude. Specifically, previous
MCMC methods \cite{Pillow2007,Mishchenko2011c} did not go beyond
a $50$ neurons, while variational approaches \cite{Fletcher2014}
did not go beyond $100$. Though the latter may be scaled for a larger
population, it is not clear if this approach can handle missing observations.

The proposed method is capable of incorporating various kinds of prior
information (Figure \ref{fig:Schematic-description}), such as the
sparsity of synaptic connections, Dale's Law, the division of neurons
into types and the fact that the probability of having a connection
between two neurons typically decreases with the distance between
two neurons. We show that each piece of information can be used to
improve estimation quality (Figure \ref{fig:Prior Results}). Another
way to significantly improve estimation quality, is to adjust the
baseline firing rate of the network using the stimulus (Figure \ref{fig:The-effect-of-stimulus}).
Specifically, we use a pulse-like stimulus, similar to that generated
by persistent ion channels \cite{Berndt2009}, which can be used in
the context of all-optical experiments with minimal cross-talk between
stimulation and recording. More sophisticated types of stimulation
schemes (e.g., \cite{Rickgauer2014}) may improve performance even
further \cite{Shababo2013}.

We conclude that, using the shotgun observation scheme, we can remove
the persistent bias resulting from the common input problem (Figure
\ref{fig: Common Input Problem}). Therefore, the limited scanning
speed of a imaging device is not a fundamental obstacle hindering
functional connectivity estimation. A complete removal of the bias
is possible only if all the neurons in the circuit are observed together
with all inputs to the circuit (a sufficient number of times). However,
in most experimental setups, some neurons would never be observed
(\emph{e.g.}, due to opacity of the brain tissue). Therefore, some
persistent bias may remain. To deal with this we will need to incorporate
any unobserved input into the circuit using a small number of latent
variable \cite{Vidne2012}. Due to intrinsic fluctuations in the neuronal
excitability \cite{Gal2010,Soudry2014d}, such a latent variable approach
can be relevant even in cases where potentially the entire circuit
could be observed (\emph{e.g.}, in zerbrafish \cite{Ahrens2013}).

\begin{figure}[H]
\begin{centering}
\begin{center}
\includegraphics[width=0.8\columnwidth]{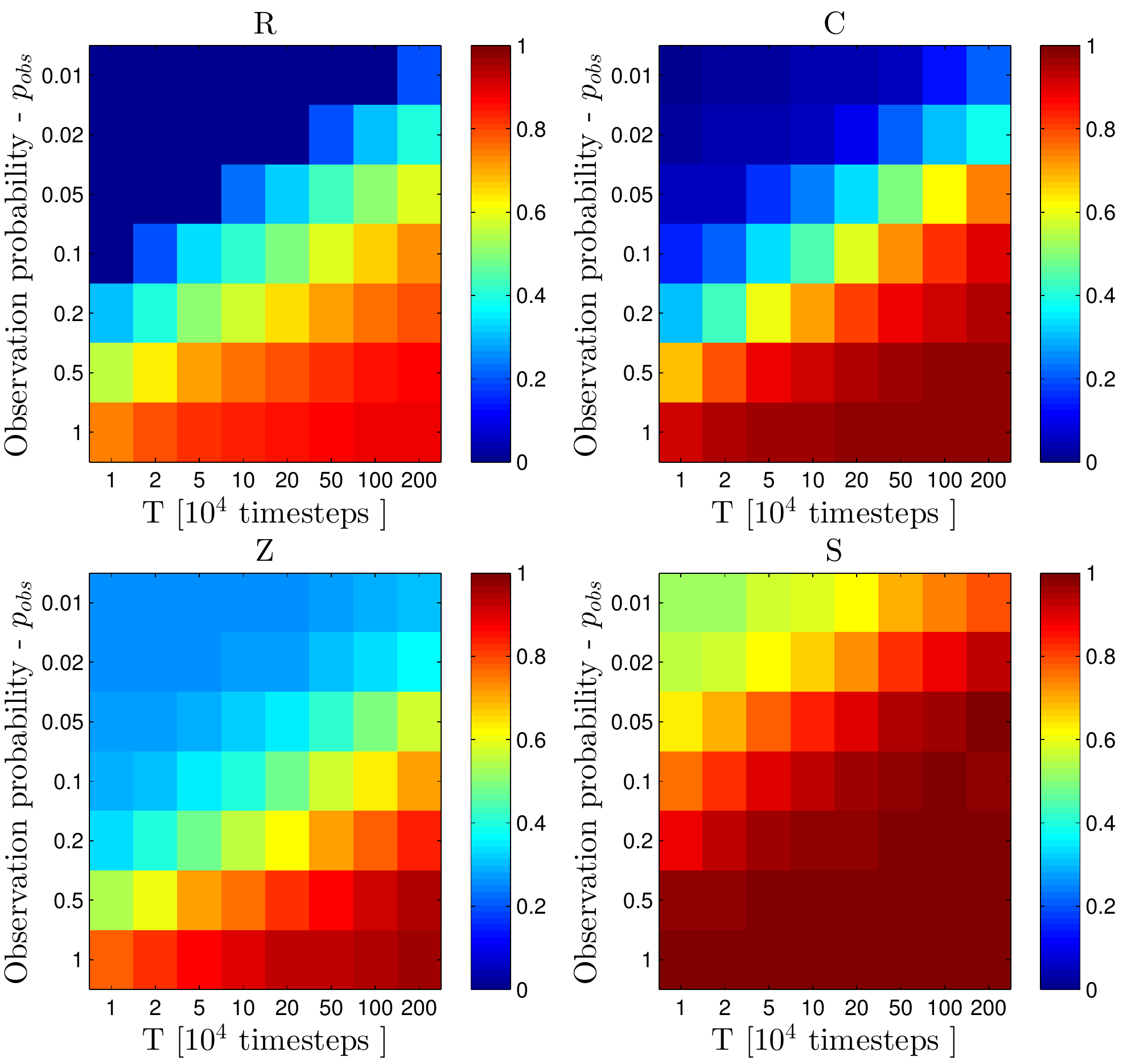}
\par\end{center}
\par\end{centering}

\protect\caption{\textbf{Parameter scans for observation time $T$ and observation
probability $p_{\mathrm{obs}}$. }Quality (based on the measures in
Eqs. \ref{eq:R}-\ref{eq:S}) of the MAP estimate of the weight matrix
with L1 prior (section \ref{sub:Sparse-prior}). Closely inspecting
these figures, we find that performance is approximately maintained
constant when $1/Tp_{\mathrm{obs}}^{2}$ is constant (but less so
when $p_{\mathrm{obs}}=1$). We used a network size of $N=500$ and
different values of\textbf{ $p_{\mathrm{obs}}$} and $T$. Performance
was averaged over three repetitions.\label{fig:Parameter-scans P_obs T}}
\end{figure}

Even if all the neuronal inputs are eventually observed, the variance
due to the unobserved inputs is still high, since, at any given time,
most of the inputs to each neuron will be unobserved (see also \cite{Mishchenko2015}).
As a result, the duration of the experiment required for accurate
inference increases quadratically with the inverse of observation
probability (Eq. \ref{eq: T to N-P scaling} and Figure \ref{fig:Parameter-scans P_obs T}).
This issue will persist for any fixed observation strategy that does
not take into account any prior information on the network connectivity.

However, there might be a significant improvement in performance if
we can focus the observations on synaptic connections which are more
probable. This way we can effectively reduce input noise from unobserved
neurons, and improve the signal to noise ratio. As a simple example,
suppose we know the network is divided in to several disconnected
components. In this case, we should scan each sub-network separately,
\emph{i.e.}, there is no point in interleaving spike observations
from two disconnected sub-networks. How should one focus observations
in the more general case? We leave this as an interesting open question
in Bayesian experimental design methods (``active learning'').

\section{Methods\label{sec:Methods}}

\subsection{Preliminaries\label{sec:Preliminaries}}

In this section we describe the statistical methods used for inferring
connectivity given data, observed using the shotgun scheme. First,
we describe the basic statistical model and framework (section \ref{sec:Preliminaries});
our main analytical results on the simplified loglikelihood (section
\ref{sec:Bayesian Inference}); how to incorporate various types of
prior with this loglikelihood (section \ref{sub:Priors}); and a few
alternative inference methods (section \ref{sec:Other-Inference-Methods}).
In the supplementary Appendix S1 we provide all the technical details
of the algorithms used.

\subsubsection{General Notation}

A boldfaced letter $\mathbf{x}$ denotes a vector with components
$x_{i}$, a boldfaced capital letter $\mathbf{X}$ denotes a matrix
with components $X_{i,j}$, $\mathbf{X}^{\left(k\right)}$ denotes
the $k$-th matrix in a list, and $\mathbf{X}_{\cdot,k}$ $\left(\mathbf{X}_{k,\cdot}\right)$
the $k$-th column (row) vector of matrix $\mathbf{X}$. For $\mathbf{X}\in\mathbb{R}^{N\times T}$
we define the empiric average and variance%
\footnote{The following expressions do not depend on $t$, despite the $t$
index, which is maintained for notational convenience.%
} 
\begin{eqnarray*}
\left\langle X_{i,t}\right\rangle _{T} & \triangleq & \frac{1}{T}\sum_{t=1}^{T}X_{i,t}\,\,;\,\,\mathrm{Var}_{T}\left(X_{i,t}\right)\triangleq\frac{1}{T}\sum_{t=1}^{T}\left(X_{i,t}-\left\langle X_{i,t}\right\rangle \right)^{2}
\end{eqnarray*}
We define the matrices $\mathbf{0}^{M\times N}$ and $\mathbf{1}^{M\times N}$
as $M\times N$ matrices in which all the components are equal to
zero or one, respectively.  For any condition $A$, we make use of
$\mathcal{I}\left\{ A\right\} $, the indicator function (\emph{i.e.},
$\mathcal{I}\left\{ A\right\} =1$ if $A$ holds, and zero otherwise).
Also, $\delta_{ij}\triangleq\mathcal{I}\left\{ i=j\right\} $, Kronecker's
delta function. If $\mathbf{x}\sim\mathcal{N}\left(\boldsymbol{\mu},\boldsymbol{\Sigma}\right)$
then $\mathbf{x}$ is Gaussian random vector with mean $\boldsymbol{\mu}$
and covariance matrix $\boldsymbol{\Sigma}$, and we denote its density
by $\mathcal{N}\left(\mathbf{x}|\boldsymbol{\mu},\boldsymbol{\Sigma}\right)$.

\subsubsection{Model }

We use a discrete-time neural network. The neurons, indexed from $i=1$
to $N$, produce spikes in time bins indexed from $t=1$ to $T$.
The spiking matrix $\mathbf{S}$ is composed of variables $S_{i,t}$
indicating the number of spikes neuron $i$ produces at time bin $t$.
We assume each neuron $i$ generates spikes $S_{i,t}\in\left\{ 0,1\right\} $
according to a Generalized Linear network Model (GLM \cite{BRIL88,Rigat06,Pillow2007}),
with a logistic probability function 
\begin{equation}
P\left(S_{i,t}|U_{i,t}\right)=\frac{e^{S_{i,t}U_{i,t}}}{1+e^{U_{i,t}}}\,.\label{eq: logistic}
\end{equation}
depending on the the input $U_{i,t}$ it receives from other neurons,
as well as from some external stimulus. The input to all the neurons
in the network is therefore 
\begin{equation}
\mathbf{U}_{\cdot,t}\triangleq\mathbf{W}\mathbf{S}_{\cdot,t-1}+\mathbf{b}+\mathbf{G}\mathbf{X}_{\cdot,t}\,,\label{eq: U}
\end{equation}
where $\mathbf{b}\in\mathbb{R}^{N}$ is the bias of neuron $i$; $\mathbf{X}\in\mathbb{R}^{D\times T}$
are the external inputs (with $D$ being the number of inputs); \textbf{$\mathbf{G}\in\mathbb{R}^{N\times D}$
}is the input gain; and $\mathbf{W}\in\mathbb{R}^{N\times N}$ is
the (unknown) network connectivity matrix. The diagonal elements $W_{i,i}$
of the connectivity matrix correspond to the post spike filter accounting
for the cell's own post-spike effects (\emph{e.g.}, refractory period),
while the off-diagonal terms $W_{i,j}$ represent the connection weights
from neuron $j$ to neuron $i$. The bias $b_{i}$ controls the mean
spike probability (firing rate) of neuron $i$. The external input
$\mathbf{X}$ can represent a direct (\emph{e.g.}, light activated
ion channels) or sensory (\emph{e.g.}, moving grating) stimulation
of neurons in the network. The input gain $\mathbf{G}$ is a spatial
filter that gives the effect of this input $\mathbf{X}$ on the network.
The input gain represents the affect of this input on the network.
We assume that spiking starts from some fixed distribution $P\left(\mathbf{S}_{\cdot,0}\right)$. 

To simplify notation we have assumed in Eq. \ref{eq: U} that $\mathbf{U}_{\cdot,t}$
is affected by spiking activity only in the previous time bin ($\mathbf{W}\mathbf{S}_{\cdot,t-1}$).
However, it is straightforward to generalize our results to the case
that the input (Eq. \ref{eq: U}) includes a longer history of the
spiking activity%
\footnote{\emph{i.e.}, more generally, we can have $\mathbf{U}_{\cdot,t}\triangleq\sum_{k=1}^{L_{1}}r^{\left(k\right)}\mathbf{U}_{\cdot,t-k}+\sum_{k=1}^{L_{2}}\mathbf{W}^{\left(k\right)}\mathbf{S}_{\cdot,t-k}+\mathbf{b}+\mathbf{G}\mathbf{X}_{\cdot,t}$,
where $\left\{ r^{\left(k\right)}\right\} _{k=1}^{L_{1}}$ and $\left\{ \mathbf{W}^{\left(k\right)}\right\} _{k=1}^{L_{2}}$
determine the neuronal timescales.%
}. It is also possible to change Eq. \ref{eq: logistic} and assume
other spiking models (\emph{e.g.}, Poisson or Binomial).

\subsubsection{Task}

\begin{figure}[H]
\begin{centering}
\begin{center}
\includegraphics[width=0.8\columnwidth]{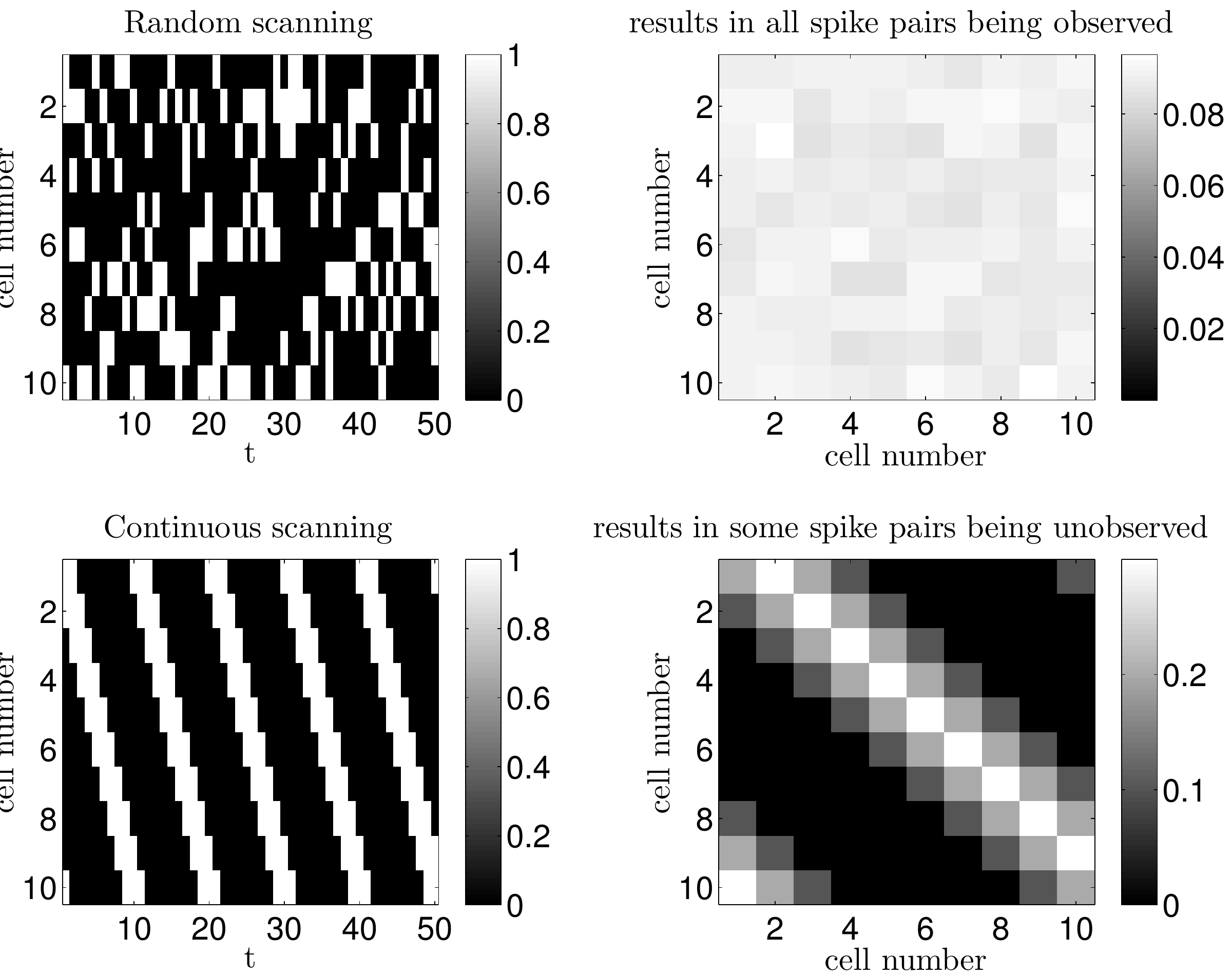}
\par\end{center}
\par\end{centering}

\protect\caption{Examples for ``good'' and ``bad'' observation schemes. We observe
three out of ten neurons in each timebin. Top\emph{:} The observations
$\mathbf{O}$ (a zero-one matrix) for a random scanning scheme (\emph{left})
and its observed pairs\textbf{ }$\left\langle O_{i,t}O_{j,t-1}\right\rangle _{T}=\frac{1}{T}\sum_{t=1}^{T}O_{i,t}O_{j,t-1}$
(\emph{right}). All possible neuron pairs are observed (\emph{i.e.},
$\left\langle O_{i,t}O_{j,t-1}\right\rangle _{T}>0$ for all pairs),
so this is a valid ``shotgun'' observation scheme. Bottom: The observation
matrix $\mathbf{O}$ for a continuous serial scanning scheme (\emph{left})
and\textbf{ }its observed pairs $\left\langle O_{i,t}O_{j,t-1}\right\rangle _{T}$.
Since we do not observe spikes pairs for which $\left|i-j\right|>3$
(\emph{i.e.}, $\left\langle O_{i,t}O_{j,t-1}\right\rangle _{T}=0$
for these pairs), this is not a valid ``shotgun'' observation scheme.
\label{fig:observation schemes}}
\end{figure}

Our goal is to infer the connectivity matrix $\mathbf{W}$, biases
$\mathbf{b}$ and the stimulus gain $\mathbb{\mathbf{G}}$. We assume
that we have some prior information on the weights (including $N$),
that we know the external input $\mathbf{X}$, and that we noiselessly%
\footnote{\emph{i.e.}, for simplicity we ignore for now the problem of inferring
the spikes from the experimental data (\emph{e.g.}, which requires
spike sorting or deconvolution of the calcium trace). The issue of
spike inference was addressed systematically in \cite{Mishchenko2011c,Lutcke2013}.%
} observe a subset of the generated spikes. We use a binary matrix
$\mathbf{O}$ to indicate which neurons were observed, so 
\[
O_{i,t}\triangleq\mathcal{I}\left[S_{i,t}\,\mathrm{was\, observed}\right]\,.
\]
We assume the observation process is uncorrelated with the spikes,
and that $\left\langle O_{i,t}\right\rangle _{T}$ and $\left\langle O_{i,t}O_{j,t-k}\right\rangle _{T}$
converge to strictly positive limits when $T\rightarrow\infty$, for
all $i,j$ and for $k\in\left\{ 0,1\right\} $. In other words, we
need to observe a large number of timebins (proportional to $T$)
from each pair of neurons $\left(i,j\right)$ - both simultaneous
spike pairs $\left(k=0\right)$ and spike pairs in which one spike
arrives one timestep after the other $\left(k=1\right)$. For example,
these assumptions are fulfilled if the spikes are uniformly and randomly
observed, or if at each time bin a random observation block is chosen
(Figure \ref{fig:observation schemes}, \emph{top}). However, some
spike pairs are never observed, if, for example, we only observe a
fixed subset of the network, or scan serially in a continuous manner
(Figure \ref{fig:observation schemes},\emph{ bottom}).

\subsection{Bayesian inference of the weights\label{sec:Bayesian Inference}}

We use a Bayesian approach in order to infer the unknown weights.
For simplicity, initially assume that all spikes are observed and
that there is no external input $\left(\mathbf{G}=0\right)$. In this
case, the log-posterior of the weights, given the spiking activity,
is

\begin{equation}
\ln P\left(\mathbf{W}|\mathbf{S},\mathbf{b}\right)=\ln P\left(\mathbf{S}|\mathbf{W},\mathbf{b}\right)+\ln P_{0}\left(\mathbf{W}\right)+C\,,\label{eq: log posterior}
\end{equation}
where $\ln P\left(\mathbf{S}|\mathbf{W},\mathbf{b}\right)$ is the
loglikelihood, $P_{0}\left(\mathbf{W}\right)$ is some prior on the
weights (we do not assume a prior on the biases $\mathbf{b}$), and
$C$ is some unimportant constant (which does not depend on $\mathbf{W}$
or $\mathbf{b}$). Our aim would be to find the Maximum A Posteriori
(MAP) estimator for $\mathbf{W}$, together with the Maximum Likelihood
(ML) estimator for $\mathbf{b}$, by solving
\begin{equation}
\max_{\mathbf{W},\mathbf{b}}\ln P\left(\mathbf{W}|\mathbf{S},\mathbf{b}\right)\,.\label{eq: ML}
\end{equation}

Next, we show that under some reasonable assumptions, the loglikelihood
can be transformed to a simple form. Importantly, this simple form
can be easily calculated even if there are some missing observations,
and we have an external stimulus $\left(\mathbf{G}\neq0\right)$.
Specifically, we consider a GLM spiking network model, as in Eq. \ref{eq: logistic},
and simplify its loglikelihood $\ln P\left(\mathbf{S}|\mathbf{W},\mathbf{b}\right)$.
Using similar techniques as in \cite{Park2011,Sadeghi12,Ramirez2013}
(and a few additional `tricks') we use the expected loglikelihood
approximation together with a generalized Central Limit Theorem (CLT)
argument \cite{Diaconis1984}, in which we approximate the neuronal
input to be Gaussian near the limit $N\rightarrow\infty$; then we
calculate the ``profile likelihood'' $\max_{\mathbf{b}}\ln P\left(\mathbf{S}|\mathbf{W},\mathbf{b}\right)$
in which the bias term has been substituted for its maximizing value
(Appendix S1, section \ref{sub:Derivation}). The end result is 
\begin{eqnarray}
\max_{\mathbf{b}}\ln P\left(\mathbf{S}|\mathbf{W},\mathbf{b}\right) & \approx & T\sum_{i=1}^{N}\left[\sum_{j=1}^{N}\left[W_{i,j}\Sigma_{i,j}^{\left(1\right)}\right]-h\left(m_{i}\right)\sqrt{1+\frac{\pi}{8}\sum_{k,j}W_{i,j}\Sigma_{k,j}^{\left(0\right)}W_{i,k}}\right]\,,\label{eq:P(S|W,B*)}
\end{eqnarray}
where we denoted
\begin{eqnarray}
m_{i} & \triangleq & \left\langle S_{i,t}\right\rangle _{T}\label{eq:m}\\
\Sigma_{i,j}^{\left(k\right)} & \triangleq & \left\langle S_{i,t}S_{j,t-k}\right\rangle _{T}-m_{i}m_{j}\label{eq: Sigma}\\
h\left(m_{i}\right) & \triangleq & -m_{i}\ln m_{i}-\left(1-m_{i}\right)\ln\left(1-m_{i}\right)\label{eq: entropy}
\end{eqnarray}
as the mean spike probability, spike covariance and the entropy function,
respectively. We make a few comments:
\begin{enumerate}
\item Importantly, this loglikelihood depends on the data only through the
sufficient statistics $\mathbf{m}$ and $\boldsymbol{\Sigma}^{\left(k\right)}$
(defined in Eqs. \ref{eq:m}-\ref{eq: Sigma}), which can be estimated
even when some observations are missing. Specifically, if we have
a only partial observation of $\mathbf{S}$, we define 
\begin{eqnarray}
\tilde{m}_{i} & \triangleq & \frac{\left\langle O_{i,t}S_{i,t}\right\rangle _{T}}{\left\langle O_{i,t}\right\rangle _{T}}\,,\label{eq: m partial}\\
\tilde{\Sigma}_{i,j}^{\left(k\right)} & \triangleq & \frac{\left\langle O_{i,t}O_{j,t-k}S_{i,t}S_{j,t-k}\right\rangle _{T}}{\left\langle O_{i,t}O_{j,t-k}\right\rangle _{T}}-\tilde{m}_{i}\tilde{m}_{j}\,.\label{eq:Sigma Partial}
\end{eqnarray}
Recall the observation process is uncorrelated with the spikes, and
that $\forall i,j$ and $\forall k\in\left\{ 0,1\right\} $, $\left\langle O_{i,t}\right\rangle _{T}$
and $\left\langle O_{i,t}O_{j,t-k}\right\rangle _{T}$ converge to
strictly positive limits when $T\rightarrow\infty$. From Slutsky's
theorem \cite{Stirzaker2001} we obtain that for $T\rightarrow\infty$
\[
\tilde{\mathbf{m}}\rightarrow\mathbf{m}\,;\,\tilde{\boldsymbol{\Sigma}}^{\left(k\right)}\rightarrow\boldsymbol{\Sigma}^{\left(k\right)}\,,
\]
given that $m_{i}$ and $\Sigma_{i,j}^{\left(k\right)}$ converge
to some limit when $T\rightarrow\infty$. Therefore, when some observations
are missing we can simply replace $\mathbf{m}^{\left(k\right)}$ with
$\tilde{\mathbf{m}}^{\left(k\right)}$ and $\mathbf{\boldsymbol{\Sigma}}^{\left(k\right)}$
with $\tilde{\boldsymbol{\Sigma}}^{\left(k\right)}$ in the profile
loglikelihood (Eq. \ref{eq:P(S|W,B*)}).
\item As we show in Appendix S1, section \ref{sec:Likelihood-Derivatives}
the profile loglikelihood (Eq. \ref{eq:P(S|W,B*)}) is concave, and
so it is easy to maximize the log-posterior and obtain the MAP estimate
of $\mathbf{W}$. Additionally, it is straightforward to calculate
the gradient, Hessian and the relevant Lipschitz constant of this
loglikelihood. Moreover, since the profile loglikelihood (Eq. \ref{eq:P(S|W,B*)})
decomposes over the rows of $\mathbf{W}$ 
\begin{equation}
\max_{\mathbf{b}}\ln P\left(\mathbf{W}|\mathbf{S},\mathbf{b}\right)=\sum_{i}\max_{\mathbf{b}}\ln P\left(\mathbf{W}_{i,\cdot}|\mathbf{S},\mathbf{b}\right)\,.\label{eq:posterior decomposition}
\end{equation}
as do many of the log-priors we will use (section \ref{sub:Priors}),
the optimization problem of finding the MAP estimate can be parallelized
over the rows of $\mathbf{W}$.
\item The ML estimator of $\mathbf{W}$ (the maximizer of Eq. \ref{eq:P(S|W,B*)})
can be derived analytically (Appendix S1, section \ref{sec:Maximum-Likelihood-Estimator})
by equating the gradient to zero. The result is 
\begin{equation}
\mathbf{W}_{\mathrm{MLE}}=\mathbf{A}^{-1}\boldsymbol{\Sigma}^{\left(1\right)}\left(\boldsymbol{\Sigma}^{\left(0\right)}\right)^{-1}\,,\label{eq: W_MLE}
\end{equation}
where $A_{i,j}\triangleq\delta_{i,j}\sqrt{\left(\frac{\pi}{8}h\left(m_{i}\right)\right)^{2}-\left(\boldsymbol{\Sigma}^{\left(1\right)}\left(\boldsymbol{\Sigma}^{\left(0\right)}\right)^{-1}\left(\boldsymbol{\Sigma}^{\left(1\right)}\right)^{\top}\right)_{i,i}\pi/8}\,.$
This estimate would coincide with any MAP estimate when $T\rightarrow\infty$.
\item Note a finite solution exists only if the last expression has a real
value.  Interestingly, this ML estimate is a rescaled version of
the ML estimate in a simple linear Gaussian neuron model. A similar
ML estimate was obtained, using similar approximations, for a Poisson
neuron model \cite{Ramirez2013}, albeit with a different re-scaling.
This may hint at the generality of this form, and its applicability
for other non-linear models.
\item Due to an approximation we make in the derivation (Eq. \ref{eq:Wang approx})
of the profile loglikelihood (Eq. \ref{eq: P(S|W,B*) appendix}),
the scale factor $A_{i,i}$ tends to be smaller then it should be.
This ``shrinkage'' affects all estimators (ML or MAP) based on the
profile loglikelihood. However, this issue can be corrected by re-fitting
the scale factors, as explained in Appendix S1, section \ref{sec:Correcting-ampitudes-and}.
\item If the neural input is small (due to weak weights, low firing rates
or a small number of neurons), the profile loglikelihood (Eq. \ref{eq: P(S|W,B*) appendix})
reduces to a simple quadratic form 
\begin{equation}
\max_{\mathbf{b}}\ln P\left(\mathbf{S}|\mathbf{W},\mathbf{b}\right)\approx T\sum_{i=1}^{N}\left[\sum_{j=1}^{N}\left[W_{i,j}\Sigma_{i,j}^{\left(1\right)}\right]-h\left(m_{i}\right)\frac{\pi}{8}\sum_{k,j}W_{i,j}\Sigma_{k,j}^{\left(0\right)}W_{i,k}\right]\,.\label{eq: Quadratic Loglikelihood}
\end{equation}
This is similar to the loglikelihood that we would have a obtained
if we would have assumed a linear Gaussian neuron model, albeit with
a different constants. Therefore, the non-linear nature of the neurons
in our model only becomes important when the neuronal input is strong.
\item Though we assumed the network does not have a stimulus ($\mathbf{G}=0$),
a stimulus can be incorporated into the inference procedure. To do
this, we treat the stimulus $\mathbf{X}_{\cdot,t}$ simply as the
activity of additional, fully observed, neurons (albeit $X_{i,t}\in\mathbb{R}$
while $S_{i,t}\in\left\{ 0,1\right\} $). Specifically, we define
a new ``spikes'' matrix $\mathbf{S}^{\mathrm{new}}\triangleq\left(\mathbf{S}^{\top},\mathbf{X}^{\top}\right)^{\top}$,
a new connectivity matrix 
\[
\mathbf{W}^{\mathrm{new}}\triangleq\left(\begin{array}{cc}
\mathbf{W} & \mathbf{G}\\
0^{D\times N} & 0^{D\times D}
\end{array}\right)\,,
\]
and a new observation matrix $\mathbf{O}^{\mathrm{new}}\triangleq\left(\mathbf{O}^{\top},1^{T\times D}\right)^{\top}.$
Repeating the derivations for $\mathbf{S}^{\mathrm{new}},\mathbf{W}^{\mathrm{new}}$
and $\mathbf{O}^{\mathrm{new}}$, we obtain the same profile loglikelihood.
Once it is used to infer $\mathbf{W}^{\mathrm{new}}$, we extract
the estimates of $\mathbf{W}$ and $\mathbf{G}$ from their corresponding
blocks in $\mathbf{W}^{\mathrm{new}}$. 
\item Formally, we need to make sure the conditions of CLT-based approximation
\cite{Diaconis1984} are fulfilled for our approximated method to
work, and this can become even more challenging with the addition
of (arbitrary) external inputs. However, as can be seen from the simulations,
such a generalized CLT-based approximation tends to work quite well
even when the neuronal input is not strictly Gaussian \cite{Teh2006,Ribeiro2011,Soudry2014}.
For example, in Figure \ref{fig:Sparsity}, on average, there are
about three non-zero spikes in the input of each neuron. In Figure
\ref{fig:The-effect-of-stimulus}, our algorithm is still accurate
even though the empirical distribution of the inputs cannot be Gaussian
- since it is bimodal due to strong the periodical pulse input.
\end{enumerate}

\subsection{Priors and optimization methods \label{sub:Priors}}

In this section we examine different priors $P_{0}\left(\mathbf{W}\right)$
on the network connectivity that can be incorporated into the posterior:
the sparsity of inter-neuron connections (section \ref{sub:Sparse-prior});
the division of neurons into several ``types'' (section \ref{sub:Stochastic-block-model});
and the fact that connection probability decreases with distance (section
\ref{sub:Distance-dependence}).

\subsubsection{Sparsity inducing prior\label{sub:Sparse-prior}}

The diagonal elements $W_{i,i}$ of the connectivity matrix are typically
negative, corresponding to the post spike filter accounting for the
cell's own (refractory) post-spike effects, while the off-diagonal
terms $W_{i,j}$ represent the connection weights from neuron $j$
to neuron $i$. As most neurons are not connected, most of the off-diagonal
$W_{i,j}$ are equal to zero, so \textbf{$\mathbf{W}$ }is a sparse
matrix.

\paragraph*{L1 norm.}

The most popular approach for incorporating prior information about
sparsity is to use a Laplace distribution as prior, which adds an
$L_{1}$ norm penalty : 
\begin{equation}
\ln P_{0}(\mathbf{W})=-\sum_{i,j}\lambda_{i,j}\left|W_{i,j}\right|+C\label{eq:L1 prior}
\end{equation}
for some set of sparsity parameters $\lambda_{i,j}$. In order to
take into account the fact that the off-diagonal terms are zero, we
set 
\begin{equation}
\lambda_{i,j}=\left(1-\delta_{i,j}\right)\lambda\,,\label{eq:lambda mask}
\end{equation}
where $\lambda>0$. This prior has a number of advantages: the resulting
log-posterior of $\mathbf{W}$ (given the full spike train $\mathbf{S}$
and the other system parameters) turns out to be a concave function
of $\mathbf{W}$, and the maximizer of this posterior, $\hat{\mathbf{W}}_{\mathrm{MAP}}$
is often sparse, i.e., many values of $\hat{\mathbf{W}}_{\mathrm{MAP}}$
are zero. Combining this prior in to the posterior (Eq. \ref{eq: log posterior}),
together with the simplified profile loglikelihood (Eq. \ref{eq:P(S|W,B*)}),
we obtain a LASSO problem for $\mathbf{W}$ \cite{Tibs96}. By solving
this objective we obtain a sparse Maximum A Posteriori (MAP) estimate
$\hat{\mathbf{W}}_{\mathrm{MAP}}$. There are many algorithms that
can be used to solve such an objective. Usually, the most efficient
\cite{Bach2011} is the FISTA algorithm \cite{Beck2009}; details
are given in Appendix S1, section \ref{sec:The-FISTA-algorithm}.
In order to set the value of $\lambda$, we assume some prior knowledge
(\emph{e.g.}, statistics from anatomical data) about the sparsity
of $\mathbf{W}$ - \emph{i.e.}, how many neurons are connected. Using
this knowledge it straightforward to set the value of $\lambda$ using
a binary search algorithm, as explained in Appendix S1, section \ref{sec:Setting lambda}.

\paragraph*{L0 norm.}

A more direct approach to sparse optimization is to replace the L1
norm with the ``L0 norm'' penalty%
\footnote{Formally, this is not an actual norm, and also the prior is not a
valid distribution.%
}

\begin{equation}
\ln P_{0}(\mathbf{W})=-\sum_{i,j}\lambda_{i,j}\mathcal{I}\left[W_{i,j}\neq0\right]+C\,,\label{eq: L0 prior}
\end{equation}
where the $\lambda_{i,j}$ are given by Eq. \ref{eq:lambda mask}.
This makes the resulting log-posterior non-concave and hard to maximize.
Fortunately, forward greedy algorithms (Appendix S1, section \ref{sec:Greedy-Algorithms})
usually work quite well with such a penalty \cite{Elad2010}. Briefly,
these algorithms initially assume that all weights are zero, and then
iteratively repeat the following steps: (1) Fix all the weights from
the previous step, and choose one (previously zero) weight which would
maximally increase the profile loglikelihood (Eq. \ref{eq:P(S|W,B*)})
if it is allowed to be non-zero. (2) Extend the support of non-zero
weights to include this weight. (3) Maximize the profile loglikelihood
(Eq. \ref{eq:P(S|W,B*)}) over this support, with all the other weights
being zero. A few comments are in order:
\begin{enumerate}
\item In our case (Eq. \ref{eq:P(S|W,B*)}), all steps can be performed
analytically. 
\item In such an approach there is no need to set regularization parameters
$\lambda_{i,j}$, since we only need to stop the algorithm when the
required level of sparsity has been reached. 
\item In the case that weights are weak and the profile loglikelihood becomes
quadratic (Eq. \ref{eq: Quadratic Loglikelihood}) we get the familiar
Orthogonal Matching Pursuit algorithm (OMP, \cite{Elad2010}), which
is typically faster.
\item Forward-backward greedy algorithms \cite{Zhang} can also be used
here, and may offer improved performance. We leave this for future
work.
\end{enumerate}
As we will show later, empirically, $L0$ and $L1$ give comparable
performance, with $L0$ being slower. However, the main advantage
of this $L0$ approach is that it can be more easily extended to include
other types of prior information, as we discuss next.

\subsubsection{Stochastic block model\label{sub:Stochastic-block-model}}

Neurons can be classified into distinct types \cite{kandel1991principles},
based on structural, synaptic, genetic and functional characteristics.
The connection strength between two neurons usually depends on the
types of the pre and post synaptic neurons \cite{Perin29032011,Seung2014,Jonas2014}. 

A well known example for this is ``Dale's Law'' - that all neurons
are either excitatory or inhibitory. Therefore, all the outgoing synaptic
connection from a neuron should have the same sign. This specific
information can be incorporated into our estimates of the weight matrix
$\mathbf{W}$, by using a greedy approach, as described in \cite[section 2.5.2]{Mishchenko2011c}.
In this approach we first maximize the posterior without considering
Dale's law, and find an estimate $\hat{\mathbf{W}}$. Then, neurons
are classified as excitatory or inhibitory according to the difference
in the count of outgoing weights of each sign $\mathcal{C}_{i}\left(\mathbf{\hat{\mathbf{W}}}\right)\triangleq|\{i:\hat{W}_{ij}>0\}|-|\{i:\hat{W}_{ij}<0\}|$.
Specifically, setting $\theta$ as some threshold (we arbitrarily
chose $\theta=0.25$), a neuron $j$ was classified as excitatory
if $\mathcal{C}_{i}\left(\mathbf{\hat{\mathbf{W}}}\right)>\theta$,
as inhibitory if $\mathcal{C}_{i}\left(\hat{\mathbf{W}}\right)<-\theta$,
and is not classified if $\left|\mathcal{C}_{i}\left(\mathbf{\hat{\mathbf{W}}}\right)\right|<\theta$.
Next, we maximize again the posterior with additional constraints
- that in all classified neurons, all outgoing weights have the proper
sign for their class. This procedure is then iterated.

However, in order to include a more general type-related information
into the connectivity prior, here we also used a ``stochastic block
model'' approach \cite{Nowicki2007,Adams14,Jonas2014}. We will assume
that all the non-zero weights are sampled from a Gaussian distribution
with a fixed variance, and a mean $V_{i,j}$ which depends only on
the pre and post synaptic types, and thus have a block structure,
where each block has a fixed value (\textbf{$\mathbf{V}$ }in\textbf{
}Figure \ref{fig:Schematic-description}). As a result, in addition
to any previous sparsity-inducing log-prior, we will add also the
following penalty
\begin{equation}
-\lambda_{V}\sum_{i,j}\mathcal{I}\left[W_{i,j}\neq0\right]\left(W_{i,j}-V_{i,j}\right)^{2}\,,\label{eq: sbm penalty}
\end{equation}
where $\lambda_{V}$ is a regularization parameter, and $\mathbf{V}$
is the aforementioned block structured matrix. If $\mathbf{V}$ is
known, we can add this penalty to the $L0$ norm penalty, and solve
using the same greedy algorithm, with a modified objective function. 

However, usually $\mathbf{V}$ is not known, and all we know (approximately)
is the number of types. Since the number of types is greater or equal
to the rank of $\mathbf{V}$, and this rank is usually much lower
then $N$, we can penalize the rank of $\mathbf{V}$. A convex relaxation
of this is $\mathbf{\left\Vert V\right\Vert _{*}}$, the nuclear norm
\cite{Fazel}, which we add to the log-posterior. Next, we describe
how to maximize the full log-posterior (which is a sum of Eqs. \ref{eq:P(S|W,B*)},
\ref{eq: sbm penalty} and $\mathbf{\lambda_{*}\left\Vert V\right\Vert _{*}}$).
We first initialize $\mathbf{V}=\boldsymbol{0}^{N\times N}$. Then
we alternate between: (1) holding $\mathbf{V}$ fixed and maximizing
the log-posterior over $\mathbf{W}$ (the sum of Eqs. \ref{eq:P(S|W,B*)}
and \ref{eq: sbm penalty} ) (2) holding $\mathbf{W}$ fixed and maximizing
the log-posterior over $\mathbf{V}$ (the sum of Eqs. \ref{eq: sbm penalty}
and $\mathbf{\lambda_{*}\left\Vert V\right\Vert _{*}}$). The first
step can be done using a forward greedy approach, as we did for the
$L0$ penalty. The second step can be done using soft-impute algorithm
\cite{Mazumder2010a}, assuming we know in advance this rank (which
determines $\lambda_{*}$). Thus, we update $\mathbf{W}$ and $\mathbf{V}$
iteratively until convergence. The value of $\lambda_{V}$ is determined
using a cross-validation set. More details are given in Appendix S1,
section \ref{sec:Greedy-Algorithms}.

\subsubsection{Distance dependence\label{sub:Distance-dependence}}

\begin{figure}[H]
\begin{centering}
\begin{center}
\includegraphics[width=0.8\columnwidth]{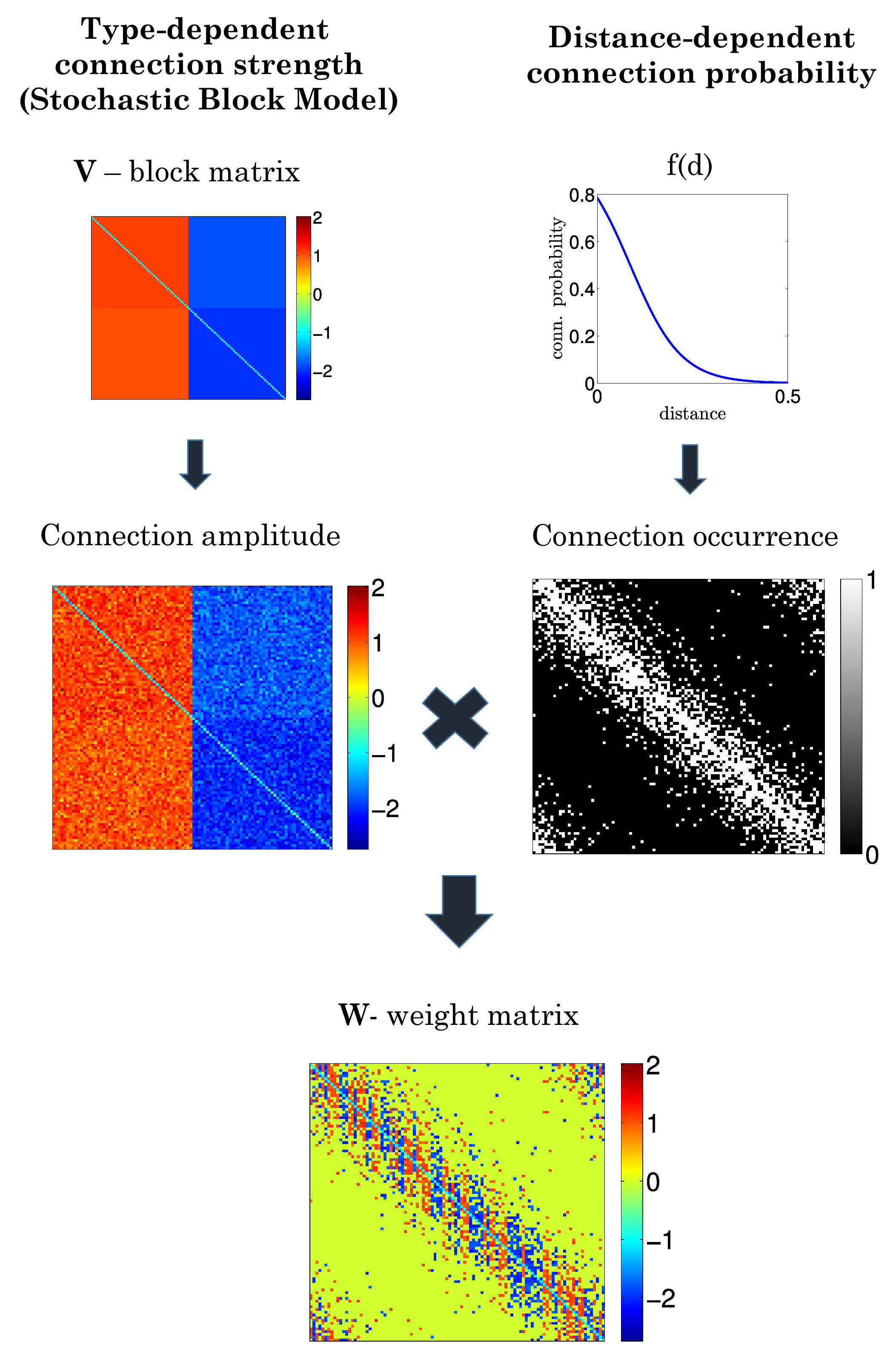}
\par\end{center}
\par\end{centering}

\protect\caption{\textbf{The prior information on the network connectivity. }Type dependence
(left): The connection strength between neurons depends on the type
of the pre- and post- synaptic neurons. This induces a block structured
matrix $\mathbf{V}$, which, when corrupted by noise, gives the connection
strength of the non-zero connections. In this figure we have 2 types
- excitatory and inhibitory. Distance dependence (right): the connection
probability between neurons declines with distance, which affects
which synaptic connections exist. We randomly set neuron locations
$x\in\left[0,1\right)$ on a ring lattice (so we have more connections
near the diagonal). Thus, after the neuronal types and the distance
dependence produce the connectivity matrix $\mathbf{W}$. Naturally,
the neuronal type also affects connection probability and distance
also affects connection amplitude, but here we did not use this information.
\label{fig:Schematic-description}}
\end{figure}

\begin{figure}[H]
\begin{centering}
\begin{center}
\includegraphics[width=0.8\columnwidth]{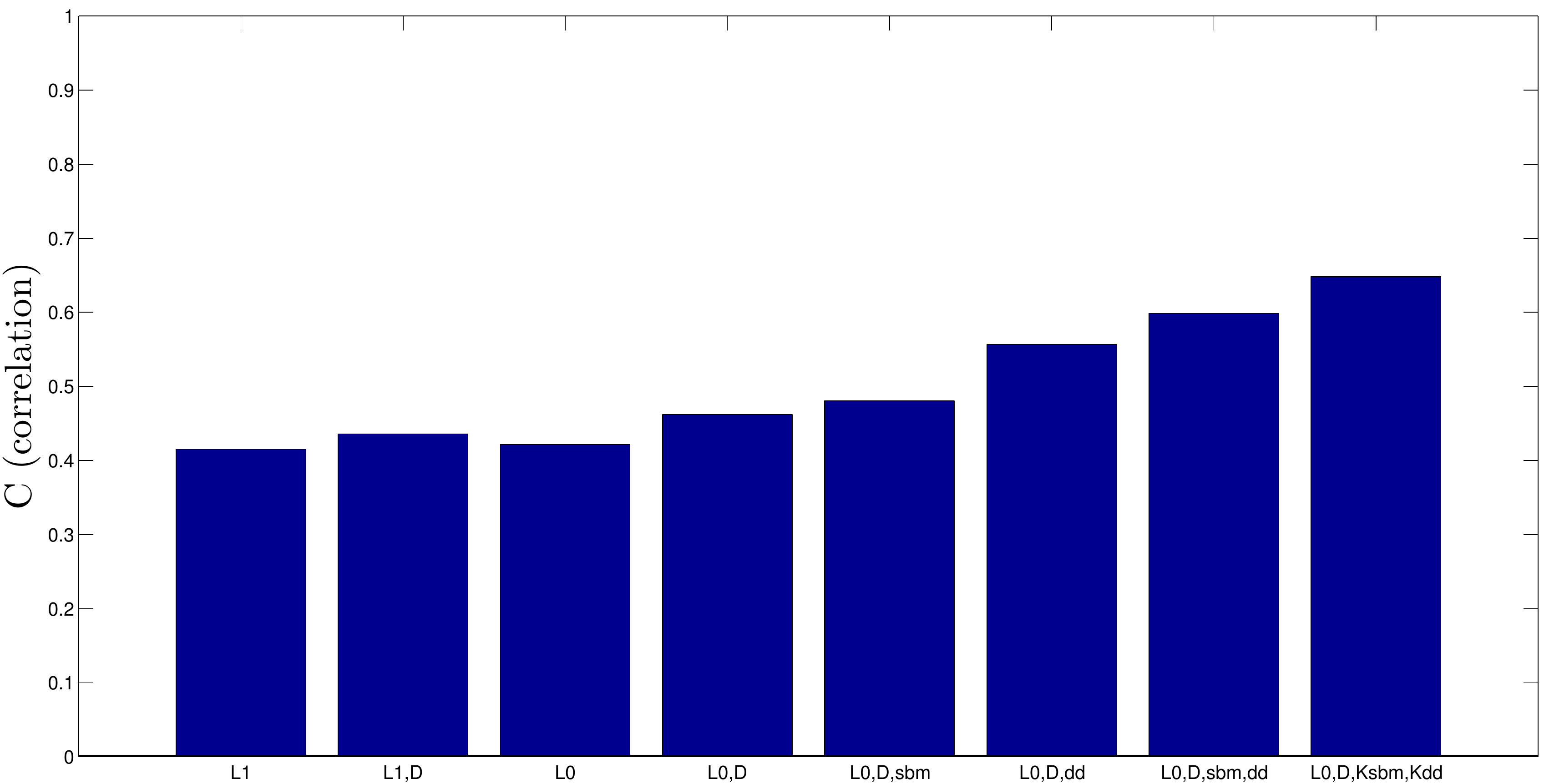}
\par\end{center}
\par\end{centering}

\textcolor{magenta}{\protect\caption{\textbf{Incorporating more prior information can improve the performance
of the MAP estimate. }We use a network with two types of neurons (excitatory
and inhibitory), with the connectivity $\mathbf{W}$ determined by
a stochastic block model following the structure of the block matrix
$\mathbf{V}$ and the distance dependence $f\left(d\right)$ (See
Figure \ref{fig:Schematic-description}). The correlation metric $C$
(Eq. \ref{eq:C}) measures quality of estimation, where we used different
priors in the MAP estimate: $L1$ (Eq. \ref{eq:L1 prior}); $L0$
(Eq. \ref{eq: L0 prior}); $D$ - Dale's Law \cite[section 2.5.2]{Mishchenko2011c};
sbm - type dependent connection strength (stochastic block model,
section \ref{sub:Stochastic-block-model}) in which we only in advance
know the number of types (two, which is also the rank of $\mathbf{V}$);
dd - distance dependent connection probability in which we $f\left(d\right)$
has the general form $1/\left(1+\mathrm{exp}\left(ad+b\right)\right)$,
but we do not know $a$ and $b$ (section \ref{sub:Distance-dependence});
Ksbm - stochastic block model with a known mean block matrix $\mathbf{V}$
;and Kdd - distance dependent connection probability with a known
function $f\left(d\right)$. We conclude that performance can significantly
improve when we incorporate stronger prior information into our estimates.
$N=100,T=10^{5},p_{\mathrm{obs}}=0.05,\,\mathrm{and}\, b_{i}\sim\mathcal{N}\left(-1.5,0.1\right)$.\textcolor{magenta}{{}
\label{fig:Prior Results}}}
}
\end{figure}

The probability of a connection between two neurons often declines
with distance \cite{Perin29032011,Seung2014,Jonas2014}. This prior
information can also be used to our advantage. Suppose first that
the probability of two neurons being connected is given by some known
function $f\left(d_{i,j}\right)$, where $d_{i,j}$ is the distance
between the neurons $i$ and $j$ (Figure \ref{fig:Schematic-description}).
A simple way to incorporate this knowledge to modify the regularization
constants $\lambda_{i,j}$ (Eq. \ref{eq:lambda mask}) in either the
L1 penalty (Eq. \ref{eq:L1 prior}) or the L0 penalty (Eq. \ref{eq: L0 prior})
so they are distance dependent
\[
\lambda_{i,j}=\frac{\left(1-\delta_{i,j}\right)\lambda}{f\left(d_{i,j}\right)}\,.
\]
This way the prior probability for a connection reflects this distance
dependence - if a connection is less probable, its regularization
constant is higher. 

If $f\left(d\right)$ is unknown, we give it some parametric form.
For example, we will use here $f\left(d\right)=\left(1+\exp\left(-ax-b\right)\right)^{-1}$.
To find $\left(a,b\right)$, we alternate between maximizing the posterior
over $\mathbf{W}$ and then over $\left(a,b\right)$. More details
are given in Appendix S1, section \ref{sec:Greedy-Algorithms}.

\subsection{Alternative inference methods - unobserved spikes as latent variables
\label{sec:Other-Inference-Methods}}

Conceptually, having missing spike observations in the data should
make it much harder to estimate $\mathbf{W}$ from data. This is because,
in this case, we need to modify the posterior (Eq. \ref{eq: log posterior}),
and replace the complete likelihood $P\left(\mathbf{S}|\mathbf{W},\mathbf{b}\right)$
with the incomplete likelihood $P\left(S_{\mathrm{obs}}|\mathbf{W},\mathbf{b}\right)$
where 
\[
S_{\mathrm{obs}}\triangleq\left\{ \forall i,t:S_{i,t}|O_{i,t}=1\right\} \,,
\]
\emph{i.e., }the set of all observed spikes from \textbf{$\mathbf{S}$}.
However, in order to obtain $P\left(S_{\mathrm{obs}}|\mathbf{W},\mathbf{b}\right)$
from $P\left(\mathbf{S}|\mathbf{W},\mathbf{b}\right)$ we need to
perform an intractable marginalization over an exponential number
of unobserved spike configurations.

The method we derived here sidesteps this problem. We infer the connectivity
matrix $\mathbf{W}$ using a MAP estimate, derived using an approximation
of the complete loglikelihood $\ln P\left(\mathbf{S}|\mathbf{W},\mathbf{b}\right)$
(Eq. \ref{eq: log posterior}). Though the complete loglikelihood
includes all the spikes (even the unobserved ones), in its approximated
form we can also handle missing observations. This is done by simply
ignoring missing observations and re-normalizing accordingly the sufficient
statistics (Eq. \ref{eq: m partial}-\ref{eq:Sigma Partial}) of the
approximated loglikelihood (Eq. \ref{eq:P(S|W,B*)}). This procedure
does not affect the asymptotic value of the sufficient statistics,
since the observations and spikes are uncorrelated.

In contrast, classical inference methods \cite{Bishop2006} treat
the unobserved spikes as latent variables, and attempt to infer them
using one of these standard approaches:
\begin{itemize}
\item Markov chain Monte Carlo (MCMC) - which samples the latent variables
and weights from the complete posterior. 
\item Expectation Maximization (EM) - which provides an estimator that locally
optimizes the posterior of $W$ given the observed data.
\item Variational Bayes (VB) - which approximates the complete posterior
using a factorized form.
\end{itemize}
We experimented with all these approaches on the posterior, without
using any of the simplifying approximations used in our main method
(section \ref{sec:Bayesian Inference}). In order to clarify the motivation
for our main method, we briefly describe these alternative approaches
below, including their advantages and disadvantages.

\subsubsection{Markov chain Monte Carlo (MCMC)\label{sub:Markov-chain-Monte}}

We assume a sparsity promoting spike and slab prior on the weights
\begin{equation}
P_{0}\left(\mathbf{W}\right)=\prod_{i,j}\left[\left(1-p_{0}\right)\delta\left(W_{i,j}\right)+p_{0}\mathcal{N}\left(W_{i,j}|\mu_{0},\sigma_{0}^{2}\right)\right]\label{eq:spike and slab}
\end{equation}
and use a Gibbs approach to sample jointly from $P(\mathbf{S},\mathbf{W}|S_{\mathrm{obs}},\mathbf{b})$:
first we sample $\mathbf{S}$ given the observed data and the current
sample from $\mathbf{W}$, and then sample $\mathbf{W}$ given $\mathbf{S}$.
Note the standard L1 prior on $\mathbf{W}$ (Eq. \ref{eq:L1 prior})
would not promote sparsity in this MCMC setting, since then every
sample of $\mathbf{W}$ would be non-zero with probability $1$.

In the first step we draw one sample from the posterior of $\mathbf{S}$
given the observed data and an estimated $\mathbf{W}$, using the
improved Metropolized Gibbs sampler \cite{Liu1996,Liu02}, as described
in Appendix S1, section \ref{sub:Sampling-the-spikes}. This sampler
is quite simple and is highly parallelizable in this setting, because
the graphical model representing the posterior $p(\mathbf{S}|S_{\mathrm{obs}},\mathbf{W},\mathbf{b})$
is local: $\mathbf{S}_{\cdot,t}$ is conditionally independent of
$\mathbf{S}_{\cdot,t+2}$ given $\mathbf{S}_{\cdot,t+1}$ since the
spiking at time $t$ only directly affects the spiking in the next
time step (\emph{i.e.}, model \ref{eq: logistic}-\ref{eq: U} can
be considered a Markov chain in $\mathbf{S}_{\cdot,t}$). Therefore
we can alternately sample the spiking vectors $\mathbf{S}_{\cdot,t}$
at all odd times $t$ completely in parallel, and then similarly for
the even times. In more general GLM models, where the neuronal input
depends on previous $k$ spikes, \emph{i.e.}, $\left\{ \mathbf{S}_{\cdot,t-k}\right\} _{l=1}^{k}$,
then the algorithm could be executed using $T/\left(k+1\right)$ parallel
computations. We note that more sophisticated sampling methods have
been developed for this type of problem \cite{Mishchenko2011b}; these
specialized methods are significantly more efficient if implemented
serially, but do not parallelize as well as the simple Gibbs-based
approach we used here. The performance of the sampler is exemplified
in Figure \ref{fig:Estimating-the-spikes}, bottom. As can be seen
in this Figure, usually (unless connectivity weights are very high)
the spike sampler can predict the spikes only in a ``small neighborhood''
of the visible spikes. 

In the second step, we sample $\mathbf{W}$ given $\mathbf{S}$. We
first note that, with the spike and slab prior the, again the posterior
factorizes: $P(\mathbf{W}|\mathbf{S},\mathbf{b})=\prod_{i}P(\mathbf{W}_{i,.}|\mathbf{S}_{i,.},\mathbf{b})$.
Thus, we can sample from each $P(\mathbf{W}_{i,.}|\mathbf{S}_{i,.},\mathbf{b})$
in parallel. To sample from $P(\mathbf{W}_{i,.}|\mathbf{S}_{i,.},\mathbf{b})$,
we simply Gibbs sample, one element $W_{i,j}$ at a time. One way
to do this is described in Appendix S1, section \ref{sub:Sampling-the-weights}.
The derivation requires the assumption that the weights are small
(empirically, it works well if $W_{i,,j}<1$). Alternatively, this
assumption is not necessary if we use instead the (slower) approach
discussed in \cite{mohamed2012bayesian}, which requires the computation
of some one-dimensional integrals. In our case these integrands are
log-concave and therefore uni-modal, and can be calculated using a
Laplace approximations. Since both methods contain some approximation,
we first need to use them to obtain proposals densities for a Metropolis-Hastings
scheme, which then generates the $W_{i,j}$ samples \cite{Liu02}.

In general, computational speed is the main disadvantage of the MCMC
method - both for the sampling the spikes and for sampling the weights.
For large networks, this approach can only be used if the sampling
scheme is completely parallelized over many processors. However, MCMC
performs rather well on small networks, similarly to our main method
(section \ref{sec:Bayesian Inference}). An important advantage of
the MCMC method over our main method is that it can be used even if
some ``spike pairs'' are never observed, \emph{i.e. }if for some
$i,j$, $\left\langle O_{i,t}O_{j,t-k}\right\rangle _{T}$ for $k=0$
or $1$. Therefore, the MCMC method might be used complement our main
method to infer the weights in this case. It should also be noted
that the MCMC approach offers somewhat simpler methods for hyperparameter
selection (via sampling from the hyperparameter posterior), and easier
incorporation into larger hierarchical models that can include richer
prior information about the network and enable proper sharing of information
across networks.

\subsubsection{Expectation Maximization (EM)}

The EM approach is similar to the methods discussed in \cite{Pillow2007,Mishchenko2011c}.
The E step requires the computation of an integral over $p(\mathbf{S}|S_{\mathrm{obs}},\mathbf{W})$.
Since this integral is high-dimensional and not analytically tractable,
we again resort to MCMC methods to sample the spikes, using the same
sampler as before. We found that in most cases it was not necessary
to take many samples from the posterior; for large enough network
sizes $N$ (and for correspondingly long experimental times $T$),
$\mathbf{S}$ is large enough that a single sample contains enough
information to adequately estimate the necessary sufficient statistics
in the E step. See \cite{Diebolt1994} for further discussion. In
the M step we perform maximization over the log-posterior averaged
over $p(\mathbf{S}|S_{\mathrm{obs}},\mathbf{W})$, with an L1 prior
\textcolor{black}{(Eq. \ref{eq:L1 prior}). This can be done using
FISTA (Appendix S1, section \ref{sec:The-FISTA-algorithm}), or any
other tool for L1-penalized GLM estimation, as in \cite{Pillow2007,Mishchenko2011c}.}
The EM approach is is still rather slow, since we need to sample from
the spikes. Moreover, the EM approach typically suffered from shrinkage
and exhibited worse performance then the MCMC approach.

\subsubsection{Variational Bayes (VB)\label{sub:Variational-Bayes-(VB)}}

Using the standard VB approach \cite{Bishop2006} and the spike and
slab prior on the weights (Eq. \ref{eq:spike and slab}) we can approximate
the posterior $P(\mathbf{S},\mathbf{W}|S_{\mathrm{obs}},\mathbf{b})$
using a fully factorized distribution 
\begin{equation}
Q\left(\mathbf{S},\mathbf{W}\right)\triangleq\prod_{i,t}q_{i,t}\left(S_{i,t}\right)\prod_{k,r}q_{k,r}\left(W_{k,r}\right)\,,\label{eq:Q factorized}
\end{equation}
in which the factors are calculated using 
\begin{eqnarray*}
\ln q_{i,t}\left(S_{i,t}\right) & = & \ln\E_{/S_{i,t}}P(\mathbf{S},\mathbf{W}|S_{\mathrm{obs}},\mathbf{b})\\
\ln q_{k,r}\left(W_{k,r}\right) & = & \ln\E_{/W_{k,r}}P(\mathbf{S},\mathbf{W}|S_{\mathrm{obs}},\mathbf{b})\,,
\end{eqnarray*}
where, for any random variable $X$, $\E_{/X}$ is an expectation
performed using the distribution $Q\left(\mathbf{S},\mathbf{W}|X\right)$.
This calculation proceeds almost identically to the derivation of
the Gibbs samplers (Appendix S1, section \ref{sec:MCMC-Approach}),
except we need to perform the expectation $\E_{/X}$ over the results.
These integrals can be calculated using the CLT and the approximations
given in \cite[sections 2.4 and 4.1]{Wang2013a}. 

The factorized form in Eq. \ref{eq:Q factorized} assumes that, approximately,
all the weights and all the spikes are independent. In this method,
which is faster then MCMC, the mean firing rates of the neurons could
be estimated reasonably well, as is exemplified in Figure \ref{fig:Estimating-the-spikes},
bottom. Unfortunately, the weight matrix $\mathbf{W}$ could not be
estimated if less than $30\%$ of the neurons were observed in each
timestep. This is in contrast to the other methods, in which the observed
fraction can be arbitrarily low. We believe this happens because the
VB approximation for the spikes ignores spike correlations - which
determine the sufficient statistics in this case (as suggested by
Eq. \ref{eq:P(S|W,B*)}). However, it is possible that more sophisticated
factorized forms (which do not assume spike independence) will still
work.

\section*{Acknowledgments}

The authors are thankful to Eftychios Pnevmatikakis, Ari Pakman and
Ben Shabado for their help and support, and to Ran Rubin, Yuriy Mishchenko
and Ron Meir for their helpful comments. This work was partially supported by the Gruss Lipper Charitable Foundation,
grant ARO MURI W911NF-12-1-0594, grant DARPA W91NF-14-1-0269, and
grant NSF CAREER IOS-0641912. Additional support was provided by the
Gatsby Foundation.

\bibliographystyle{plos2009}

\begin{thebibliography}{10}
\providecommand{\url}[1]{\texttt{#1}}
\providecommand{\urlprefix}{URL }
\expandafter\ifx\csname urlstyle\endcsname\relax
  \providecommand{\doi}[1]{doi:\discretionary{}{}{}#1}\else
  \providecommand{\doi}{doi:\discretionary{}{}{}\begingroup
  \urlstyle{rm}\Url}\fi
\providecommand{\bibAnnoteFile}[1]{%
  \IfFileExists{#1}{\begin{quotation}\noindent\textsc{Key:} #1\\
  \textsc{Annotation:}\ \input{#1}\end{quotation}}{}}
\providecommand{\bibAnnote}[2]{%
  \begin{quotation}\noindent\textsc{Key:} #1\\
  \textsc{Annotation:}\ #2\end{quotation}}
\providecommand{\eprint}[2][]{\url{#2}}

\bibitem{Stevenson2011}
Stevenson IH, Kording KP (2011) {How advances in neural recording affect data
  analysis.}
\newblock Nature neuroscience 14: 139--42.
\input{Stevenson2011}

\bibitem{Alivisatos2012}
Alivisatos AP, Chun M, Church GM, Greenspan RJ, Roukes ML, et~al. (2012) {The
  Brain Activity Map Project and the Challenge of Functional Connectomics}.
\newblock Neuron 74: 970--974.
\input{Alivisatos2012}

\bibitem{NYK03b}
Nykamp DQ (2003) {Reconstructing stimulus-driven neural networks from spike
  times}.
\newblock NIPS 15: 309--316.
\input{NYK03b}

\bibitem{Paninski2004}
Paninski L (2004) {Maximum likelihood estimation of cascade point-process
  neural encoding models}.
\newblock Network: Computation in Neural Systems 15: 243--262.
\input{Paninski2004}

\bibitem{PILL05a}
Pillow JW, Paninski L, Uzzell V, Simoncelli E, Chichilnisky EJ (2005)
  {Prediction and Decoding of Retinal Ganglion Cell Responses with a
  Probabilistic Spiking Model}.
\newblock Journal of Neuroscience 25: 11003--11013.
\input{PILL05a}

\bibitem{Rigat06}
Rigat F, de~Gunst M, van Pelt J (2006) {Bayesian modelling and analysis of
  spatio-temporal neuronal networks}.
\newblock Bayesian Analysis 1: 733--764.
\input{Rigat06}

\bibitem{Nykamp2007}
Nykamp DQ (2007) {A mathematical framework for inferring connectivity in
  probabilistic neuronal networks.}
\newblock Mathematical biosciences 205: 204--51.
\input{Nykamp2007}

\bibitem{Pillow2007}
Pillow JW, Latham P (2007) {Neural characterization in partially observed
  populations of spiking neurons}.
\newblock NIPS .
\input{Pillow2007}

\bibitem{Pillow2008}
Pillow JW, Shlens J, Paninski L, Sher A, Litke AM, et~al. (2008)
  {Spatio-temporal correlations and visual signalling in a complete neuronal
  population.}
\newblock Nature 454: 995--9.
\input{Pillow2008}

\bibitem{Mishchenko2011c}
Mishchenko Y, Vogelstein JT, Paninski L (2011) {A Bayesian approach for
  inferring neuronal connectivity from calcium fluorescent imaging data}.
\newblock The Annals of Applied Statistics 5: 1229--1261.
\input{Mishchenko2011c}

\bibitem{Stetter2012}
Stetter O, Battaglia D, Soriano J, Geisel T (2012) {Model-free reconstruction
  of excitatory neuronal connectivity from calcium imaging signals.}
\newblock PLoS computational biology 8: e1002653.
\input{Stetter2012}

\bibitem{Paninski2012}
Paninski L, Vidne M, DePasquale B, Ferreira DG (2012) {Inferring synaptic
  inputs given a noisy voltage trace via sequential Monte Carlo methods.}
\newblock Journal of computational neuroscience 33: 1--19.
\input{Paninski2012}

\bibitem{Pakman2012}
Pakman A, Huggins J, Smith C, Paninski L (2014) {Fast penalized state-space
  methods for inferring dendritic synaptic connectivity}.
\newblock Journal of computational neuroscience 36: 415--443.
\input{Pakman2012}

\bibitem{Gerhard2013}
Gerhard F, Kispersky T, Gutierrez GJ, Marder E, Kramer M, et~al. (2013)
  {Successful reconstruction of a physiological circuit with known connectivity
  from spiking activity alone.}
\newblock PLoS computational biology 9: e1003138.
\input{Gerhard2013}

\bibitem{Lutcke2013}
L\"{u}tcke H, Gerhard F, Zenke F, Gerstner W, Helmchen F (2013) {Inference of
  neuronal network spike dynamics and topology from calcium imaging data.}
\newblock Frontiers in neural circuits 7: 201.
\input{Lutcke2013}

\bibitem{Memmesheimer2014}
Memmesheimer RM, Rubin R, Olveczky BP, Sompolinsky H (2014) {Learning precisely
  timed spikes.}
\newblock Neuron 82: 925--38.
\input{Memmesheimer2014}

\bibitem{Fletcher2014}
Fletcher AK, Rangan S (2014) {Scalable Inference for Neuronal Connectivity from
  Calcium Imaging}.
\newblock In: NIPS.
\newblock \eprint{1409.0289}.
\input{Fletcher2014}

\bibitem{Nykamp2008}
Nykamp DQ (2008) {Pinpointing connectivity despite hidden nodes within
  stimulus-driven networks}.
\newblock Physical Review E 78: 021902.
\input{Nykamp2008}

\bibitem{Vidne08}
Vidne M, Ahmadian Y, Shlens J, Pillow JW, Kulkarni J, et~al. (2012) {Modeling
  the impact of common noise inputs on the network activity of retinal ganglion
  cells}.
\newblock J Computational Neuroscience 33: 97--121.
\input{Vidne08}

\bibitem{Mishchenko2011b}
Mishchenko Y, Paninski L (2011) {Efficient methods for sampling spike trains in
  networks of coupled neurons}.
\newblock The Annals of Applied Statistics : 1--26.
\input{Mishchenko2011b}

\bibitem{Turaga2013a}
Turaga S, Buesing L, Packer AM, Dalgleish H, Pettit N, et~al. (2013) {Inferring
  neural population dynamics from multiple partial recordings of the same
  neural circuit}.
\newblock In: NIPS.
\input{Turaga2013a}

\bibitem{Romano2014}
Romano L, Opper M (2014) {Inferring hidden states in a random kinetic Ising
  model: replica analysis}.
\newblock arXiv preprint arXiv:14054164 .
\input{Romano2014}

\bibitem{Tyrcha2014}
Tyrcha J, Hertz J (2014) {Network inference with hidden units.}
\newblock Mathematical biosciences and engineering : MBE 11: 149--65.
\input{Tyrcha2014}

\bibitem{Ahrens2013}
Ahrens MB, Orger MB, Robson DN, Li JM, Keller PJ (2013) {Whole-brain functional
  imaging at cellular resolution using light-sheet microscopy.}
\newblock Nature methods 10: 413--20.
\input{Ahrens2013}

\bibitem{Venter1998}
Venter JC, Adams MD, Sutton GG, Kerlavage AR, Smith HO, et~al. (1998) {Shotgun
  sequencing of the human genome}.
\newblock Science 280: 1540--2+.
\input{Venter1998}

\bibitem{RKFS08}
Reddy GD, Kelleher K, Fink R, Saggau P (2008) {Three-dimensional random access
  multiphoton microscopy for functional imaging of neuronal activity}.
\newblock Nature neuroscience 11: 713--720.
\input{RKFS08}

\bibitem{Grewe2010}
Grewe BF, Langer D, Kasper H, Kampa BM, Helmchen F (2010) {High-speed in vivo
  calcium imaging reveals neuronal network activity with near-millisecond
  precision.}
\newblock Nature methods 7: 399--405.
\input{Grewe2010}

\bibitem{Hochbaum2014}
Hochbaum DR, Zhao Y, Farhi SL, Klapoetke N, Werley Ca, et~al. (2014)
  {All-optical electrophysiology in mammalian neurons using engineered
  microbial rhodopsins.}
\newblock Nature methods 11.
\input{Hochbaum2014}

\bibitem{Mishchenko2011a}
Mishchenko Y, Paninski L (2011) {Efficient methods for sampling spike trains in
  networks of coupled neurons}.
\newblock The Annals of Applied Statistics : 1--26.
\input{Mishchenko2011a}

\bibitem{Park2011}
Park IM, Pillow JW (2011) {Bayesian Spike-Triggered Covariance Analysis.}
\newblock NIPS .
\input{Park2011}

\bibitem{Sadeghi12}
Sadeghi K, {Gauthier J L}, Field GD, Greschner M, Agne M, et~al. (2012) {Monte
  Carlo methods for localization of cones given multielectrode retinal ganglion
  cell recordings}.
\newblock Network : 1--25.
\input{Sadeghi12}

\bibitem{Ramirez2013}
Ramirez AD, Paninski L (2013) {Fast inference in generalized linear models via
  expected log-likelihoods.}
\newblock Journal of computational neuroscience : 1--29.
\input{Ramirez2013}

\bibitem{Diaconis1984}
Diaconis P, Freedman D (1984) {Asymptotics of graphical projection pursuit}.
\newblock The annals of statistics 12: 793--815.
\input{Diaconis1984}

\bibitem{Song2005}
Song S, Sj\"{o}str\"{o}m PJPJ, Reigl M, Nelson SB, Chklovskii DB (2005) {Highly
  nonrandom features of synaptic connectivity in local cortical circuits}.
\newblock PLoS biology 3: e68.
\input{Song2005}

\bibitem{Lichtman2008}
Lichtman JW, Livet J, Sanes JR (2008) {A technicolour approach to the
  connectome.}
\newblock Nature reviews Neuroscience 9: 417--22.
\input{Lichtman2008}

\bibitem{Chung2013}
Chung K, Wallace J, Kim S, Kalyanasundaram S, Andalman AS, et~al. (2013)
  {Structural and molecular interrogation of intact biological systems}.
\newblock Nature 497: 332--337.
\input{Chung2013}

\bibitem{Jonas2014}
Jonas E, Kording K (2014) {Automatic discovery of cell types and microcircuitry
  from neural connectomics}.
\newblock arXiv preprint arXiv:14074137 : 1--19.
\input{Jonas2014}

\bibitem{Berndt2009}
Berndt A, Yizhar O, Gunaydin LA, Hegemann P, Deisseroth K (2009) {Bi-stable
  neural state switches.}
\newblock Nature neuroscience 12: 229--34.
\input{Berndt2009}

\bibitem{Wallace2008}
Wallace D, Borgloh SzA, Astori S (2008) {Single-spike detection in vitro and in
  vivo with a genetic Ca2+ sensor}.
\newblock Nature \ldots .
\input{Wallace2008}

\bibitem{BRIL88}
Brillinger D (1988) {Maximum likelihood analysis of spike trains of interacting
  nerve cells}.
\newblock Biological Cyberkinetics 59: 189--200.
\input{BRIL88}

\bibitem{Mishchenko2015}
Mishchenko Y {Consistency of the complete neuronal population connectivity
  reconstructions using shotgun imaging}.
\newblock In prep .
\input{Mishchenko2015}

\bibitem{Rickgauer2014}
Rickgauer JP, Deisseroth K, Tank DW (2014) {Simultaneous cellular-resolution
  optical perturbation and imaging of place cell firing fields}.
\newblock Nature Neuroscience 17: 1816--1824.
\input{Rickgauer2014}

\bibitem{Shababo2013}
Shababo B, Brooks P, Pakman A, Paninski L (2013) {Bayesian Inference and Online
  Experimental Design for Mapping Neural Microcircuits}.
\newblock Advances in Neural Information Processing Systems .
\input{Shababo2013}

\bibitem{Vidne2012}
Vidne M, Ahmadian Y, Shlens J, Pillow JW, Kulkarni J, et~al. (2012) {Modeling
  the impact of common noise inputs on the network activity of retinal ganglion
  cells}.
\newblock Journal of Computational Neuroscience 33: 97--121.
\input{Vidne2012}

\bibitem{Gal2010}
Gal A, Eytan D, Wallach A, Sandler M, Schiller J, et~al. (2010) {Dynamics of
  Excitability over Extended Timescales in Cultured Cortical Neurons}.
\newblock Journal of Neuroscience 30: 16332--16342.
\input{Gal2010}

\bibitem{Soudry2014d}
Soudry D, Meir R (2014) {The neuronal response at extended timescales:
  long-term correlations without long-term memory}.
\newblock Frontiers in computational neuroscience .
\input{Soudry2014d}

\bibitem{Stirzaker2001}
Stirzaker D, Grimmett D (2001) {Probability and random processes}.
\newblock Oxford, 3rd edition.
\input{Stirzaker2001}

\bibitem{Teh2006}
Teh Y, Newman D, Welling M (2006) {A collapsed variational Bayesian inference
  algorithm for latent Dirichlet allocation}.
\newblock Advances in neural \ldots .
\input{Teh2006}

\bibitem{Ribeiro2011}
Ribeiro F, Opper M (2011) {Expectation propagation with factorizing
  distributions: a Gaussian approximation and performance results for simple
  models.}
\newblock Neural computation 23: 1047--69.
\input{Ribeiro2011}

\bibitem{Soudry2014}
Soudry D, Hubara I, Meir R (2014) {Expectation Backpropagation: parameter-free
  training of multilayer neural networks with real and discrete weights}.
\newblock In: NIPS (in press). Montreal, pp. 1--9.
\input{Soudry2014}

\bibitem{Tibs96}
Tibshirani R (1996) {Regression Shrinkage and Selection via the Lasso}.
\newblock Journal of the Royal Statistical Society Series B 58: 267--288.
\input{Tibs96}

\bibitem{Bach2011}
Bach F, Jenatton R (2011) {Convex optimization with sparsity-inducing norms}.
\newblock Optimization for Machine Learning : 1--35.
\input{Bach2011}

\bibitem{Beck2009}
Beck A, Teboulle M (2009) {A Fast Iterative Shrinkage-Thresholding Algorithm
  for Linear Inverse Problems}.
\newblock SIAM Journal on Imaging Sciences 2: 183--202.
\input{Beck2009}

\bibitem{Elad2010}
Elad M (2010) {Sparse and redundant representations: from theory to
  applications in signal and image processing}.
\newblock New York, NY: Springer New York.
\input{Elad2010}

\bibitem{Zhang}
Zhang T (2009) {Adaptive forward-backward greedy algorithm for sparse learning
  with linear models}.
\newblock Advances in Neural Information Processing Systems : 1--8.
\input{Zhang}

\bibitem{kandel1991principles}
Kandel ER, Schwartz JH, Jessell TM (2000) {Principles of neural science},
  volume~3.
\newblock McGraw Hill, 4th edition.
\input{kandel1991principles}

\bibitem{Perin29032011}
Perin R, Berger TK, Markram H (2011) {A synaptic organizing principle for
  cortical neuronal groups}.
\newblock Proceedings of the National Academy of Sciences 108: 5419--5424.
\input{Perin29032011}

\bibitem{Seung2014}
Seung H, S\"{u}mb\"{u}l U (2014) {Neuronal Cell Types and Connectivity: Lessons
  from the Retina}.
\newblock Neuron 83: 1262--1272.
\input{Seung2014}

\bibitem{Nowicki2007}
Nowicki K, Snijders T (2001) {Estimation and prediction for stochastic
  blockstructures}.
\newblock Journal of the American Statistical Association 96: 1077--1087.
\input{Nowicki2007}

\bibitem{Adams14}
Linderman S, Adams R (2014) {Discovering Latent Network Structure in Point
  Process Data}.
\newblock Arxiv 1402.0914.
\input{Adams14}

\bibitem{Fazel}
Fazel M, Hindi H, Boyd SP (2001) {A rank minimization heuristic with
  application to minimum order system approximation}.
\newblock American Control Conference 6: 4734--4739.
\input{Fazel}

\bibitem{Mazumder2010a}
Mazumder R, Hastie T, Tibshirani R (2010) {Spectral regularization algorithms
  for learning large incomplete matrices}.
\newblock The Journal of Machine Learning \ldots 11: 2287--2322.
\input{Mazumder2010a}

\bibitem{Bishop2006}
Bishop CM (2006) {Pattern recognition and machine learning}.
\newblock Singapore: Springer.
\input{Bishop2006}

\bibitem{Liu1996}
Liu JS (1996) {Metropolized independent sampling with comparisons to rejection
  sampling and importance sampling}.
\newblock Statistics and Computing 6: 113--119.
\input{Liu1996}

\bibitem{Liu02}
Liu J (2002) {Monte Carlo Strategies in Scientific Computing}.
\newblock Springer.
\input{Liu02}

\bibitem{mohamed2012bayesian}
Mohamed S, Heller K, Ghahramani Z (2012) {Bayesian and L1 approaches to sparse
  unsupervised learning}.
\newblock Proceedings of the 29th International Conference on Machine Learning
  (ICML-12) .
\input{mohamed2012bayesian}

\bibitem{Diebolt1994}
Diebolt J, Ip E, Olkin I (1994) {A stochastic EM algorithm for approximating
  the maximum likelihood estimate}.
\newblock Technical report, Stanford University.
\input{Diebolt1994}

\bibitem{Wang2013a}
Wang S, Manning CD (2013) {Fast dropout training}.
\newblock ICML 28.
\input{Wang2013a}

\end{thebibliography}

\section*{Figure Legends}

\begin{figure}[H]
\begin{centering}
\begin{center}
\includegraphics[width=0.8\columnwidth]{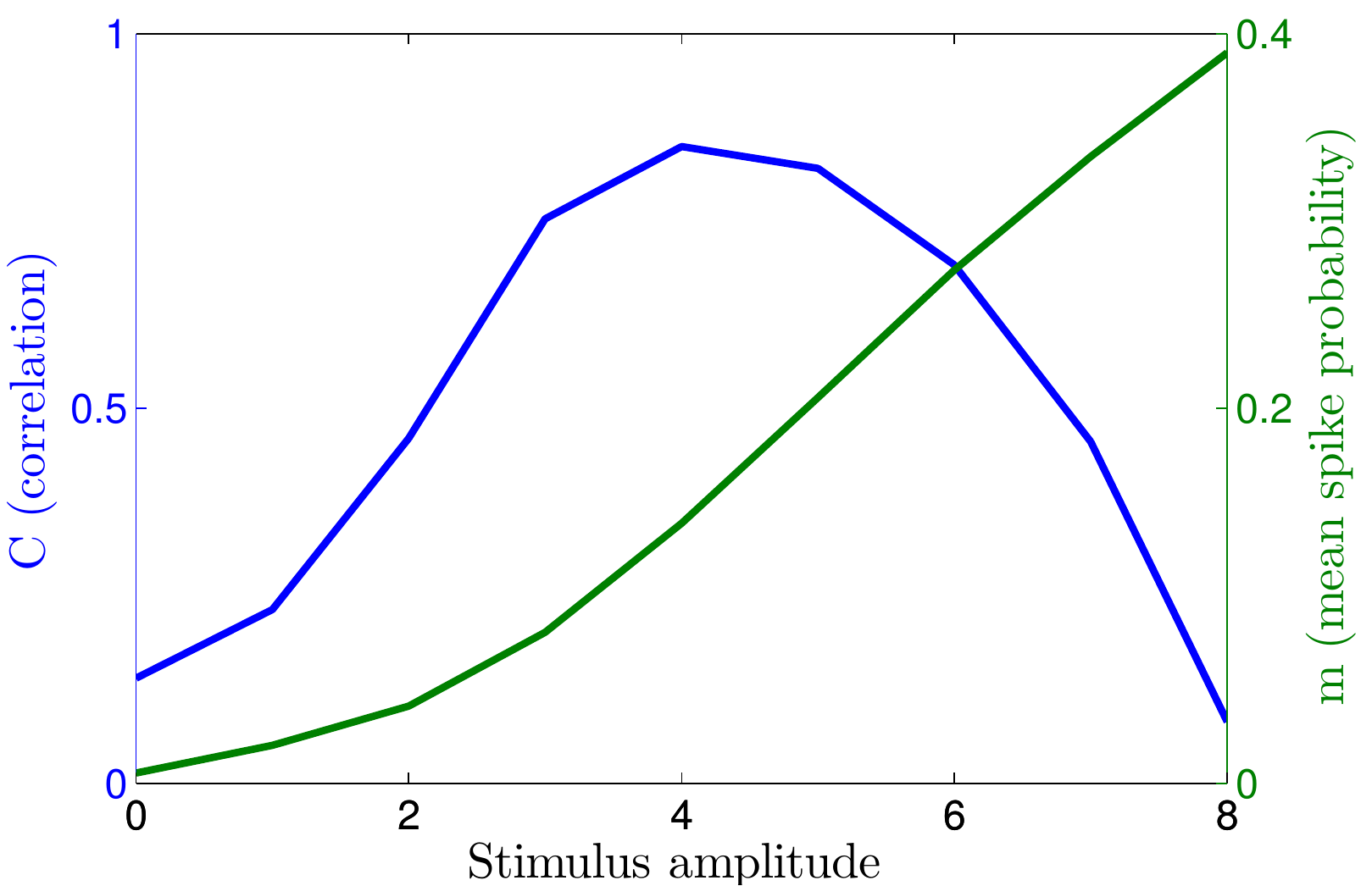}
\par\end{center}
\par\end{centering}

\protect\caption{\textbf{The optimal stimulus magnitude. }We examine the effect of
a periodic pulse stimulus on mean spike probability $m$ and reconstruction
correlation $C$ (Eq. \ref{eq:C}). See section \ref{fig:The-effect-of-stimulus})
for more details. Parameters: $N=50,T=10^{4},p_{\mathrm{obs}}=0.2,\,\mathrm{and}\, b_{i}\sim\mathcal{N}\left(-5.2,0.1\right)$,
a low firing rate regime. We find that there is an optimal stimulus
magnitude near mean spike probability of $m=0.13$. \label{fig:The-effect-of-stimulus}.}
\end{figure}

\textcolor{magenta}{}

\begin{figure}[H]
\begin{centering}
\begin{center}
\includegraphics[width=0.8\columnwidth]{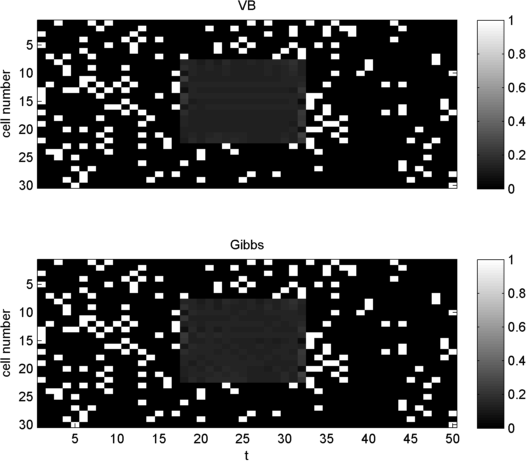}
\par\end{center}
\par\end{centering}

\protect\caption{Estimating the spikes using a latent-variable approach with a known
network connectivity ($\mathbf{W},\mathbf{b}$). Each raster plot
shows the spikes activity in a circular network with local connectivity
(spikes are in white). For neurons $10-20$ and times $18-33$ the
spiking activity is unobserved. We try to estimate this activity using
Variational Bayes method (VB\emph{, Top}), or using Gibbs sampling
(\emph{bottom}). In the unobserved rectangle, the shade indicates
the probability of having a spike - between $0$ (black) and $1$
(white). Each neuron in the network has excitatory connections to
its neighbors and self-inhibition. Given this connectivity, we can
see both methods give reasonable estimates near the edges of unobserved
rectangle. Further from the edges, the estimation becomes less certain
and converges to the mean spike probability for that neuron. \label{fig:Estimating-the-spikes}}
\end{figure}

\section*{Supporting Information Captions}

\textbf{Appendix S1.} Mathematical appendix with full derivations
and algorithmic details.

\section*{Tables}

\begin{table}[H]
\protect\caption{Basic notation\label{tab:Basic-notation}}

\centering %
\begin{tabular}{ll}
\hline & \tabularnewline
$N$  & Total number of neurons\tabularnewline
$T$  & Total number of time bins\tabularnewline
$p_{\mathrm{obs}}$ & Observation probability - the fraction of neurons observed at each
timebin\tabularnewline
$p_{\mathrm{conn}}$ & Network sparsity - the average probability that two neurons are directly
connected\tabularnewline
$\mathbf{S}$  & $N\times T$ matrix of spike activity\tabularnewline
$\mathbf{W}$  & $N\times N$ matrix of synaptic connection weights\tabularnewline
\textbf{$\mathbf{U}$} & $N\times T$ matrix of neuronal inputs\tabularnewline
$\mathbf{b}$  & $N\times1$ vector of neuronal biases\tabularnewline
$D$  & Total number of external inputs\tabularnewline
$\mathbf{X}$ & $D\times T$ matrix of external input to network\tabularnewline
$\mathbf{G}$  & $N\times D$ matrix of input gain\tabularnewline
$\mathbf{O}$ & $N\times T$ binary matrix denoting when spikes are observed\tabularnewline
$C$ & Unimportant constant\tabularnewline
\end{tabular}
\end{table}

\appendix
\newpage{}

\part*{Supporting Information - Appendix S1}

Code will be available on author's website after publication.

\section{Derivation of the simplified loglikelihood (Eq. \ref{eq:P(S|W,B*)})
\label{sub:Derivation}}

\begin{figure}[H]
\begin{centering}
\includegraphics[width=0.5\columnwidth]{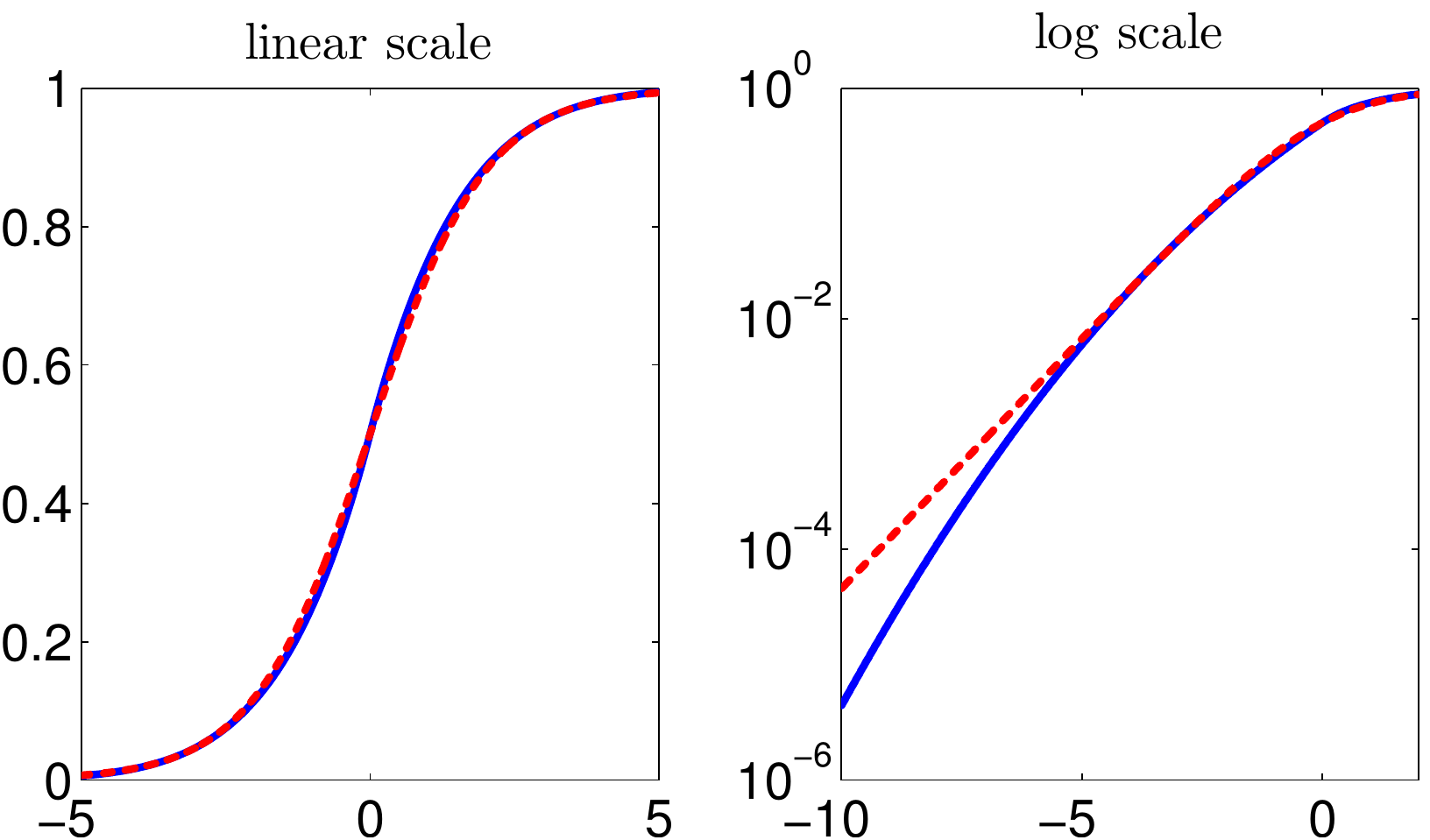}
\par\end{centering}

\protect\caption{\textbf{Visualization of the approximation in Eq. \ref{eq:Wang approx}.}
We compare of right hand side (red) and left hand side (blue) of the
approximation. Note this approximation breaks if $\left|x\right|$
is too large (\emph{e.g.}, $\left|x\right|>6$). This is because the
asymptotic forms of both sides of Eq. \ref{eq:Wang approx} are different.
For example, it is easy to show that, as $x\rightarrow-\infty$, the
left hand side converges as $e^{-\frac{1}{2}x^{2}}/\sqrt{-x}$ while
the right hand side converges as $e^{\left(\sqrt{\pi/8}\right)x}$.
\label{fig:Visualization-of-the-WangApprox}}
\end{figure}

Recall Eq. \ref{eq: logistic} 
\[
P\left(S_{i,t}|U_{i,t}\right)=\frac{e^{S_{i,t}U_{i,t}}}{1+e^{U_{i,t}}}\,.
\]
and Eq. \ref{eq: U} with $\mathbf{G}=0$ 
\[
\mathbf{U}_{\cdot,t}\triangleq\mathbf{W}\mathbf{S}_{\cdot,t-1}+\mathbf{b}\,.
\]
Combining both together for times $t=1,\cdots,T$ and neurons $i=1,\dots,N$,
we obtain

\begin{eqnarray}
 &  & \ln P\left(\mathbf{S}|\mathbf{W},\mathbf{b}\right)\nonumber \\
 & = & \sum_{i=1}^{N}\sum_{t=1}^{T}\ln\left[\frac{e^{S_{i,t}U_{i,t}}}{1+e^{U_{i,t}}}\right]\nonumber \\
 & = & \sum_{i=1}^{N}\sum_{t=1}^{T}\left[S_{i,t}U_{i,t}-\ln\left(1+e^{U_{i,t}}\right)\right],\nonumber \\
 & = & T\sum_{i=1}^{N}\left[\left\langle S_{i,t}U_{i,t}\right\rangle _{T}-\left\langle \ln\left(1+e^{U_{i,t}}\right)\right\rangle _{T}\right]\nonumber \\
 & \overset{\left(1\right)}{\approx} & T\sum_{i=1}^{N}\left[\left\langle S_{i,t}U_{i,t}\right\rangle _{T}-\int\ln\left(1+e^{x}\right)\mathcal{N}\left(x|\left\langle U_{i,t}\right\rangle _{T},\mathrm{Var}_{T}\left(U_{i,t}\right)\right)dx\right]\nonumber \\
 & \overset{\left(2\right)}{\approx} & T\sum_{i=1}^{N}\left[\left\langle S_{i,t}U_{i,t}\right\rangle _{T}-\sqrt{1+\pi\mathrm{Var}_{T}\left(U_{i,t}\right)/8}\ln\left(1+\exp\left(\frac{\left\langle U_{i,t}\right\rangle _{T}}{\sqrt{1+\pi\mathrm{Var}_{T}\left(U_{i,t}\right)/8}}\right)\right)\right]\nonumber \\
 & \overset{\left(3\right)}{\approx} & T\sum_{i=1}^{N}\sum_{j=1}^{N}W_{i,j}\Sigma_{k,j}^{\left(1\right)}+m_{i}b_{i}\label{eq: P(S|W,b) simplified}\\
 & - & \sqrt{1+\pi\sum_{k,j}W_{i,j}\Sigma_{k,j}^{\left(0\right)}W_{i,k}/8}\ln\left(1+\exp\left(\frac{\sum_{k=1}^{N}W_{i,k}m_{k}+b_{i}}{\sqrt{1+\pi\sum_{k,j}W_{i,j}\Sigma_{k,j}^{\left(0\right)}W_{i,k}/8}}\right)\right)\,,\nonumber 
\end{eqnarray}
where we used the following
\begin{enumerate}
\item The neuronal input, as a sum of $N$ variables, tends to converge
to a Gaussian distribution, in the limit of large $N$, under rather
general conditions \cite{Diaconis1984}.
\item Eq. 8 from \cite{Wang2013a}, an approximation which builds upon the
following ``empiric'' observation (Figure \ref{fig:Visualization-of-the-WangApprox})
\begin{equation}
\int_{-\infty}^{x}\mathcal{N}\left(z|1,0\right)dz\approx\frac{1}{1+e^{-\left(\sqrt{\pi/8}\right)x}}\,.\label{eq:Wang approx}
\end{equation}

\end{enumerate}

\begin{enumerate}
\item Eq. \ref{eq: U} for the neuronal input and Eqs. \ref{eq:m}-\ref{eq: Sigma}
for the spike statistics, which yields 
\begin{eqnarray*}
\left\langle U_{i,t}\right\rangle _{T} & = & \sum_{k=1}^{N}W_{i,k}m_{k}+b_{i}\\
\mathrm{Var}_{T}\left(U_{i,t}\right) & = & \sum_{k=1}^{N}\sum_{j=1}^{N}W_{i,j}\Sigma_{k,j}^{\left(0\right)}W_{i,k}\,
\end{eqnarray*}

\end{enumerate}
Though in Eq. \ref{eq: P(S|W,b) simplified} the loglikelihood has
already become tractable (and depends only on the sufficient statistics
from Eqs. \ref{eq:m}-\ref{eq: Sigma}), we can simplify it further
by maximizing it over $\mathbf{b}$. To do so, we equate to zero the
derivative of the simplified loglikelihood (Eq. \ref{eq: P(S|W,b) simplified})
\begin{eqnarray*}
\frac{d}{db_{\alpha}}\ln P\left(\mathbf{S}|\mathbf{W},\mathbf{b}\right) & = & 0\,.
\end{eqnarray*}
Solving this equation, we obtain 
\[
m_{\alpha}=\left(1+\exp\left(-\frac{\sum_{k=1}^{N}W_{\alpha,k}m_{k}+b_{\alpha}}{\sqrt{1+\pi\sum_{k,j}W_{\alpha,j}\Sigma_{k,j}^{\left(0\right)}W_{\alpha,k}/8}}\right)\right)^{-1}
\]
so
\[
b_{\alpha}=\sqrt{1+\pi\sum_{k,j}W_{\alpha,j}\Sigma_{k,j}^{\left(0\right)}W_{\alpha,k}/8}\ln\left(\frac{m_{\alpha}}{1-m_{\alpha}}\right)-\sum_{k=1}^{N}W_{\alpha,k}m_{k}\,.
\]
Substituting this maximizer into Eq. \ref{eq: P(S|W,b) simplified},
we obtain 

\begin{eqnarray}
\max_{\mathbf{b}}\ln P\left(\mathbf{S}|\mathbf{W},\mathbf{b}\right) & = & T\sum_{i=1}^{N}\left[\sum_{j=1}^{N}\left[W_{i,j}\Sigma_{i,j}^{\left(1\right)}\right]-h\left(m_{i}\right)\sqrt{1+\pi\sum_{k,j}W_{i,j}\Sigma_{k,j}^{\left(0\right)}W_{i,k}/8}\right]\label{eq: P(S|W,B*) appendix}
\end{eqnarray}
which is Eq. \ref{eq:P(S|W,B*)}.

\section{Likelihood derivatives\label{sec:Likelihood-Derivatives}}

We examine the profile loglikelihood in Eq. \ref{eq:P(S|W,B*)}, divided
by $T$ for convenience
\[
L\left(\mathbf{W}\right)\triangleq\frac{1}{T}\max_{\mathbf{b}}\ln P\left(\mathbf{S}|\mathbf{W},\mathbf{b}\right)\approx\sum_{i=1}^{N}\left[\sum_{j=1}^{N}\left[W_{i,j}\Sigma_{i,j}^{\left(1\right)}\right]-h\left(m_{i}\right)\sqrt{1+\frac{\pi}{8}\sum_{k,j}W_{i,j}\Sigma_{k,j}^{\left(0\right)}W_{i,k}}\right]\,,
\]
Its gradient is
\begin{eqnarray}
\frac{\partial}{\partial W_{\alpha\beta}}L\left(\mathbf{W}\right) & = & \Sigma_{\alpha,\beta}^{\left(1\right)}-\frac{\pi}{8}h\left(m_{\alpha}\right)\frac{\sum_{k}W_{\alpha,k}\Sigma_{k,\beta}^{\left(0\right)}}{\sqrt{1+\pi\sum_{k,j}W_{\alpha,j}\Sigma_{k,j}^{\left(0\right)}W_{\alpha,k}/8}}\,.\label{eq: Grad}
\end{eqnarray}
Its Hessian is 
\begin{eqnarray}
\frac{\partial^{2}L\left(\mathbf{W}\right)}{\partial W_{\alpha\beta}\partial W_{\gamma\eta}} & = & \frac{\frac{\pi}{8}h_{2}\left(m_{\alpha}\right)}{\left[1+\pi\sum_{k,j}W_{\alpha,j}\Sigma_{k,j}^{\left(0\right)}W_{\alpha,k}/8\right]^{3/2}}\label{eq:Hessian}\\
 & \cdot & \delta_{\gamma,\alpha}\left[\frac{\pi}{8}\sum_{k,j}W_{\alpha,k}W_{\alpha,j}\left(\Sigma_{k,\beta}^{\left(0\right)}\Sigma_{j,\eta}^{\left(0\right)}-\Sigma_{k,j}^{\left(0\right)}\Sigma_{\eta,\beta}^{\left(0\right)}\right)-\Sigma_{\eta,\beta}^{\left(0\right)}\right]\,.\nonumber 
\end{eqnarray}

\paragraph*{Concavity }

In order show that $L\left(\mathbf{W}\right)$ is concave, we need
to prove that for any $\mathbf{W}$ and any matrix $\mathbf{Z}\in\mathbb{R}^{N\times N}$
\[
\sum_{\alpha,\beta,\gamma,\eta}Z_{\alpha\beta}Z_{\gamma\eta}\frac{\partial^{2}L\left(\mathbf{W}\right)}{\partial W_{\alpha\beta}\partial W_{\gamma\eta}}\leq0\,.
\]
Therefore, we need to calculate the sign of 
\begin{eqnarray*}
 &  & \sum_{\alpha,\beta,\eta}Z_{\alpha\beta}Z_{\alpha\eta}\left[\frac{\pi}{8}\sum_{k,j}W_{\alpha,k}W_{\alpha,j}\left(\Sigma_{k,\beta}^{\left(0\right)}\Sigma_{j,\eta}^{\left(0\right)}-\Sigma_{k,j}^{\left(0\right)}\Sigma_{\eta,\beta}^{\left(0\right)}\right)-\Sigma_{\eta,\beta}^{\left(0\right)}\right]\\
 & = & \sum_{\alpha}\left[\frac{\pi}{8}\left(\mathbf{W}_{\alpha,\cdot}^{\top}\boldsymbol{\Sigma}^{\left(0\right)}\mathbf{Z}_{\alpha,\cdot}\right)^{2}-\left(1+\frac{\pi}{8}\mathbf{W}_{\alpha,\cdot}^{\top}\boldsymbol{\Sigma}^{\left(0\right)}\mathbf{W}_{\alpha,\cdot}\right)\mathbf{Z}_{\alpha,\cdot}^{\top}\boldsymbol{\Sigma}^{\left(0\right)}\mathbf{Z}_{\alpha,\cdot}\right]\,.
\end{eqnarray*}
Since (given $T$ is large enough) $\boldsymbol{\Sigma}^{\left(0\right)}$
is positive definite (since $\E\boldsymbol{\Sigma}^{\left(0\right)}$
is a covariance matrix), we can decompose $\boldsymbol{\Sigma}^{\left(0\right)}=\boldsymbol{\Lambda}\boldsymbol{\Lambda}^{\top}$.
Denoting $\boldsymbol{u}_{\alpha}\triangleq\boldsymbol{\Lambda}^{\top}\mathbf{W}_{\alpha,\cdot}$
and $\mathbf{v}_{\alpha}\triangleq\boldsymbol{\Lambda}^{\top}\mathbf{Z}_{\alpha,\cdot}$
we get that the last line can be written as 
\begin{eqnarray*}
 &  & \sum_{\alpha}\left[\frac{\pi}{8}\left(\left(\boldsymbol{u}_{\alpha}^{\top}\boldsymbol{v}_{\alpha}\right)^{2}-\boldsymbol{u}_{\alpha}^{\top}\boldsymbol{u}_{\alpha}\boldsymbol{v}_{\alpha}^{\top}\boldsymbol{v}_{\alpha}\right)-\boldsymbol{v}_{\alpha}^{\top}\boldsymbol{v}_{\alpha}\right]\\
 & \leq & -\sum_{\alpha}\boldsymbol{v}_{\alpha}^{\top}\boldsymbol{v}_{\alpha}\\
 & \leq & 0\,,
\end{eqnarray*}
where in the second line we used the Cauchy-Schwarz inequality $\left(\boldsymbol{u}^{\top}\boldsymbol{v}\right)^{2}\leq\boldsymbol{u}^{\top}\boldsymbol{u}\boldsymbol{v}^{\top}\boldsymbol{v}$,
and in the last we used the fact that $\boldsymbol{x}^{\top}\boldsymbol{x}\geq0$
for any vector $\mathbf{x}$. Therefore, $L\left(\mathbf{W}\right)$
is concave.

\paragraph*{Lipschitz constant}

For the FISTA algorithm (appendix \ref{sec:The-FISTA-algorithm})
we are required to calculate the Lipschitz constant of $\nabla L\left(\mathbf{W}\right)$.
A Lipschitz constant $l$ of a function $f$ is defined through 
\[
\forall\mathbf{x},\mathbf{y}:\,\,\,\left\Vert f\left(\mathbf{x}\right)-f\left(\mathbf{y}\right)\right\Vert \leq l\left\Vert \mathbf{x}-\mathbf{y}\right\Vert \,.
\]
We obtain that, in our case, 
\begin{equation}
l=\frac{\pi}{8}\max_{\alpha}h_{2}\left(m_{\alpha}\right)\lambda_{\mathrm{max}}\left[\boldsymbol{\Sigma}^{\left(0\right)}\right]\label{eq: Lipshitz constant}
\end{equation}
is the Lipschitz constant by observing that 
\begin{eqnarray*}
 &  & \left\Vert \frac{\partial}{\partial W_{\alpha\beta}}L\left(\mathbf{W}\right)-\frac{\partial}{\partial W_{\alpha\beta}^{\prime}}L\left(\mathbf{W}^{\prime}\right)\right\Vert ^{2}\\
 & = & \sum_{\alpha}\left[-\frac{\pi}{8}h\left(m_{\alpha}\right)\frac{\sum_{k}W_{\alpha,k}\Sigma_{k,\beta}^{\left(0\right)}}{\sqrt{1+\pi\sum_{k,j}W_{\alpha,j}\Sigma_{k,j}^{\left(0\right)}W_{\alpha,k}/8}}-\Sigma_{\alpha,\beta}^{\left(1\right)}-\frac{\pi}{8}h\left(m_{\alpha}\right)\frac{\sum_{k}W_{\alpha,k}^{\prime}\Sigma_{k,\beta}^{\left(0\right)}}{\sqrt{1+\pi\sum_{k,j}W_{\alpha,j}^{\prime}\Sigma_{k,j}^{\left(0\right)}W_{\alpha,k}^{\prime}/8}}\right]^{2}\\
 & \leq & \sum_{\alpha,\beta}\left[-\frac{\pi}{8}h\left(m_{\alpha}\right)\sum_{k}\Sigma_{k,\beta}^{\left(0\right)}\left(W_{\alpha,k}-W_{\alpha,k}^{\prime}\right)\right]^{2}\\
 & \leq & \left[\frac{\pi}{8}\max_{\alpha}h\left(m_{\alpha}\right)\lambda_{\mathrm{max}}\left[\boldsymbol{\Sigma}^{\left(0\right)}\right]\right]^{2}\sum_{\alpha,k}\left(W_{\alpha,k}-W_{\alpha,k}^{\prime}\right)^{2}\,,
\end{eqnarray*}
where $\lambda_{\mathrm{max}}\left[\mathbf{X}\right]$ is the maximal
eigenvalue of $\mathbf{X}$.

\section{Maximum Likelihood Estimator\label{sec:Maximum-Likelihood-Estimator}}

We wish to find $\mathbf{W}$ that solves
\[
0=\frac{\partial}{\partial W_{\alpha\beta}}\max_{\mathbf{b}}\ln P\left(\mathbf{S}|\mathbf{W},\mathbf{b}\right)\,.
\]
Using Eq. \ref{eq: Grad}, we obtain 
\begin{equation}
0=\Sigma_{\alpha,\beta}^{\left(1\right)}-\frac{\pi}{8}h\left(m_{\alpha}\right)\frac{\sum_{k}W_{\alpha,k}\Sigma_{k,\beta}^{\left(0\right)}}{\sqrt{1+\pi\sum_{k,j}W_{\alpha,j}\Sigma_{k,j}^{\left(0\right)}W_{\alpha,k}/8}}\,.\label{eq: Grad L =00003D0}
\end{equation}
We define

\begin{equation}
A_{\alpha,\beta}\triangleq\frac{\frac{\pi}{8}h\left(m_{\alpha}\right)\delta_{\alpha,\beta}}{\sqrt{1+\pi\sum_{k,j}W_{\alpha,j}\Sigma_{k,j}^{\left(0\right)}W_{\alpha,k}/8}}\label{eq:Q}
\end{equation}
so we can write Eq. \ref{eq: Grad L =00003D0} as
\[
0=\boldsymbol{\Sigma}^{\left(1\right)}-\mathbf{A}\mathbf{W}\boldsymbol{\Sigma}^{\left(0\right)}\,,
\]
which is solved by 
\[
\mathbf{W}=\mathbf{A}^{-1}\boldsymbol{\Sigma}^{\left(1\right)}\left(\boldsymbol{\Sigma}^{\left(0\right)}\right)^{-1}\,.
\]
Substituting this into Eq. \ref{eq:Q}, we have
\[
A_{\alpha,\alpha}=\frac{\frac{\pi}{8}h\left(m_{\alpha}\right)}{\sqrt{1+\left(\boldsymbol{\Sigma}^{\left(1\right)}\left(\boldsymbol{\Sigma}^{\left(0\right)}\right)^{-1}\left(\boldsymbol{\Sigma}^{\left(1\right)}\right)^{\top}\right)_{\alpha,\alpha}\pi A_{\alpha,\alpha}^{-2}/8}}
\]
so
\[
A_{\alpha,\alpha}=\sqrt{\left(\frac{\pi}{8}h\left(m_{\alpha}\right)\right)^{2}-\left(\boldsymbol{\Sigma}^{\left(1\right)}\left(\boldsymbol{\Sigma}^{\left(0\right)}\right)^{-1}\left(\boldsymbol{\Sigma}^{\left(1\right)}\right)^{\top}\right)_{\alpha,\alpha}\pi/8}\,.
\]

\section{Correcting amplitudes and biases\label{sec:Correcting-ampitudes-and}}

We noticed empirically that using Eq. \ref{eq: P(S|W,B*) appendix}
as the simplified loglikelihood tends to cause some inaccuracy in
our estimates of the weight gains and biases. To correct for this
error, we re-estimate the gains and biases in the following way. Suppose
we obtain a MAP estimate $\hat{\mathbf{W}}$ (Eq. \ref{eq: ML}) after
using the above profile likelihood. Next, we examine again the original
likelihood (without any approximation)
\[
\ln P\left(\mathbf{S}|\mathbf{b},\mathbf{W}\right)=\sum_{i=1}^{N}\left[\sum_{t=1}^{T}S_{i,t}\left(\sum_{j=1}^{N}W_{i,j}S_{j-1}+b_{j}\right)-\sum_{t=1}^{T}\ln\left(1+\exp\left(\sum_{j=1}^{N}W_{i,j}S_{j-1}+b_{j}\right)\right)\right]+C\,.
\]
We assume that the MAP estimate is accurate, up to a scaling constant
in each row, so $W_{i,j}=$$a_{i}\hat{W}_{i,j}$, and we obtain 
\begin{eqnarray}
 &  & \sum_{i=1}^{N}\left[\sum_{t=1}^{T}S_{i,t}\left(a_{i}\sum_{j=1}^{N}\hat{W}_{i,j}S_{j-1}+b_{j}\right)-\sum_{t=1}^{T}\ln\left(1+\exp\left(a_{i}\sum_{j=1}^{N}\hat{W}_{i,j}S_{j,t-1}+b_{j}\right)\right)\right]+C\nonumber \\
 & = & T\sum_{i=1}^{N}\left[a_{i}\sum_{j=1}^{N}\hat{W}_{i,j}\left\langle S_{i,t}S_{j,t-1}\right\rangle _{T}+b_{j}\left\langle S_{i,t}\right\rangle _{T}-\left\langle \ln\left(1+\exp\left(a_{i}\sum_{j=1}^{N}\hat{W}_{i,j}S_{j,t-1}+b_{j}\right)\right)\right\rangle _{T}\right]\label{eq: L(a,b)}\\
 & \approx & T\sum_{i=1}^{N}\left[a_{i}\sum_{j=1}^{N}\hat{W}_{i,j}\left(\tilde{\Sigma}_{i,j}^{\left(1\right)}+\tilde{m}_{i}\tilde{m}_{j}\right)+b_{j}\tilde{m}_{i}-\left\langle \ln\left(1+\exp\left(a_{i}z_{i,t}+b_{j}\right)\right)\right\rangle _{T}\right]+C\,,
\end{eqnarray}
where in the last line we used the expected loglikelihood approximation
with CLT again, and denoted (Recall Eqs. \ref{eq:m}-\ref{eq: Sigma})
\begin{equation}
z_{i,t}\sim N\left(\sum_{j=1}^{N}\hat{W}_{i,j}\tilde{m}_{j},\sum\limits _{j,k=1}^{N}\hat{W}_{i,j}\hat{W}_{i,k}\tilde{\Sigma}_{j,k}^{\left(0\right)}\right)\,.\label{eq:z_i}
\end{equation}
Sampling $z_{i,t}$ from \ref{eq:z_i} (we found that about $10^{3}$
samples was usually enough), we can calculate the expectation in (Eq.
\ref{eq: L(a,b)}). Next, we can now maximize the likelihood in (Eq.
\ref{eq: L(a,b)}) for each gain $a_{i}$ and bias $b_{i}$, separately
$\forall i$, by solving an easy 2D unconstrained optimization problem.
The new gains can now be used to adjust our estimation of $\mathbf{W}$.

\section{The FISTA algorithm\label{sec:The-FISTA-algorithm}}

We define the proximal operator \cite{Bach2011} 
\begin{eqnarray}
\left(\mathcal{T}_{\mu}\left(\mathbf{u}\right)\right)_{i} & \triangleq & \max\left(1-\frac{\mu}{\left|u_{i}\right|},0\right)u_{i}\,.\label{eq:Proximal 1}
\end{eqnarray}
FISTA solves the following minimization problem
\[
\min_{\mathbf{W}}\left[f\left(\mathbf{W}\right)+\lambda\left\Vert \mathbf{W}\right\Vert _{1}\right]\,.
\]
From \cite[Eqs. 4.1-4.3]{Beck2009}, we have the following algorithm
\ref{alg:The-FISTA-algorithm.}, where the component of the gradient
$\nabla f\left(\mathbf{W}\right)$ are given by Eq. \ref{eq: Grad},
and the Lipschitz constant $l$ is given by Eq. \ref{eq: Lipshitz constant}.

\begin{algorithm}[H]
\begin{description}
\item [{Input}] Initial point $\mathbf{W}^{\left(k\right)}$,$\nabla f$
and $l$ (Lipschitz constant of $\nabla f$).
\item [{Initialize}] $\mathbf{Y}^{\left(0\right)}=\mathbf{X}^{\left(0\right)}$,
$t_{1}=1$,$\mu=\lambda/l$.
\item [{Repeat}] For $k\geq1$ compute 
\begin{eqnarray*}
\mathbf{W}^{\left(k\right)} & = & \mathcal{T}_{\mu}\left[\mathbf{Y}^{\left(k\right)}-\frac{1}{l}\nabla f\left(\mathbf{Y}^{\left(k\right)}\right)\right]\\
t_{k+1} & = & \left(1+\sqrt{1+4t_{k}^{2}}\right)/2\\
\mathbf{Y}^{\left(k+1\right)} & = & \mathbf{W}^{\left(k\right)}+\left(\frac{t_{k}-1}{t_{k+1}}\right)\left(\mathbf{W}^{\left(k\right)}-\mathbf{W}^{\left(k-1\right)}\right)\,.
\end{eqnarray*}

\end{description}
\protect\caption{The FISTA algorithm. \label{alg:The-FISTA-algorithm.}}
\end{algorithm}

\section{Setting $\lambda$ using the sparsity constraint\label{sec:Setting lambda}}

As explained in section \ref{sub:Sparse-prior}, we use a sparsity
promoting prior (Eqs. \ref{eq:L1 prior}-\ref{eq:lambda mask}), which
depends on a regularization constant $\lambda$, to generate and estimate
$\hat{\mathbf{W}}\left(\lambda\right)$ of the connectivity matrix.
Though this constant is unknown in advance, we can set it using the
sparsity level of \textbf{$\mathbf{\hat{\mathbf{W}}\left(\lambda\right)}$},
defined as
\[
\mathrm{spar}\left(\lambda\right)\triangleq\frac{1}{N}\sum_{i,j}\mathcal{I}\left(\hat{W}_{i,j}\left(\lambda\right)\neq0\right)\,.
\]
We aim for this to be approximately equal to some target sparsity
level $p_{\mathrm{conn}}$. This is done using a fast binary search
algorithm (Algorithm \ref{alg:A-binary-search algorithm}), that exploits
the fact that $\mathrm{spar}\left(\lambda\right)$ is non-increasing
in $\lambda$. This monotonic behavior can be observed from the fixed
point of the FISTA algorithm
\[
\mathbf{W}=\max\left(1-\frac{\lambda/L}{\left|\mathbf{W}-\frac{1}{L}\nabla f\left(\mathbf{W}\right)\right|},0\right)\left[\mathbf{W}-\frac{1}{L}\nabla f\left(\mathbf{W}\right)\right]\,,
\]
for which the number of zeros components is clearly non-decreasing
with $\lambda$.

\begin{algorithm}[H]
\begin{description}
\item [{Input}] Target sparsity level - $p_{\mathrm{conn}}$, tolerance
level - $\epsilon$, measured sparsity - $\mathrm{spar}\left(\lambda\right)$.
\item [{Initialize}] Initial points $\lambda_{H}$ and $\lambda_{L}$,
where $\lambda_{H}\gg1\gg\lambda_{L}$. 
\item [{Repeat}]~

\begin{description}
\item [{$\lambda=\left(\lambda_{H}+\lambda_{L}\right)/2$}]~
\item [{If}] $\left|\mathrm{spar}\left(\lambda\right)-\theta\right|<\epsilon p_{\mathrm{conn}}$

\begin{description}
\item [{Return}] $\lambda$
\end{description}
\item [{end}]~
\item [{If}] \textbf{$\mathrm{spar}\left(\lambda\right)<p_{\mathrm{conn}}$}

\begin{description}
\item [{$\lambda_{H}=\lambda$}]~
\end{description}
\item [{else}]~

\begin{description}
\item [{$\lambda_{L}=\lambda$}]~
\end{description}
\item [{end}]~
\end{description}
\end{description}
\protect\caption{A binary search algorithm for setting $\lambda$. \label{alg:A-binary-search algorithm}}
\end{algorithm}

\section{Greedy algorithms \label{sec:Greedy-Algorithms}}

In this appendix we explain how greedy forward algorithms can be used
to solve some of the optimization problems described in section \ref{sub:Priors}.
In this section, we use the notation $X_{\alpha,Q}\triangleq\left\{ X_{\alpha,\beta}|\beta\in Q\right\} $.

\subsection{L0 penalty}

In this appendix we explain in detail how to greedily maximize $\max_{\mathbf{b}}\log P(\mathbf{S}|\mathbf{W},\mathbf{b})$
exactly for each row in the weight matrix $\mathbf{W}$ by incrementally
extending the support of non-zero weights. Since the problem is separable
(and parallelizable) over the rows (Eq. \ref{eq:posterior decomposition}),
we do this on a single row $\alpha$ - $\mathbf{W}_{\alpha,\cdot}$.
For that row, the normalized profile loglikelihood is 
\begin{equation}
L\left(\mathbf{W}_{\alpha,\cdot}\right)\triangleq\frac{1}{T}\max_{\mathbf{b}}\ln P\left(\mathbf{S}|\mathbf{W}_{\alpha,\cdot},\mathbf{b}\right)\approx\sum_{j=1}^{N}\left[W_{\alpha,j}\Sigma_{\alpha,j}^{\left(1\right)}\right]-h\left(m_{\alpha}\right)\sqrt{1+\frac{\pi}{8}\sum_{k,j}W_{\alpha,j}\Sigma_{k,j}^{\left(0\right)}W_{\alpha,k}}\,,\label{eq: L(W_alpha)}
\end{equation}

First we define the support of our estimate $\hat{\mathbf{W}}_{\alpha,\cdot}$
(the set of non-zero components) to be $Q\subset\{1,...,N\}$, and
we initialize $Q=\emptyset$. Then, we define $Q^{C}$, the complement
of the support (the set of zero weights), and initialize $Q^{C}=\{1,...,N\}$.
Lastly, we define $\hat{W}_{\alpha,Q}$, the set of non-zero weights.
Since $Q=\emptyset$, then we also initialize $\hat{W}_{\alpha,Q}=\emptyset$. 

Given $Q$ and $\hat{W}_{\alpha,Q}$ from the previous step, we extend
the support, by finding for each potential new weight index $\beta\in Q^{C}$
the best weight

\begin{equation}
W_{\alpha,\beta}^{*}=\mathrm{argmax}{}_{W_{\alpha,\beta}}L(W_{\alpha,\beta}|\hat{W}_{\alpha,Q}),\label{eq: greedy step 1}
\end{equation}
where $L(W_{\alpha,\beta}|\hat{W}_{\alpha,Q})$ is simply $L\left(\mathbf{W}_{\alpha,\cdot}\right)$
under the constraint that $\forall i\in Q:$ $W_{\alpha,i}=\hat{W}_{\alpha,i}$,
and $\forall i\in Q^{C}\backslash\beta:$ $W_{\alpha,i}=0.$ Substituting
these constraints into Eq. \ref{eq: L(W_alpha)}, we obtain 
\begin{eqnarray}
L(W_{\alpha,\beta}|\hat{W}_{\alpha,Q}) & = & \hat{W}_{\alpha,Q}^{T}\Sigma_{\alpha,Q}^{(1)}+\Sigma_{\alpha,Q}^{(1)}W_{\alpha,\beta}\label{eq: L(W_a,b|W_a,Q)}\\
 & - & h(m_{\alpha})\sqrt{1+\frac{\pi}{8}(\hat{W}_{\alpha,Q}^{T}\Sigma_{Q,Q}^{(0)}\hat{W}_{\alpha,Q}+2\hat{W}_{\alpha,Q}^{T}\Sigma_{Q,\beta}^{(0)}W_{\alpha,\beta}+\Sigma_{\beta,\beta}^{(0)}(W_{\alpha,\beta})^{2}}\,.\nonumber 
\end{eqnarray}
We can find the maximizing values for $W_{\alpha,\beta}$ exactly,
as the equation 
\begin{eqnarray*}
\frac{\partial}{\partial W_{\alpha,\beta}}L(W_{\alpha,\beta}|\hat{W}_{\alpha,Q}) & = & 0
\end{eqnarray*}
 is quadratic in $W_{\alpha,\beta}$:

\[
aW_{\alpha,\beta}^{^{2}}+bW_{\alpha,\beta}+c=0,
\]
with

\[
a\triangleq\left(\frac{\Sigma_{\alpha,\beta}^{(1)}}{\frac{\pi}{8}h(m_{\alpha})}\right)^{2}\frac{\pi}{8}\Sigma_{\beta,}^{(0)}-\left(\Sigma_{\beta,\beta}^{(0)}\right)^{2}
\]

\[
b\triangleq\left(\frac{\Sigma_{\alpha,\beta}^{(1)}}{\frac{\pi}{8}h(m_{\alpha})}\right)^{2}\frac{\pi}{4}\hat{W}_{\alpha,Q}^{T}\Sigma_{Q,\beta}^{(0)}-2\hat{W}_{\alpha,Q}^{T}\Sigma_{Q,\beta}^{(0)}\Sigma_{\beta,\beta}^{(0)}
\]

\[
c\triangleq\left(\frac{\Sigma_{\alpha,\beta}^{(1)}}{\frac{\pi}{8}h(m_{\alpha})}\right)^{2}\left(1+\frac{\pi}{8}\hat{W}_{\alpha,Q}^{T}\Sigma_{Q,Q}^{(0)}\hat{W}_{\alpha,Q}\right)-\left(\hat{W}_{\alpha,Q}^{T}\Sigma_{Q,\beta}^{(0)}\right)^{2}\,.
\]
Once we have updated the support by finding $W_{\alpha,\beta}^{*}$,
the weight that maximizes this log-likelihood, we can update the support
\begin{equation}
Q\leftarrow Q\cup\mathrm{argma}x_{\beta\in Q^{C}}L(W_{\alpha,\beta}^{*}|\hat{W}_{\alpha,Q})\,.\label{eq: greedy step 2}
\end{equation}
Then, we find the maximum likelihood estimate of weights, constrained
to the new support, to be

\begin{equation}
W_{\alpha,Q}=\underset{\mathbf{W}_{\alpha,\cdot}:W_{\alpha,Q^{c}}=0}{\mathrm{argmax}}L(\mathbf{W}_{\alpha,\cdot})=\frac{1}{A_{\alpha,\alpha}}\Sigma_{Q,Q}^{(0)^{-1}}\Sigma_{\alpha,Q}^{(1)}\,,\label{eq: greedy step 3}
\end{equation}
where $A_{\alpha,\alpha}=\sqrt{\left(\frac{\pi}{8}h(m_{\alpha})\right)^{2}-\frac{\pi}{8}\Sigma_{\alpha,Q}^{(1)^{T}}\Sigma_{Q,Q}^{(0)^{-1}}\Sigma_{\alpha,Q}^{(1)}}$
and we used similar derivations as in Appendix \ref{sec:Maximum-Likelihood-Estimator}.
We repeat this process until $|Q|/N$ is within tolerance of our sparsity
constraint ($p_{\mathrm{conn}}$).

\subsection{Stochastic block model}

In the stochastic block model (section \ref{sub:Stochastic-block-model}),
we add a quadratic penalty to the profile loglikelihood (Eq. \ref{eq: sbm penalty}),
which penalizes non-zero weights which are far from some block matrix
$\mathbf{V}$, which represent the mean value of the non-zero connection
strength between the different types of neurons. We do this by first
replacing $L(W_{\alpha,\beta}|\hat{W}_{\alpha,Q})$ in Eqs. \ref{eq: greedy step 1}-\ref{eq: greedy step 2}
with
\begin{equation}
L(W_{\alpha,\beta}|\hat{W}_{\alpha,Q})-\lambda_{V}(W_{\alpha,\beta}-V_{\alpha,\beta})^{2}\,,\label{eq:modified penalty}
\end{equation}
where $\mathbf{V}$ represents the underlying block matrix. In order
to maximize this, we again differentiate and equate to zero

\[
\Sigma_{\alpha,\beta}^{(1)}-\frac{\pi}{8}h(m_{\alpha})\frac{\sum_{k}W_{\alpha,k}\Sigma_{k,\beta}^{(0)}}{\sqrt{1+\frac{\pi}{8}\sum_{k,j}W_{\alpha,j}\Sigma_{k,j}^{(0)}W_{\alpha,k}}}-2\lambda(W_{\alpha,\beta}-V_{\alpha,\beta})=0\,.
\]
This equation, when expanded out, takes the form
\[
\left(b^{2}e\right)W_{\alpha,\beta}^{4}+\left(b^{2}d+2abe\right)W_{\alpha,\beta}^{3}+\left(b^{2}c+2abd+a^{2}e-g^{2}\right)W_{\alpha,\beta}^{2}+\left(2abc+2a^{2}d-2fg\right)W_{\alpha,\beta}+\left(a^{2}c-f^{2}\right)=0,
\]
where we defined 
\begin{eqnarray*}
a & \triangleq & \frac{\Sigma_{\alpha,\beta}^{(1)}+2\lambda V_{\alpha,\beta}}{\frac{\pi}{8}h(m_{\alpha})};\quad b\triangleq\frac{-2\lambda}{\frac{\pi}{8}h(m_{\alpha})};\quad c\triangleq1+\frac{\pi}{8}\left(W_{\alpha,Q}^{T}\Sigma_{Q,Q}^{(0)}W_{\alpha,Q}\right)\\
d & \triangleq & \frac{\pi}{4}W_{\alpha,Q}^{T}\Sigma_{Q,\beta}^{(0)};\quad e\triangleq\frac{\pi}{8}\Sigma_{\beta,\beta}^{(0)};\quad f\triangleq W_{\alpha,Q}^{T}\Sigma_{Q,\beta}^{(0)};\quad g\triangleq\Sigma_{\beta,\beta}^{(0)}\,.
\end{eqnarray*}
Since the polynomial is quartic, we can find the optimal weight by
running a standard polynomial solver (the roots() function in MATLAB).
Once the support for the row has the desired sparsity, we include
this penalty in the calculation of the optimal set of weights, so
instead of Eq. \ref{eq: greedy step 3}, we have
\begin{eqnarray}
\hat{W}_{\alpha,Q} & = & \underset{\mathbf{W}_{\alpha,\cdot}:W_{\alpha,Q^{c}}=0}{\mathrm{argmax}}L(\mathbf{W}_{\alpha,\cdot})-\lambda_{V}\sum_{i\in Q}(W_{\alpha,i}-V_{\alpha,i})^{2}\label{eq:modified_W_est}\\
 & = & \left(\Sigma_{\alpha,Q}^{(1)}+2\lambda_{V}V_{\alpha,Q}\right)\left(A_{\alpha,\alpha}\Sigma_{Q,Q}^{(0)}+2\lambda_{V}I\right)^{-1}\,.\nonumber 
\end{eqnarray}

\subsection{Distance-dependent connectivity}

As explained in section \ref{sub:Distance-dependence}, we can also
incorporate into the model a prior on $f(d_{ij})\triangleq P(W_{i,j}\neq0|d_{ij})$,
the probability of a connection between neurons $i,j$ as a function
of the distance between them $d_{i,j}$. To promote the selection
of more probable connections using this prior, we subtract the penalty
$\lambda_{d}/f(d_{\alpha,\beta})$ from $L(W_{\alpha,\beta}|\hat{W}_{\alpha,Q})$
in Eqs. \ref{eq: greedy step 1}-\ref{eq: greedy step 2}, or Eq.
\ref{eq:modified penalty} if we want also to include the stochastic
block model prior. While this penalty modifies the model selection
step of extending the support, it has no effect on the values of the
weights, determined by Eq. \ref{eq:modified_W_est}.

\subsection{Inferring unknown penalty parameters}

In many situations, we will not know what is the true block model
matrix or distance-dependent connectivity function. However, we can
infer these parameters from estimates of the weight matrix, as we
explain next.

\subsubsection{Inferring $\mathbf{V}$ using the soft-impute algorithm}

Recall section \ref{sub:Stochastic-block-model}. We wish to estimate
$\mathbf{V}$ from $\mathbf{\hat{W}}$ by solving (see Eq. \ref{eq: sbm penalty}
and explanation below) 
\begin{equation}
\hat{\mathbf{V}}=\underset{_{\mathbf{V}}}{\mathrm{argmin}}\sum_{i,j}\mathcal{I}\left[\hat{W}_{i,j}\neq0\right]\left(\hat{W}_{i,j}-V_{i,j}\right)^{2}+\lambda_{*}\left\Vert \mathbf{V}\right\Vert _{*}\,,\label{eq: low rank estimation}
\end{equation}
where the last nuclear norm penalty $\lambda_{*}\left\Vert \mathbf{V}\right\Vert _{*}$
promotes a low rank solution $\hat{\mathbf{V}}$. Suppose we know
the target $r_{V}\triangleq\mathrm{rank}\left(\mathbf{V}\right)$,
\emph{i.e.}, the minimal number of neuron types affecting neuronal
connectivity. Then we can use noisy low-rank matrix completion techniques
to estimate $\hat{\mathbf{V}}$ from $\mathbf{\hat{W}}$. Specifically,
we found that soft-impute (\cite{Mazumder2010a}, described here in
Algorithm \ref{alg:Soft-Impute}) works well in this estimation task,
where all components $\left(i,j\right)$ for which $\hat{W}_{i,j}=0$,
are considered unobserved. In the algorithm we iterate over different
values of $\lambda_{*}$, starting with a small value (so initially
$\mathbf{\hat{V}}$ has a larger rank than desired) and slowly incrementing,
until $\hat{\mathbf{V}}$ is of the desired rank $r_{V}$.

\subsubsection{Inferring $f\left(d\right)$}

Recall section \ref{sub:Distance-dependence}. We assume the distance-dependent
connectivity relationship to be of a sigmoidal form
\[
P(W_{i,j}\neq0|d_{ij};a,b)=f(d_{ij};a,b)=\frac{1}{1+\exp(-(ad_{ij}+b))}\,.
\]
If $f$ is unknown, we can run logistic regression on the binarized
inferred weight matrix to estimate $\hat{f}$, \emph{i.e.}
\begin{eqnarray*}
(\hat{a},\hat{b}) & = & \mathrm{argmax}{}_{(a,b)}\prod_{i,j=1,...,N}P(W_{i,j}\neq0|d_{ij};a,b)\\
 & = & \mathrm{argmax}{}_{(a,b)}\prod_{i,j=1,...,N}\frac{\exp(\mathcal{I}\left[\hat{W}_{ij}\neq0\right](ad_{ij}+b))}{1+\exp((ad_{ij}+b))}\\
\hat{f} & = & f(d_{ij};\hat{a},\hat{b})\,.
\end{eqnarray*}

\subsubsection{Full Model}

Finally, we can combine distance-dependent connectivity and a low-rank
mean matrix and infer both $\mathbf{\hat{V}}$ and $\hat{f}$ from
$\hat{\mathbf{W}}$ at the same time. Furthermore, once we have an
estimate for these parameters, we can use them to find a new estimate
$\hat{\mathbf{W}},$ which in turn, allows us to re-estimate both
$\mathbf{\hat{V}}$ and $\hat{f}$. Thus, we can iterate estimating
$\mathbf{\hat{W}}$ and the penalty parameters $\mathbf{\hat{V}}$
and $\hat{f}$ until we reach convergence (Algorithm \ref{alg:Infer_W_unknown_params}).
If our model does not include a low-rank mean matrix or distance-dependent
connectivity, we can just set the corresponding penalty coefficient
to zero. The penalty coefficients $\lambda_{V}$ and $\lambda_{D}$
are selected by finding each penalty coefficient that maximizes the
correlation between $\hat{\mathbf{W}}$ and $\mathbf{W}$ while the
other coefficient is set to 0. On actual data, where the ground truth
is unknown, one may instead maximize prediction of observed spikes
(Figure \ref{fig:Estimating-the-spikes}) or spike correlations \cite{Turaga2013a}.
We found it is also possible to set the regularization parameters
in the case $\mathbf{\hat{V}}$ and $\hat{f}$ are inferred from the
data, to their optimal value in the case where $\mathbf{V}$ and $f$
are known - multiplied by $\frac{1}{2}$ to account for uncertainty. 

\begin{algorithm}[H]
\protect\caption{Infer\_W\_known\_params($\boldsymbol{\Sigma}^{(0)},\boldsymbol{\Sigma}^{(1)},p_{\mathrm{conn}},\lambda_{V},\mathbf{V},\lambda_{d},f,d)$}

Define $L(W_{\alpha,\beta}|W_{\alpha,Q})$ as in Eq. \ref{eq: L(W_a,b|W_a,Q)}.

Initialize solution $\hat{\mathbf{W}}=0^{N\times N}$.

\textbf{For $\alpha=1,...,N$}

\qquad{}Initialize support $Q=\emptyset$

\qquad{}While \textbf{$|Q|/N<p_{\mathrm{conn}}$}

\qquad{}\qquad{}Find the optimal penalized weight index:
\[
\beta^{*}=\underset{\beta\in Q^{C}}{\mathrm{argmax}}\max_{W_{\alpha,\beta}}\left[L(W_{\alpha,\beta}|W_{\alpha,Q})-\lambda_{M}(W_{\alpha,\beta}-V_{\alpha,\beta})^{2}-\frac{\lambda_{d}}{f(d_{\alpha,\beta})}\right]
\]

\qquad{}\qquad{}Update the support: 
\[
Q=Q\cup\beta^{*}
\]

\qquad{}\qquad{}Update the weights: 
\[
\hat{W}_{\alpha,Q}=\left(\Sigma_{\alpha,Q}^{(1)}+2\lambda_{V}V_{\alpha,Q}\right)\left(A_{\alpha,\alpha}\Sigma_{Q,Q}^{(0)}+2\lambda_{V}I\right)^{-1}
\]

\textbf{Return }$\hat{\mathbf{W}}$
\end{algorithm}

\begin{algorithm}[H]
\protect\caption{Soft-Impute($\mathbf{A},r)$\label{alg:Soft-Impute}}

Define $S_{\lambda}(\mathbf{A})\triangleq\mathbf{U}\mathbf{D}_{\lambda}\mathbf{V}^{T}\,,$
where $\mathbf{UD}\mathbf{V}^{T}$ being the singular value decomposition
of $\mathbf{A}$, with $\mathbf{D}=\text{diag}(d_{1},...,d_{r})$,
and $\mathbf{D}_{\lambda}\triangleq\text{diag}((d_{1}-\lambda)_{+},...,(d_{r}-\lambda)_{+})),$
with $(x)_{+}=\max\left(0,x\right)$. 

Initialize $\mathbf{Z}=\boldsymbol{0}^{N\times N}$, and some decreasing
set \textbf{$\lambda_{1}>\lambda_{2}>...>\lambda_{k}$.}

\textbf{For $\lambda_{1}>\lambda_{2}>...>\lambda_{k}$}

\qquad{}$\forall i,j:$ $Y_{i,j}=A_{i,j}+\mathcal{I}\left[A_{i,j}=0\right]Z_{i,j}\,.$

\qquad{}$\mathbf{Z}=S_{\lambda_{k}}(\mathbf{Y})$

\qquad{}\textbf{If $\mathrm{rank}\left(\mathbf{Z}\right)=r$, return
$\mathbf{Z}$}

\textbf{End}
\end{algorithm}

\begin{algorithm}[H]
\protect\caption{Infer\_W\_unknown\_params($\boldsymbol{\Sigma}^{(0)},\boldsymbol{\Sigma}^{(1)},d,p_{\mathrm{conn}},\lambda_{V},\lambda_{d},r_{V})$\label{alg:Infer_W_unknown_params}}

$\hat{f}^{0}(d)=\infty$

$\mathbf{\hat{V}}^{0}=\boldsymbol{0}^{N\times N}$

\textbf{For $t=1,...,k$}

\qquad{}$\hat{\mathbf{W}}^{t}=$Infer\_W\_known\_params($\Sigma^{(0)},\Sigma^{(1)},p_{\mathrm{conn}},\lambda_{M},\hat{\mathbf{V}}^{t-1},\lambda_{d},\hat{f}^{t-1},d)$

\qquad{}$(\hat{a},\hat{b})=$$\mathrm{argmax}{}_{(a,b)}\prod_{i,j=1,...,N}\frac{\exp(\mathcal{I}\left[\hat{W}_{ij}\neq0\right](ad_{ij}+b))}{1+\exp((ad_{ij}+b))}$

\qquad{}$\hat{f}^{t}(d)=\frac{1}{1+\exp(-(\hat{a}d+\hat{b}))}$

\qquad{}$\mathbf{\hat{V}}^{t}$=Soft-Impute($\hat{\mathbf{W}^{t}},r_{V}$)

\textbf{Return $\hat{\mathbf{W}^{t}}$}
\end{algorithm}

\section{An MCMC approach for inferring connectivity\label{sec:MCMC-Approach}}

In this section we give the details of the MCMC approach for inferring
the weights (summarized in section \ref{sub:Markov-chain-Monte}).
To do this we alternate between sampling the spikes (section \ref{sub:Sampling-the-spikes}),
and sampling the weights (section \ref{sub:Sampling-the-weights}). 

First, recall again Eq. \ref{eq: logistic} 
\[
P\left(S_{i,t}|U_{i,t}\right)=\frac{e^{S_{i,t}U_{i,t}}}{1+e^{U_{i,t}}}\,.
\]
and Eq. \ref{eq: U} (with $\mathbf{G}=0$) 
\[
\mathbf{U}_{\cdot,t}\triangleq\mathbf{W}\mathbf{S}_{\cdot,t-1}+\mathbf{b}\,.
\]

\subsection{Sampling the spikes\label{sub:Sampling-the-spikes}}

We denote here $S_{/\left(i,t\right)}$ to be all the component of
$\mathbf{S}$ without the $S_{i,t}$ component. In order to do Gibbs
sampling, we need to calculate

\[
\ln P\left(S_{i,t}|S_{/\left(i,t\right)},\mathbf{W},\mathbf{b}\right)=\ln P\left(S_{i,t}|\mathbf{S}_{\cdot,t-1},\mathbf{W},\mathbf{b}\right)+\ln P\left(\mathbf{S}_{\cdot,t+1}|\mathbf{S}_{t},\mathbf{W},\mathbf{b}\right)+C\,,
\]
 where we can neglect any additive constant that does not depend on
$S_{i,t}$. On the right hand side, the first term is 
\begin{eqnarray*}
\ln P\left(S_{i,t}|\mathbf{S}_{\cdot,t-1},\mathbf{W},\mathbf{b}\right) & = & \ln\left[\frac{e^{S_{i,t}U_{i,t}}}{1+e^{U_{i,t}}}\right]\\
 & = & S_{i,t}U_{i,t}+C\,,\\
 & = & S_{i,t}\left(\sum_{k=1}^{N}W_{i,k}S_{k,t-1}+b_{i}\right)\,,
\end{eqnarray*}
while the second term is
\begin{eqnarray*}
 &  & \ln P\left(\mathbf{S}_{\cdot,t+1}|\mathbf{S}_{\cdot,t},\mathbf{W},\mathbf{b}\right)\\
 & = & \sum_{j}\ln\left[\frac{e^{S_{j,t+1}U_{j,t+1}}}{1+e^{U_{j,t+1}}}\right]\\
 & = & \sum_{j}\left[S_{j,t+1}U_{j,t+1}-\ln\left(1+e^{U_{j,t+1}}\right)\right]\\
 & = & \sum_{j}\left[S_{j,t+1}\left(\sum_{k=1}^{N}W_{j,k}S_{k,t}+b_{j}\right)-\ln\left(1+\exp\left(\sum_{k=1}^{N}W_{j,k}S_{k,t}+b_{j}\right)\right)\right]+C\\
 & = & C+\sum_{j}S_{j,t+1}S_{i,t}W_{j,i}\\
 & - & \sum_{j}\left[\ln\left(1+\exp\left(\sum_{k\neq i}^{N}W_{j,k}S_{k,t}+b_{j}+W_{j,i}\right)\right)-\ln\left(1+\exp\left(\sum_{k\neq i}^{N}W_{j,k}S_{k,t}+b_{j}\right)\right)\right]S_{i,t}\,.
\end{eqnarray*}
Therefore, we can sample the spikes from 
\begin{eqnarray*}
P\left(S_{i,t}|S_{/\left(i,t\right)},\mathbf{W},\mathbf{b}\right) & \propto & \exp\left(\alpha_{i,t}S_{i,t}\right)\,,
\end{eqnarray*}
where
\begin{eqnarray*}
\alpha_{i,t} & \triangleq & b_{i}+\sum_{k=1}^{N}W_{i,k}S_{k,t-1}\\
 & + & \sum_{j}\left[S_{j,t+1}W_{j,i}-\ln\left(1+\exp\left(\sum_{k\neq i}^{N}W_{j,k}S_{k,t}+b_{j}+W_{j,i}\right)+\ln\left(1+\exp\left(\sum_{k\neq i}^{N}W_{j,k}S_{k,t}+b_{j}\right)\right)\right)\right].
\end{eqnarray*}
Note that, for a given $i$, $\alpha_{i,t}$ depends only on spikes
from time $t-1$ and $t+1$. Therefore, $S_{i,t}$ samples generated
at odd times $t$ are independent from samples $S_{i,t^{\prime}}$
generated at even times $t'$. Therefore, we can sample $S_{i,t}$
simultaneously for all odd times $t$, and then sample simultaneously
at all even times $t$. Such a simple block-wise Gibbs sampling scheme
can be further accelerated by using the Metropolized Gibbs method
\cite{Liu1996}, in which we propose a ``flip'' of our previous
sample. So if $S_{i,t}$ is out previous sample and $S_{i,t}^{\prime}$
is our new sample, we propose that $S_{i,t}^{\prime}=1-S_{i,t}$ and
then accept this proposal with probability 
\[
\min\left(1,\frac{1-P\left(S_{i,t}|S_{/\left(i,t\right)},\mathbf{W},\mathbf{b}\right)}{1-P\left(S_{i,t}^{\prime}|S_{/\left(i,t\right)},\mathbf{W},\mathbf{b}\right)}\right)\,.
\]
If the proposal is not accepted, we keep our previous sample $S_{i,t}$.

\subsection{Sampling the weights\label{sub:Sampling-the-weights}}

We denote here $W_{/\left(i,j\right)}$ to be all the components of
$\mathbf{W}$ without the $W_{i,j}$ component, and 
\[
f\left(x\right)\triangleq\frac{1}{1+e^{-x}}\,.
\]
In order to do Gibbs sampling, we need to calculate
\[
\ln P\left(W_{i,j}|\mathbf{S},W_{/\left(i,j\right)},\mathbf{b}\right)=\ln P\left(\mathbf{S}|\mathbf{W},\mathbf{b}\right)+\ln P_{0}\left(W_{i,j}\right)+C\,,
\]
where, as before, we can neglect on the right hand side any additive
constant that does not depend on $W_{i,j}$. The first term on the
right hand side is 
\begin{eqnarray*}
 &  & \ln P\left(\mathbf{S}|\mathbf{W},\mathbf{b}\right)\\
 & = & \sum_{i}\sum_{t}\ln\left[\frac{e^{S_{i,t}U_{i,t}}}{1+e^{U_{i,t}}}\right]\\
 & = & \sum_{i}\sum_{t}\left[S_{i,t}U_{i,t}-\ln\left(1+e^{U_{i,t}}\right)\right],\\
 & = & \sum_{i}\sum_{t}\left[W_{i,j}S_{i,t}S_{j,t-1}-\ln\left(1+\exp\left(\sum_{k=1}^{N}W_{i,k}S_{k,t-1}+b_{i}\right)\right)\right]+C\\
 & = & \sum_{i}\sum_{t}\left[W_{i,j}S_{i,t}S_{j,t-1}-\ln\left(1+\exp\left(W_{i,j}S_{j,t-1}\right)\exp\left(\sum_{k\neq j}^{N}W_{i,k}S_{k,t-1}+b_{i}\right)\right)\right]+C\\
 & \approx & \sum_{i}\sum_{t}\left[W_{i,j}S_{i,t}S_{j,t-1}-\ln\left(1+\left(1+W_{i,j}S_{j,t-1}+\frac{1}{2}W_{i,j}^{2}S_{j,t-1}\right)\exp\left(\sum_{k\neq j}^{N}W_{i,k}S_{k,t-1}+b_{i}\right)\right)\right]+C\\
 & = & \sum_{i}\sum_{t}\left[W_{i,j}S_{i,t}S_{j,t-1}-\ln\left(1+f\left(\sum_{k\neq j}^{N}W_{i,k}S_{k,t-1}+b_{i}\right)\left(W_{i,j}S_{j,t-1}+\frac{1}{2}W_{i,j}^{2}S_{j,t-1}\right)\right)\right]+C\\
 & \approx & \sum_{i}\sum_{t}\left[W_{i,j}S_{j,t-1}\left[S_{i,t}-f\left(\sum_{k\neq j}^{N}W_{i,k}S_{k,t-1}+b_{i}\right)\right]\right.\\
 & - & \left.\frac{1}{2}W_{i,j}^{2}S_{j,t-1}\left[f\left(\sum_{k\neq j}^{N}W_{i,k}S_{k,t-1}+b_{i}\right)-f^{2}\left(\sum_{k\neq j}^{N}W_{i,k}S_{k,t-1}+b_{i}\right)\right]\right]\,,
\end{eqnarray*}
where in both approximations we used the fact that a single weight
is typically small $W_{i,j}\ll1$. Therefore, denoting

\begin{eqnarray*}
\omega_{i,j} & \triangleq & \sum_{t}\left[S_{i,t}-f\left(\sum_{k\neq j}^{N}W_{i,k}S_{k,t-1}+b_{i}\right)\right]S_{j,t-1}\\
\epsilon_{i,j} & \triangleq & \sum_{t}\left[f\left(\sum_{k\neq j}^{N}W_{i,k}S_{k,t-1}+b_{i}\right)-f^{2}\left(\sum_{k\neq j}^{N}W_{i,k}S_{k,t-1}+b_{i}\right)\right]S_{j,t-1}\,,
\end{eqnarray*}
we can write 
\[
P\left(\mathbf{S}|\mathbf{W},\mathbf{b}\right)\propto\exp\left(W_{i,j}\omega_{i,j}-\frac{1}{2}W_{i,j}^{2}\epsilon_{i,j}\right)\,.
\]
Therefore, 
\[
P\left(W_{i,j}|\mathbf{S},W_{/\left(i,j\right)},\mathbf{b}\right)\propto\exp\left(W_{i,j}\omega_{i,j}-\frac{1}{2}W_{i,j}^{2}\epsilon_{i,j}\right)P_{0}\left(W_{i,j}\right)\,.
\]
Assuming spike-and-slab prior

\begin{equation}
P_{0}\left(W_{i,j}\right)=f\left(-h_{0}\right)\delta\left(W_{i,j}\right)+f\left(h_{0}\right)\mathcal{N}\left(W_{i,j}|\mu_{0},\sigma_{0}^{2}\right)\,,\label{eq:prior-1}
\end{equation}
we can use standard Gaussian completion we do spike-and-slab completion
- re-normalizing to obtain a proper spike-and-slab distribution. This
gives

\begin{equation}
P\left(W_{i,j}|\mathbf{S},W_{/\left(i,j\right)},\mathbf{b}\right)=f\left(-h_{i,j}\right)\delta\left(W_{i,j}\right)+f\left(h_{i,j}\right)\mathcal{N}\left(W_{i,j}|\mu_{i,j},\sigma_{i,j}^{2}\right)\,,\label{eq: W_ij distribution for Gibbs}
\end{equation}
with 
\begin{eqnarray}
\sigma_{i,j}^{2} & \triangleq & \frac{\sigma_{0}^{2}}{1+\epsilon_{i,j}\sigma_{0}^{2}}\\
\mu_{i,j} & \triangleq & \frac{\sigma_{i,j}^{2}}{\sigma_{0}^{2}}\left[\mu_{0}+\sigma_{0}^{2}\omega_{i,j}\right]\\
h_{i,j} & \triangleq & h_{0}+\frac{1}{2}\ln\left(\frac{\sigma_{i,j}^{2}}{\sigma_{0}^{2}}\right)+\frac{\mu_{i,j}^{2}}{2\sigma_{i,j}^{2}}-\frac{\mu_{0}^{2}}{2\sigma_{0}^{2}}\,.
\end{eqnarray}
We can than proceed and sample $\mathbf{W}_{\cdot,j}$ from this spike-and-slab
distribution (in Eq. \ref{eq: W_ij distribution for Gibbs}) - sampling
$W_{i,j}$ simultaneously for all $i$. Now, since we used the approximation
that assuming the weights are weak, so this sampling is not exact.
Therefore, even if the approximation is very good, we can not use
direct sampling, or the error will accumulate over time catastrophically.
To correct his, we use this approximation as a proposal distribution
in a Metropolis Hastings simulation scheme \cite{Liu02}. 

\newpage{}
\end{document}